\documentclass[a4,12pt]{article}

\usepackage{theorem,amsmath,enumerate,amsfonts,xspace}
\usepackage{latexsym}
\usepackage{amssymb,amsbsy,amsmath,amsfonts}
\usepackage{graphicx}
\usepackage{color}
\usepackage[font=small]{caption, subfig}
\usepackage{natbib}
\usepackage{JASA_manu}

\newcommand{\patha}{./}
\newcommand{\pathb}{./}

\newcommand{\subfigure}{\subfloat}
\newcommand{\half}{\frac 1 2}
\newcommand{\tr}{\mbox{tr}}
\newcommand{\diag}{\mbox{diag}}
\newcommand{\Var}{\mbox{Var}}
\newcommand{\Cov}{\mbox{Cov}}
\newcommand{\E}{\mbox{E}}
\newcommand{\logit}{\mbox{logit}}
\newcommand{\ve}[1]{\mbox{\boldmath ${#1}$}}
\newcommand{\vesub}[2]{\mbox{{\boldmath ${#1}$}$_{#2}$}}
\newcommand{\vesup}[2]{\mbox{{\boldmath ${#1}$}$^{#2}$}}
\newcommand{\vess}[3]{\mbox{{\boldmath ${#1}$}$_{#2}^{#3}$}}
\newcommand{\hve}[1]{\hat{\ve{#1}}}

\newcommand{\matrixbb}[4]{\left[ \begin{array}{cc} #1 & #2 \\ #3 & #4 \end{array} \right]}

\newcommand{\CD}{\mathcal D}
\newcommand{\CT}{\mathcal T}
\newcommand{\D}{\mbox{${\cal D}$}}

\newcommand{\CF}{\mathcal F}
\newcommand{\CN}{N}
\newcommand{\cmu}{\mathcal U}

\newcommand{\jian}{}
\newcommand{\hC}{\boldsymbol{\hat C}_{nn}}

\begin{document}

\title{Generalized Gaussian Process Regression Model for Non-Gaussian Functional Data}

\author{Bo Wang \\
Department of Mathematics \\
University of Leicester \\
Leicester LE1 7RH, UK \\
E-mail: \texttt{bw77@leicester.ac.uk} \\ \\
Jian Qing Shi\\
School of Mathematics and Statistics  \\
Newcastle University  \\
Newcastle NE1 7RU, UK  \\
E-mail: \texttt{j.q.shi@ncl.ac.uk}
} 
\date{}
\maketitle

\newpage
\begin{center}
\textbf{Author's Footnote}
\end{center}
B. Wang is Lecturer in Statistics, Department of Mathematics, University of Leicester, Leicester LE1 7RH, UK (e-mail: \texttt{bw77@leicester.ac.uk}). J. Q. Shi is Reader in Statistics, School of Mathematics and Statistics, Newcastle University, Newcastle NE1 7RU, UK (e-mail: \texttt{j.q.shi@ncl.ac.uk}). The authors thank the Associate Editor and the reviewers for their constructive suggestions and helpful comments.

\newpage
\setcounter{page}{1}

\begin{abstract} In this paper we propose a generalized Gaussian process \jian{concurrent} regression model for functional data  where the functional response variable has a binomial, Poisson or other non-Gaussian distribution from an exponential family while the covariates are mixed functional and scalar variables. The proposed model offers a nonparametric generalized \jian{concurrent} regression method for functional data with multi-dimensional covariates, and provides a natural framework on modeling common mean structure and covariance structure simultaneously  for repeatedly observed functional data. The mean structure provides an overall information about the observations, while the covariance structure can be used to catch up the characteristic of each individual batch. The prior specification of covariance kernel enables us to accommodate a wide class of nonlinear models. The definition of the model, the inference and the implementation as well as its asymptotic properties are discussed. Several numerical examples with different non-Gaussian response variables are presented. Some technical details and more numerical examples as well as an extension of the model are provided as supplementary materials.

\bigskip

\end{abstract}

\newpage

\section{Introduction}

A functional regression model with functional response variable can be defined by
\begin{equation}
y_m(t)=f(\vesub{x}{m}(t), \vesub{u}{m})+\epsilon_m(t), \label{basic}
\end{equation}
where $y_m(t)$ ($m=1, \ldots, M$) stands for $M$ batches (or curves) of functional data, $f(\cdot)$ is an unknown nonlinear function, depending on a set of functional covariates $\vesub{x}{m}(t)$ and a set of scalar covariates $\vesub{u}{m}$, and $\epsilon_m(t)$ is the random error. A special case of such model is the following concurrent regression model with functional covariates $\vesub{x}{m}(t)$ \citep[see e.g.][]{Ramsay05}
\[
y_m(t)=\vess{x}{m}{T}(t) \ve{\beta}(t) +\epsilon_m(t). 
\]
However, when the relationship between the response and the covariates cannot be justified as linear, it is intractable to model the function $f(\cdot)$ nonparametrically for multi-dimensional $\vesub{x}{m}(t)$ since most nonparametric regression models suffer from the \emph{curse of dimensionality}. A variety of alternative approaches with special model structures have been proposed to overcome the problem; examples include dimension reduction methods, the additive model \citep[see e.g.][]{Hastie90}, varying-coefficient model \citep[see e.g.][]{Fan00, Fan03, Senturk08}, and the neural network model \citep[see e.g.][]{Cheng94}. \cite{Shi07g} proposed a Gaussian process functional regression (GPFR) model, which is defined by
\begin{equation}
f(\vesub{x}{m}(t), \vesub{u}{m})=\mu_m(t)+\tau_m(\vesub{x}{m}(t)), \label{gpfr}
\end{equation}  
where $\mu_m(t)$ is the mean structure of the functional data and $\tau_m(\vesub{x}{m}(t))$ represents a Gaussian process regression (GPR) model having zero mean and covariance kernel $k(\cdot,\cdot; \ve \theta)$ \citep[for the detailed definition of Gaussian process regression models, see][]{Rasmussen06, Shi11}. This nonparametric concurrent functional regression model can address the regression problem with multi-dimensional functional covariates and model the mean structure and covariance structure simultaneously; see the detailed discussion in \cite{Shi07g}. 

The aim of this paper is to extend the concurrent GPFR model \eqref{gpfr}  to situations where the response variable, denoted by $z(t)$, is known to be non-Gaussian. The work is motivated by the following example, concerning data collected during standing-up manoeuvres of paraplegic patients. The outputs are the human body's standing-up phases during rising from sitting position to standing position. Specifically, $z(t)$ takes value of either 0, 1 or 2, corresponding to the phases of `sitting', `seat unloading and ascending' or `stablising' respectively, required for feeding back to a simulator control system. Since it is usually difficult to measure the body position in practice, the aim of the example is to develop a model for reconstructing the position of the human body by using some easily measured quantities such as motion kinematic, reaction forces and torques, which are functional covariates denoted by $\ve{x}(t)$. This is to investigate the regression  relationship between the non-Gaussian functional response variable $z(t)$ and a set of functional covariates $\ve{x}(t)$. Since the standing-up phases are irreversible, $z(t)$ is an ordinal response variable, taking value from three ordered categories. If we assume that there exists an unobservable latent process 
$\eta(t)$ associated with $\ve{x}(t)$ and the response variable $z(t)$ depends on this latent process, then by using a probit link function, we can define a model as follows:
$$z(t)=j \;\;\mbox{ if } b_{j} < \eta(t) \leq b_{j+1}, \quad j\in \{0,1,2\}, $$
where $b_0=-\infty$, $b_3=\infty$, and $b_1,b_2\in\mathbb{R}$ are the thresholds. Now the problem becomes how to model $\eta(\cdot)$ by the functional covariates $\ve{x}(t)$, or how to find a function $f$ such that $\eta(t)=f(\ve{x}(t))$.  More discussion of this example is given in Section \ref{para_data} \jian{and Appendix G of the supplementary materials}. 

Generally, letting $h^{-1}(\cdot)$ be a given link function, a generalized linear regression model is defined as
$E(z_m(t))=h(\vess{x}{m}{T}(t) \ve{\beta})$.
\cite{Breslow93} proposed a generalized linear mixed model to  deal with heterogeneity:
$
E(z_m(t)|\vesub{\gamma}{2})=h(\vess{x}{m1}{T}(t) \vesub{\gamma}{1}+ \vess{x}{m2}{T}(t)
\vesub{\gamma}{2}),
$ 
where $\vesub{\gamma}{1}$ is the coefficient for the fixed effect and $\vesub{\gamma}{2}$ is a random vector representing random effect. 
However, if we have little practical knowledge on the relationship between the response variable and the covariates (such as the case in the above Paraplegia example), it is more sensible to use a nonparametric model. In this paper, we propose to use a Gaussian process regression model to define such a nonparametric model, namely  a concurrent generalized Gaussian process functional regression (GGPFR) model. 
Similar to GPFR model \citep{Shi07g}, the advantages of this model include: (1) it offers a nonparametric generalized concurrent regression model for functional data with functional response and multi-dimensional functional covariates; (2) it provides a natural framework on modeling mean structure and covariance structure simultaneously and the latter can be used to model the individual characteristic for each batch; and (3) the prior specification of covariance kernel enables us to accommodate a wide class of nonlinear functions. 

This paper is organized as follows. Section 2 proposes the GGPFR model and describes how to estimate the hyper-parameters and how to calculate prediction, for which the implementation is mainly based on Laplace approximation. The asymptotic properties, focusing on the information consistency, are discussed in Section 3.  Several numerical examples are reported in Section 4. Discussion and further development are given in Section 5.  \jian{Some technical details and more numerical examples are provided as the supplementary materials.}

\section{Generalized Gaussian process functional regression model}

\subsection{The Model}

Let $\{z_m(t), t\in \CT\}$ be a functional or longitudinal response variable for the $m$-th subject, namely the $m$-th batch. We assume that $z_m(t)$'s are independent for different batches $m=1, \ldots, M$, but within the batch, $z_m(t_i)$ and $z_m(t_j)$ are dependent at different   points.  We suppose that $z_m(t)$ has a distribution from an exponential family with the following density function
\begin{equation}
p(z_m(t)|\alpha_m(t), \phi_m(t)) = \exp \left \{ \frac{z_m(t) \alpha_m(t) -
b(\alpha_m(t))}{a(\phi_m(t))} + c(z_m(t), \phi_m(t)) \right \} \label{exfamily}
\end{equation}
where $\alpha_m(t)$ and $\phi_m(t)$ are canonical parameter and
dispersion parameter respectively, both functional. We
have $\E(z_m(t))=b'(\alpha_m(t))$ and
$\Var(z_m(t))=b''(\alpha_m(t))a(\phi_m(t))$, where $b'(\alpha)$ and
$b''(\alpha)$ are the first two derivatives of $b(\alpha)$ with respect to $\alpha$.  

Suppose that $\vesub{x}{m}(t)$ is a $Q$-dimensional vector of functional
covariates. Nonparametric concurrent generalized
Gaussian process functional regression (GGPFR) 
models are defined by \eqref{exfamily} and the following
\begin{eqnarray}
\E(z_m(t)|\tau_m(t))& =& h(\mu_m(t)+\tau_m(t)), \nonumber \\ 
\tau_m(t)= \tau_m(\vesub{x}{m}(t)) & \sim & GPR(0, k(\cdot,\cdot;\ve \theta)|\vesub{x}{m}(t)).  \label{ggpfr}
\end{eqnarray}
Here, the unobserved latent variable $\tau_m(t)$ is modeled by a nonparametric GPR model via a Gaussian process prior, depending on the functional covariates $\vesub{x}{m}(t)$. The GPR model is specified by a covariance kernel $k(\cdot,\cdot;\ve \theta)$, and
by the Karhunen-Lo\`eve expansion
$$\tau(\ve{x}) = \sum_{j=1}^{\infty} r_j\psi_j(\ve{x}), $$
where $r_j\sim N(0,\lambda_j)$, $\lambda_1\geq \lambda_2 \geq \cdots \geq 0$ are the eigenvalues and 
$\psi_1(\ve{x}), \psi_2(\ve{x}), \ldots$ are the associated eigenfunctions of the covariance kernel.
One example of $k(\cdot,\cdot;\ve \theta)$ is the following squared exponential covariance 
function with a nonstationary linear term:
\begin{align}
&\Cov(\tau(t_i),\tau(t_j)) = k(\ve{x}(t_i),\ve{x}(t_j); \ve \theta) \nonumber \\
= & v_1\exp\Big(-\half\sum^Q_{q=1}
w_q(x_{q}(t_i)-x_{q}(t_j))^2\Big) +
a_1\sum^Q_{q=1}x_{q}(t_i)x_{q}(t_j) , 
\label{covfun0}
\end{align}
where $\ve{\theta} =(w_1,\ldots,w_Q,v_1,a_1)$ is a set of hyper-parameters involved in the Gaussian process prior. The hyper-parameter $w_q$ corresponds to the smoothing parameters in spline and other nonparametric models. More specifically, $w_q^{-1}$ is called the length-scale. The decrease in length-scale produces more rapidly fluctuating functions and a very large length-scale means that the underlying curve is expected to be essentially flat. More information on the relationship between smoothing splines and Gaussian processes can be found in \cite{Seeger99}.   We can use generalized cross-validation (GCV) or empirical Bayesian method to choose the value of $\ve \theta$. When $Q$ is large, GCV approach is usually inefficient. We will use the empirical Bayesian method in this paper; the details are given in the next subsection. Some other covariance kernels such as powered exponential and Mat\'{e}rn covariance functions can also be used; see more discussion on the choice of covariance function in \cite{Rasmussen06} and \cite{Shi11}.

In the model given by \eqref{ggpfr} the response variable $z_m(t)$ depends on $\vesub{x}{m}(t)$ at the current time only, therefore the proposed model can be regarded as a generalization of the concurrent functional linear model discussed in \cite{Ramsay05}. In this model the common mean structure across $M$ batches is given by $\mu_m(t)$. If we use a linear mean function which depends on a set of $p$ scalar covariates $\vesub{u}{m}$ only, \eqref{ggpfr} can be expressed as 
\begin{equation}
\E(z_m(t)|\tau_m(t))=h(\mu_m(t)+\tau_m(t))=h(\vess{u}{m}{T} \ve{\beta}(t) +
\tau_m(t)). \label{gegpfr}
\end{equation}
In this case the regression relationship between the functional response $z_m(t)$ and the functional covariates $\vesub{x}{m}(t)$ is 
modeled by the covariance structure $\tau_m(\vesub{x}{m})$. 
Other mean structures, including concurrent form of functional covariates, can also be used. 

The proposed model has some features worth noting. In addition to those discussed in Section 1, we highlight that the GGPFR model is actually very flexible. It can model the regression relationship between the non-Gaussian functional response and the multi-dimensional functional covariates nonparametrically.  Moreover, if we had known some prior information between $z_m(t)$ (or $\E(z_m(t))$) and some of the functional covariates, we could easily integrate it by adding a parametric mean part. For example we may define 
\[
\mu_m(t)= \vess{u}{m}{T} \ve{\beta}(t)+\vess{x}{m1}{T}(t) \vesub{\gamma}{1}+ \vess{x}{m2}{T}(t)\vesub{\gamma}{2},
\]
 i.e. including a term in the GGPFR similar to the generalized linear mixed model \citep{Breslow93}\jian{; an example of such models is provided in Appendix G}. The nonparametric part can still be modeled by $\tau_m(t)$ via a GPR model. Other nonparametric covariance structure can also be considered; some examples can be found in \cite{Rice91}, \cite{Hall08} and \cite{Leng09}.  However, most of these methods are limited to small (usually one) dimensional $\ve x(t)$ or the covariance matrix with a special structure. 

As an example of the GGPFR model, we consider a special case of binary data (e.g. for classification problem with two classes). In this case, $z_m(t) \sim Bin(1, \pi_m(t))$. If we use the logit link function, the density function is given by
\begin{equation}
p(z_m(t)|\tau_m(t)) =  \frac{\exp \{ [\vess{u}{m}{T} \ve{\beta}(t)+\tau_m(t)] z_m(t) \} } {
1+ \exp \{\vess{u}{m}{T} \ve{\beta}(t)+\tau_m(t) \} }.
\label{binden}
\end{equation}
The marginal density function
of $z_m(t)$ is therefore given by
\[
p(z_m(t))=\int p(z_m(t)| \tau_m(t), \ve{\beta}(t)) p(\tau_m(t)| \ve\theta) d \tau_m(t),
\]
where $p(\tau_m(t)| \ve \theta)$ is the density function of $\tau_m(t)$, which is
a multivariate normal distribution for any given   points $\{t_1,\ldots,t_N\}$ and depends on the functional covariates $\vesub{x}{m}(t)$ and the
unknown hyper-parameter $\ve{\theta}$. 

The density functions for other distributions from the exponential families can be obtained similarly.  

\subsection{Empirical Bayesian Learning}\label{mle_laplace}

Now suppose that we have $M$ batches of data from $M$ subjects or experimental units. In
the $m$-th batch, $N_m$ observations are collected at
$\vesub{T}{m}=\{t_{m1},\ldots,t_{mN_m}\}$. We denote $z_m(t_{mi})$,
$\tau_m(t_{mi})$ and $\ve{x}_m(t_{mi})$ by $z_{mi}$, $\tau_{mi}$ and
$\ve{x}_{mi}$, respectively, for $i=1,\ldots,N_m$ and
$m=1,\ldots,M$. Collectively, we denote
$\vesub{Z}{m}=(z_{m1},\ldots,z_{mN_m})^T$ and
$\ve{Z}=\{\vesub{Z}{1},\ldots,\vesub{Z}{M}\}$,
and denote $\ve{\tau}_{m}$, $\ve{\tau}$, $\vesub{X}{m}$ and
$\ve{X}$ in the same way. They are the realizations of $z_m(t)$, $\tau_m(t)$ and
$\ve{x}_m(t)$ at $\vesub{T}{m}$. A discrete GGPFR model is therefore given by
\begin{eqnarray}
z_{mi}|\alpha_{mi}, \phi, \tau_{mi} & \sim &  \ EF(\alpha_{mi}, \phi),\quad i=1, \ldots, N_m, \label{exmi} \\
\E(z_{mi}|\tau_{mi})& = & b'(\alpha_{mi})=h(\vess{u}{m}{T}\ve{\beta}(t_i) + \tau_{mi}),  \label{hmi} \\
\vesub{\tau}{m} & = & (\tau_{m1}, \ldots, \tau_{mN_m})^T \ \sim \ N(0,\vesub{C}{m})     \label{gpr}
\end{eqnarray}
for $m=1, \ldots, M$, where $EF(\cdot, \cdot)$ is a distribution from the exponential
family \eqref{exfamily} and $\alpha_{mi}=\alpha_m(t_{i})$. $\vesub{\tau}{m}$ has an $N_m$-variate normal distribution with zero mean and covariance matrix
$\vesub{C}{m}=(C^{ij}_{m})$ for $i,j=1,\ldots,N_m$.
Here we assume a fixed dispersion parameter $\phi$, but the method
developed in this paper can be applied to more general cases.

 We consider the estimation of $\ve{\beta}(t)$ first. To estimate the
functional coefficient $\ve{\beta}(t)$, we expand it by a set of
basis functions \citep[see e.g.][]{Ramsay05}. In this paper, we use B-spline
approximation. Let
$\ve{\Phi}(t)=(\Phi_1(t), \ldots, \Phi_D(t))^T$ be the B-spline
basis functions, then the functional coefficient $\ve{\beta}(t)$
can be represented as $\ve{\beta}(t)=\vesup{B}{T} \ve{\Phi}(t)$,
where the $j$-th column of $\ve{B}$, $B_j=(B_1^j,\ldots,B_D^j)^T$, is the
B-spline coefficients for $\beta_j(t)$. Thus, at the observation  
point $\vesub{T}{m}$, we have
$\vesub{\mu}{m} = \vesub{\Phi}{m}\ve{B} \vesub{u}{m}$, where
$\vesub{\Phi}{m}$ is an $N_m \times D$ matrix with the $(i,d)$-th
element $\Phi_d(t_{mi})$. In practice, the performance of the B-spline approximation may
strongly depend on the choice of the knots locations and the number of basis functions. 
There are three widely-used methods for locating the knots: equally spaced method, equally
spaced sample quantiles method and model selection based method. 
{The guidance on which method is to use in different situations can be found in \cite{Wu2006}. 
The first method is used in our numerical examples in Section 4 which all have equally-spaced time points 
and the second is adopted in the PBC data in the supplementary materials.} 
The number of basis functions
can be determined by generalized cross-validation or AIC (BIC) methods. 
More details on this issue can be found in \cite{Wu2006}.

The covariance matrix $\vesub{C}{m}=(C^{ij}_{m})$ of $\vesub{\tau}{m}$ depends on $\vesub{X}{m}$ and the unknown hyper-parameter $\ve\theta$. If we use covariance kernel  (\ref{covfun0}),
its element $C^{ij}_{m}$ is given by
\begin{equation}
C^{ij}_{m} = v_1\exp\Big(-\half\sum^Q_{q=1} w_q(x_{miq}-x_{mjq})^2\Big) + a_1\sum^Q_{q=1}x_{miq}x_{mjq} .
\label{cmij} \end{equation}
The covariance matrix involves the hyper-parameter $\ve{\theta}=\{w_1,\ldots,w_Q,v_1,a_1\}$, whose value is given based on the prior knowledge in conventional Bayesian analysis. As discussed in  \cite{Shi11}, empirical Bayesian learning method is preferable for GPR models
when the dimension of $\ve{\theta}$ is large.  

The idea of empirical Bayesian learning is to choose the value of the hyper-parameter $\ve \theta$ by maximizing the  marginal density function. Thus, $\ve \theta$ as well as the unknown parameter $\ve B$ can be estimated at the same time by maximizing the following marginal density
\begin{eqnarray*}
p(\ve Z|\ve{B},\ve{\theta}, \ve X)&=&\prod^M_{m=1}p(\vesub{Z}{m}|\ve{B},\ve{\theta},\vesub{X}{m}) \nonumber \\
&=& \prod^M_{m=1}\int p(\vesub{Z}{m}|\ve{\tau}_m,\ve{B})p(\ve{\tau}_m|\ve{\theta},\vesub{X}{m})d\ve{\tau}_m   \nonumber \\
&=& \prod^M_{m=1}\int \Big\{ \prod^{N_m}_{i=1}p(z_{mi}|\tau_{mi},\ve{B})\Big\}
p(\ve{\tau}_m|\ve{\theta},\vesub{X}{m})d\ve{\tau}_m  
\end{eqnarray*}
or the marginal log-likelihood 
\begin{align}
& l(\ve{B},\ve{\theta}) = \sum^M_{m=1}\log\{p(\vesub{Z}{m}|\ve{B},\ve{\theta},\vesub{X}{m})\}   \nonumber \\
=& \sum^M_{m=1}\log \int \Big\{ \prod^{N_m}_{i=1}p(z_{mi}|\tau_{mi},\ve{B})\Big\} 
(2\pi)^{-\frac{N_m}{2}}|\vesub{C}{m}|^{-\frac 1 2}\exp\Big\{-\half\ve{\tau}^T_m\vess{C}{m}{-1}\ve{\tau}_m\Big\}
d\ve{\tau}_m, \label{eq:loglike}
\end{align}
where $p(z_{mi}|\tau_{mi},\ve{B})$ is derived from the exponential family as defined in \eqref{exmi}. For binomial distribution, it is given in \eqref{binden}.  Obviously the integral involved in the above marginal density is analytically intractable 
unless $p(z_{mi}|\tau_{mi},\ve{B})$ has a special form such as the density function of normal distribution. One method to address this problem is to use Laplace approximation. We denote
\begin{equation}
\Psi(\ve{\tau}_m) = \sum^{N_m}_{i=1}\log\big\{p(z_{mi}|\tau_{mi},\ve{B})\big\} - \frac{N_m}{2}\log (2\pi) - \half\log|\vesub{C}{m}| -\half \ve{\tau}_m^T\vess{C}{m}{-1}\ve{\tau}_m,
\label{def_Psi}
\end{equation}
then the log-likelihood \eqref{eq:loglike} can be rewritten as
\[
l(\ve{B},\ve{\theta}) = \sum^M_{m=1}\log \int
\exp\{\Psi(\ve{\tau}_m)\}d\ve{\tau}_m.
\]
Let $\ve{\hat\tau}_m$ be the maximiser of $\Psi(\ve{\tau}_m)$,
then by Laplace approximation we have
\begin{equation}
\int \exp\{\Psi(\ve{\tau}_m)\}d\ve{\tau}_m = \exp\Big\{
\Psi(\ve{\hat\tau}_m) + \frac{N_m}{2}\log (2\pi) -
\half\log|\vess{C}{m}{-1}+\vesub{K}{m}| \Big\}, \label{laplace}
\end{equation}
where $\vesub{K}{m}$ is the second order derivative of
$\sum^{N_m}_{i=1}\log\big\{p(z_{mi}|\tau_{mi},\ve{B})\big\}$ with
respect to $\ve{\tau}_m$ and evaluated at $\ve{\hat\tau}_m$ \citep[see,
for example,][]{BNC1989, ES2000}. The procedure of finding the
maximiser $\ve{\hat\tau}_m$ can be carried out by the
Newton-Raphson iterative method and is given in Appendix A of the supplementary materials.

However, as pointed out in Section 4.1 in \cite{Rue09}, the error rate of the approximation \eqref{laplace} may be $O(1)$ since the dimension of $\ve{\tau}_m$ increases with the sample size $N_m$. 
A better method is to approximate $p(\vesub{Z}{m}|\ve{B},\ve{\theta})$ in \eqref{eq:loglike} (here and in the rest of the section the conditioning on $\vesub{X}{m}$ is omitted for simplicity) by   
\begin{equation}
\tilde{p}(\vesub{Z}{m}|\ve \Theta) {\triangleq} \left . \frac{p(\vesub{\tau}{m},\vesub{Z}{m}|\ve \Theta)}{ \tilde{p}_G(\vesub{\tau}{m}|\vesub{Z}{m},\ve \Theta)} \right |_{\vesub{\tau}{m}=\vesub{\tilde\tau}{m}(\Theta)}, 
\label{nest}
\end{equation}
where $\ve \Theta=(\ve B, \ve \theta)$, $\tilde{p}_G(\vesub{\tau}{m}|\vesub{Z}{m},\ve \Theta)$ is the Gaussian approximation to the full conditional density ${p}(\vesub{\tau}{m}|\vesub{Z}{m},\ve \Theta)$, and \vesub{\tilde\tau}{m} is the mode of the full conditional density of $\vesub{\tau}{m}$ for a given $\ve \Theta$. Here, 
\begin{eqnarray*}
p(\vesub{\tau}{m},\vesub{Z}{m}|\ve \Theta) & = & p(\vesub{Z}{m}|\vesub{\tau}{m},\ve \Theta) p(\vesub{\tau}{m}|\ve \Theta) \\
&=& \exp \Big\{ \log p(\vesub{\tau}{m}|\ve \theta) + \sum_{i=1}^{N_m} \log p(z_{mi}| \tau_{mi}, \ve B) \Big\}. 
\end{eqnarray*}
We approximate $g_{mi}(\tau_{mi})\triangleq\log p(z_{mi}| \tau_{mi}, \ve B)$ by Taylor expansion to the second order
\[
g_{mi}(\tau_{mi}) \approx g_{mi}(\tau_{mi}^{(0)}) + a_{mi} \tau_{mi}-\half d_{mi} \tau_{mi}^2, 
\]
where $a_{mi}$ and $d_{mi}$ depend on the first two derivatives of $g_{mi}(\tau_{mi})$ respectively and evaluated at $\tau_{mi}^{(0)}$. Thus, 
\[
p(\vesub{\tau}{m},\vesub{Z}{m}|\ve \Theta) \propto \exp \big\{-\half \vess{\tau}{m}{T} \vess{C}{m}{-1} \vesub{\tau}{m}  -\half \vess{\tau}{m}{T} \vesub{D}{m} \vesub{\tau}{m}  + \vess{a}{m}{T} \vesub{\tau}{m} \big\},
\] 
where $\vesub{D}{m}=\mbox{diag}(d_{m1}, \ldots, d_{mN_m})$ and $\vess{a}{m}{T}=(a_{m1}, \ldots, a_{mN_m})$. 
We can then use the following Fisher scoring algorithm \citep{Fahrmeir01} to find the Gaussian approximation. Starting with $\tau_{mi}^{(0)}$,  the $k$-th iteration is given by
\begin{itemize}
\item[(i)] Find the solution $\vess{\tau}{m}{(k)}$ from 
$
( \vess{C}{m}{-1}+ \vesub{D}{m}) \vess{\tau}{m}{(k)} = \vesub{a}{m},
$
\item[(ii)] Update $\vesub{a}{m}$ and $ \vesub{D}{m}$ using $\vess{\tau}{m}{(k)}$ and repeat (i). 
\end{itemize}
After the process converges, say at $\vesub{\tilde \tau}{m}$, we get the Gaussian approximation $\tilde{p}_G(\vesub{\tau}{m}|\vesub{Z}{m},\ve \Theta)$ which is the density function of the normal distribution 
$N \left ( \vesub{\tilde \tau}{m},  (\vess{C}{m}{-1}+ \vesub{D}{m})^{-1} \right ).
$
We can then calculate $\hve \Theta=(\hve B, \hve\theta)$ by maximizing \eqref{eq:loglike} using the approximation \eqref{nest}. 
  
\subsection{Prediction}

Now we consider two types of prediction problems. First suppose that we have already
observed some data for a subject, say $N$ observations
in the $k$-th batch, and want to obtain prediction at other   points. 
This can be for one of the batches
$1,\ldots,M$ or a completely new one. The observations are denoted by
$\vesub{Z}{k}=\{z_{ki}, i=1, \ldots, N\}$ which are collected at
$\{t_{k1}, \ldots, t_{kN}\}$. The corresponding input vectors are
$\vesub{X}{k}=\{\vesub{x}{k1}, \ldots, \vesub{x}{kN}\}$, and we also know the
subject-based covariate $\vesub{u}{k}$. It is of interest to
predict $z^*$ at a new   point $t^*$ for the $k$-th
subject given the test input $\vesup{x}{*}=\ve{x}_k(t^*)$. Secondly we
will assume there are no data observed from the subject of interest except the subject-based
covariate and want to predict $z^*$ at a new   point $t^*$ with the input $\vesup{x}{*}$. We use ${\mathcal D}$ to denote all
the training data and assume that the model itself has been trained
(i.e. all the unknown parameters have been estimated) by the method
discussed in the previous section. The main purpose in this
section is to calculate $\E(z^*|\D)$ and $\Var(z^*|\D)$, which are
used as the prediction and the predictive variance of $z^*$.

We now consider the first type of prediction.  Let $\tau^*=\tau_k(t^*)$ be the
underlying latent variable at   $t^*$, then $\tau^*$ (for convenience we ignore the subscript) and $\vesub{\tau}{k}=(\tau_{k1},\ldots,\tau_{kN})^T$ satisfy \eqref{gpr}, and the expectation of $z^*$ 
conditional on $\tau^*$ is given by (\ref{hmi}):
\begin{equation}
\E(z^*|\tau^*,\D) = h(\vess{u}{k}{T}\hve{B}^T\ve{\Phi}(t^*)+\tau^*). \label{cmean}
\end{equation}
It follows that
\begin{equation}
\E(z^*|\D) = \E \big[\E(z^*|\tau^*,\D)\big] = \int
h(\vess{u}{k}{T}\hve{B}^T\ve{\Phi}(t^*) + \tau^*) p(\tau^*|\D)
d\tau^*.  
 \label{pred_mean0}
\end{equation}
A simple method to calculate the above expectation is to approximate $p(\tau^*|\D)$ using a Gaussian approximation $\tilde{p}_G(\vesub{\tau}{k}|\D)$ as discussed around equation \eqref{nest}, that is,
\begin{equation}
p(\tau^*|\D) = \int p(\tau^*|\vesub{\tau}{k},\D) {p}(\vesub{\tau}{k}|\D) d \vesub{\tau}{k}  \approx \int p(\tau^*|\vesub{\tau}{k}) \tilde{p}_G(\vesub{\tau}{k}|\D) d \vesub{\tau}{k}. 
\label{tausd} \end{equation}
Since it is assumed that both $\vesub{\tau}{k}$ and $\tau^*$ come from the same Gaussian process with covariance kernel $k(\cdot, \cdot; \ve \theta)$, we have
\[
(\vess{\tau}{k}{T}, \tau^*)^T  \sim N\left ( \ve{0}, \vesub{C}{N+1, N+1} \right ), \ \ 
\vesub{C}{N+1, N+1} = \matrixbb{\vesub{C}{N,N}}{\vess{C}{N}{*}}{\vess{C}{N}{*T}}{k(\vesup{x}{*}, \vesup{x}{*};\hat{\ve\theta})}
\]  
where $\vesub{C}{N,N}$ is the covariance matrix of $\vesub{\tau}{k}$, and $\vess{C}{N}{*}$ is a vector of the covariances between $\vesub{\tau}{k}$ and $\tau^*$. Thus, 
$
p(\tau^*|\vesub{\tau}{k}) = N(\vesup{a}{T} \vesub{\tau}{k}, \sigma^{*2}),$ where  $ \vesup{a}{T}=\vess{C}{N}{*T} \vess{C}{N,N}{-1} $ and $\sigma^{*2}=k(\vesup{x}{*}, \vesup{x}{*};\hat{\ve\theta})- \vess{C}{N}{*T} \vess{C}{N,N}{-1}\vess{C}{N}{*}$. 
From the discussion given in the last paragraph in Section 2.2, we have
\[
\tilde{p}_G(\vesub{\tau}{k}|\D) = N(\vesub{\tilde{\tau}}{k}, \ve{\Omega}), \ \ \ve{\Omega} \triangleq (\vess{C}{N,N}{-1} + \vesub{D}{k})^{-1}. 
\]

The integrand in \eqref{tausd} is therefore the product of two normal density functions. It is not difficult to prove (see the details in Appendix B of the supplementary materials) that $p(\tau^*|\D)$ is still a normal density function 
\begin{equation}
p(\tau^*|\D) = N(\vesup{a}{T} \vesub{\tilde{\tau}}{k},\vesup{a}{T}\ve{\Omega}\ve{a}+\sigma^{*2}).
\label{taus} \end{equation}
Then \eqref{pred_mean0} can be evaluated by numerical integration.

To calculate $\Var(z^*|\D)$, we use the formula:
\begin{equation}
\Var(z^*|\D)=\E[\Var(z^*|\tau^*,\D)] + \Var[\E(z^*|\tau^*,\D)].
\label{var1} \end{equation}
From the model definition, we have
\begin{align}
&\Var[\E(z^*|\tau^*,\D)] = \E\Big[\E(z^*|\tau^*,\D)\Big]^2 - \Big(\E[\E(z^*|\tau^*,\D)]\Big)^2  \nonumber \\
=& \int \big [ h(\vess{u}{k}{T}\hat{B}^T \ve{\Phi}(t^*) + \tau^*)\big]^2 p(\tau^*|\D) d\tau^* - \big[\E(z^*|\D)\big]^2,
\label{var2} \end{align}
and
\begin{equation}
\E[\Var(z^*|\tau^*,\D)] = \int \Var(z^*|\tau^*,\D) p(\tau^*|\D) d\tau^* = \int b''(\hat{\alpha}^*) a(\phi) p(\tau^*|\D) d\tau^* ,
\label{var3} \end{equation}
where $\hat{\alpha}^*$ is a function of $h(\vess{u}{k}{T}\hat{B}^T \ve{\Phi}(t^*)+ \tau^*)$, and $p(\tau^*|\D)$ is given by \eqref{taus}. Thus \eqref{var2} and \eqref{var3} can also be evaluated by numerical integration.

The posterior density $p(\tau^*|\D)$ in \eqref{taus} is obtained based on the Gaussian approximation $\tilde{p}_G(\vesub{\tau}{k}|\D)$ to ${p}(\vesub{\tau}{k}|\D) $. It usually gives quite accurate results. The methods to improve Gaussian approximation were discussed in \cite{Rue09}. They can also be used to calculate  $p(\tau^*|\D)$ from \eqref{tausd}.  

An alternative way is to use the first integral in \eqref{tausd} to replace $p(\tau^*|\D)$ in \eqref{pred_mean0} and perform a multi-dimensional integration using, for example, Laplace approximation; see Appendix C of the supplementary materials for the details.

The second type of prediction is to predict a completely new batch with subject-based covariate $\vesup{u}{*}$. We want to predict $z^*$ at $(t^*, \vesup{x}{*})$. In this case, the training data $\D$ are the data collected from the batches $1, \ldots, M$. Since we have not observed any data for this new batch, we cannot directly use the predictive mean and variance discussed above. A simple way is to predict $z^*$ by using  $h(\vesup{u}{*T}\hve{B}^T\ve{\Phi}(t^*))$, i.e. ignoring $\tau^*$ in \eqref{cmean}. This approach however does not use the information of $\vesup{x}{*}$, the observed functional covariates. Alternatively as argued in   \cite{Shi07g}, batches $1$ to $ M$ actually provide an empirical distribution of the set of all possible subjects. A similar idea is used here. We assume that, for $m=1, \ldots, M$,
$$
\omega_m=P(z^* \mbox{ belongs to the $m$-th batch}).
$$
If we assume that the new batch or $z^*$ belongs to the $m$-th batch, we can  calculate the conditional predictive mean by \eqref{cmean}, formulated by
\[ \E(z^*|\tau^*,\D) = h(\vesup{u}{*T}\hve{B}^T\ve{\Phi}(t^*)+\tau^*). \]
The predictive mean $\E(z^*_m|\D)$ in \eqref{pred_mean0} and
the predictive variance $\Var(z^*_m|\D)$ in \eqref{var1} can be calculated  as if the test data belong
to the $m$-th batch. Here both $\vesup{u}{*}$ and $\vesup{x}{*}$ are used.

Based on the above empirical assumption, the prediction of the response for the test input $\vesup{x}{*}$ at $t^*$ in a completely new subject is
\begin{equation}
\E(z^*|\D)= \sum_{m=1}^M \omega_m \E(z^*_m|\D),
\label{newpmean}
\end{equation}
and the predictive variance is
\begin{equation}
\Var(z^*|\D)=\sum_{m=1}^M \omega_m \Var(z^*_m|\D) + \sum_{m=1}^M \omega_m \big[\E(z^*_m|\D) \big]^{2} - \big[\E(z^*|\D)\big]^2.
\label{newpvar}
\end{equation}
We usually use the equal weights, i.e. $\omega_m=1/M$ for $m=1, \ldots, M$. In general these $M$ batches may not provide equal information to the new batch. In this case varying weights may be considered; see more discussion in \cite{Shi08}.

\medskip
\section{Consistency}
The consistency of Gaussian process functional regression method involves two issues. One is related to the common mean $\vess{u}{m}{T} \ve{\beta}(t)$ in \eqref{gegpfr} and the other is related to the curve $z_k(t)$ itself ($k=1,\ldots,M$ or a new one). The common mean structure is estimated from the data collected from all $M$ subjects, and has been proved to be consistent in many functional linear models under suitable regularity conditions \citep[see][]{Ramsay05, Yao05b}. 

This paper focuses on the second issue, the consistency of $\hat{z}_k(\cdot)$ to $z_k(\cdot)$, one of the key features in nonparametric regression.  This kind of problems for GPR related models  have received increasing attention in recent years, see for example \cite{Choi05}, \cite{Ghosal06} and \cite{Seeger08}. \cite{Choi05} considered the posterior consistency of Gaussian process prior for normal response, \cite{Ghosal06} proved the posterior consistency of Gaussian process prior for nonparametric binary regression no matter what the mean function of Gaussian process prior is set to, and \cite{Pillai07} extended the result to Poisson distribution. But the consistency for general exponential family distributions is yet to be investigated. Meanwhile, \cite{Seeger08} proved the information consistency  via a regret bound on cumulative log loss. Generally speaking, if the sample size of the data collected from a certain curve is sufficiently large and the covariance function satisfies certain regularity conditions, the prediction based on a GPR model is consistent to the real curve, and the consistency does not depend on the common mean structure or the choice of the values of hyper-parameters involved in covariance function; see more detailed discussion in \cite{Shi11}. 

In this section, we discuss the information consistency and extend the result of \cite{Seeger08} to a more general context such as $z_k$ following Poisson distribution which has not been covered in the literature. 

Similar to other GPR related models, the consistency of  $\hat{z}_k(\cdot)$ to $z_k(\cdot)$ depends on the observations collected from the $k$-th curve only. We assume that the underlying mean function for the $k$-th curve, denoted by $\mu_k(t)$, is known. The case where the mean function is estimated from data is discussed in the supplementary materials. For ease of presentation we omit the subscript $k$ in the rest of the section and denote the data by $\ve{z}_n=\{z_1,\ldots,z_n\}$  at the points $t_{1},\ldots,t_{n}$, and the corresponding covariate values $\ve{X}_n=\{\ve{x}_1,\ldots,\ve{x}_n\}$ where $\ve{x}_i\in\mathcal{X}\subset\mathbb{R}^Q$ are independently drawn from a distribution $\cmu(\ve{x})$. Let  $\CD_n=\{(\ve{x}_i,z_i),\; i=1,\ldots,n\}$. We assume that $\ve{z}_n$ is a set of samples taking values in $\mathcal{Z}$ and follows a distribution in exponential family, $E(z_i|\tau) = h({\mu}(t_i)+\tau(\ve{x}_i))$ for an inverse link function $h(\cdot)$, and the underlying process $\tau(\cdot)\sim GPR(0,k(\cdot, \cdot;\ve\theta))$. Therefore, the stochastic process $\tau(\cdot)$ induces a measure on space $\mathcal{F}=\{f(\cdot):\mathcal{X} \mapsto \mathbb{R}\}$.  

Suppose that the hyper-parameter $ \ve\theta$ in the covariance function of $\tau(\cdot)$ is estimated by empirical Bayesian method and the estimator is denoted by $\hat{\ve\theta}_n$. Let $\tau_0(\cdot)$ be the true underlying function, i.e. the true mean of $z_i$ is given by $h({\mu}(t_i)+\tau_0(\ve{x}_i))$. Denote
\begin{align*}
p_{gp}(\ve{z}_n)&=\int_{\mathcal{F}} p(z_1,\cdots,z_n|\tau(\ve{X}_n))dp_n(\tau),  \\
p_0(\ve{z}_n) &= p(z_1,\cdots,z_n|\tau_0(\ve{X}_n)), 
\end{align*}
then $p_{gp}(\ve{z}_n)$ is the Bayesian predictive distribution of $\ve{z}_n$ based on the GPR model. Note that $p_n(\tau)$ depends on $n$ since the hyper-parameter of $\tau(\cdot)$ is estimated from the data. It is said that  $p_{gp}(\ve{z}_n)$ achieves {\it information consistency} if
\begin{equation}
\frac 1 n E_{\ve{X}_n}\Big( D[p_0(\ve{z}_n),p_{gp}(\ve{z}_n)] \Big) \rightarrow 0 
\quad\text{ as } n\rightarrow \infty,  \label{inf:con}
\end{equation}
where $E_{\ve{X}_n}$ denotes the expectation under the distribution of $\ve{X}_n$ and $D[p_0(\ve{z}_n),p_{gp}(\ve{z}_n)]$ is the Kullback-Leibler divergence between $p_0(\cdot)$ and $p_{gp}(\cdot)$, i.e.
$$D[p_0(z),p_{gp}(z)]=\int p_0(z)\log \frac{p_0(z)}{p_{gp}(z)}dz.$$

{\bf Theorem 1:} Under the GGPFR models \eqref{exfamily} and \eqref{ggpfr} and the conditions given in Lemma 1 of the supplementary materials, the prediction $\hat{z}(\cdot)$ is information consistent to the true curve $z_0(\cdot)$ if the RKHS norm $\lVert \tau_0\rVert_k$ is bounded and the expected regret term $E_{\ve{X}_n}(\log|\ve{I} + \delta \vesub{C}{nn}|)=o(n)$.  
The error bound is specified in \eqref{errorb}. 

\medskip
{The proof of the theorem is given in Appendix D of the supplementary materials.}

\medskip

{\bf Remark 1.} The condition $|b''(\alpha)|\leq e^{\kappa\alpha}$ in Lemma 1 can be satisfied by a wide range of distributions, such as normal distribution where $b(\alpha)=\alpha^2/2$, binomial distribution (with the number of trials $m$) where $b(\alpha)= m\log (1+e^\alpha)$ and Poisson distribution where $b(\alpha)=e^\alpha$.

{\bf Remark 2.}
The regret term $R = \log|\ve I + \delta \vesub{C}{nn}|$  depends on the covariance function $k(\cdot,\cdot; \ve\theta)$ and the covariate distribution $\cmu(\ve{x})$. It can be shown that for some widely used covariance functions, such as linear, squared exponential and Mat\'{e}rn class, the expected regret terms are of order $o(n)$; see \cite{Seeger08} for the detailed discussion.

{\bf Remark 3.} Lemma 1 requires that the estimator of the hyperparameter $ \ve\theta$ is consistent. In Appendix D of the supplementary materials we proved that the estimator by maximizing the marginal likelihood based on Laplace approximation \eqref{laplace} satisfies this condition when the number of curves and the number of observations on each curve are sufficiently large. This implies that the information consistency in GGPFR models is achieved for the covariance functions listed in Remark 2. 
A more general asymptotic analysis is to study the convergence rates of both the mean function estimation and the individual curves prediction when the number of curves and/or the number of observations on each curve tend to infinity, as discussed in \cite{Nie07} for the maximum likelihood estimators of the parameters in mixed-effects models. The research along this direction is worth further development.

\section{Numerical examples}

In this section we demonstrate the proposed method with serveral examples. We
first use simulated data and then consider the paraplegia data discussed in Section 1. More simulated and real examples are provided in the supplementary materials.

\subsection{Simulated Examples}
%

{\bf (i) Simulation study.} The true model used to generate the latent process is $y_{mi}(t_{mi}) = 0.8\sin(0.5t_{mi})^3 + \tau_{mi}$, where, for each $m$,  $t_{mi}$'s are equally spaced points in $(-4,4)$ and $\{ \tau_{mi} \}$  is a Gaussian process with zero mean and the squared exponential covariance function defined in \eqref{covfun0} 
with $v_1=0.04$, $w_1=1.0$ and $a_1=0.1$. In this example, the covariate $x(t)$ is the same as $t$. The observations $z_{mi}$ follow a binomial distribution $Bin(1,\pi_{mi})$ with 
$
\pi_{mi}= {1}/(1+\exp(-y_{mi})).
$

\begin{figure}
\begin{center}
\subfigure[\label{Bin_fig11} $N_m$=20]{\includegraphics[width=0.43\linewidth]{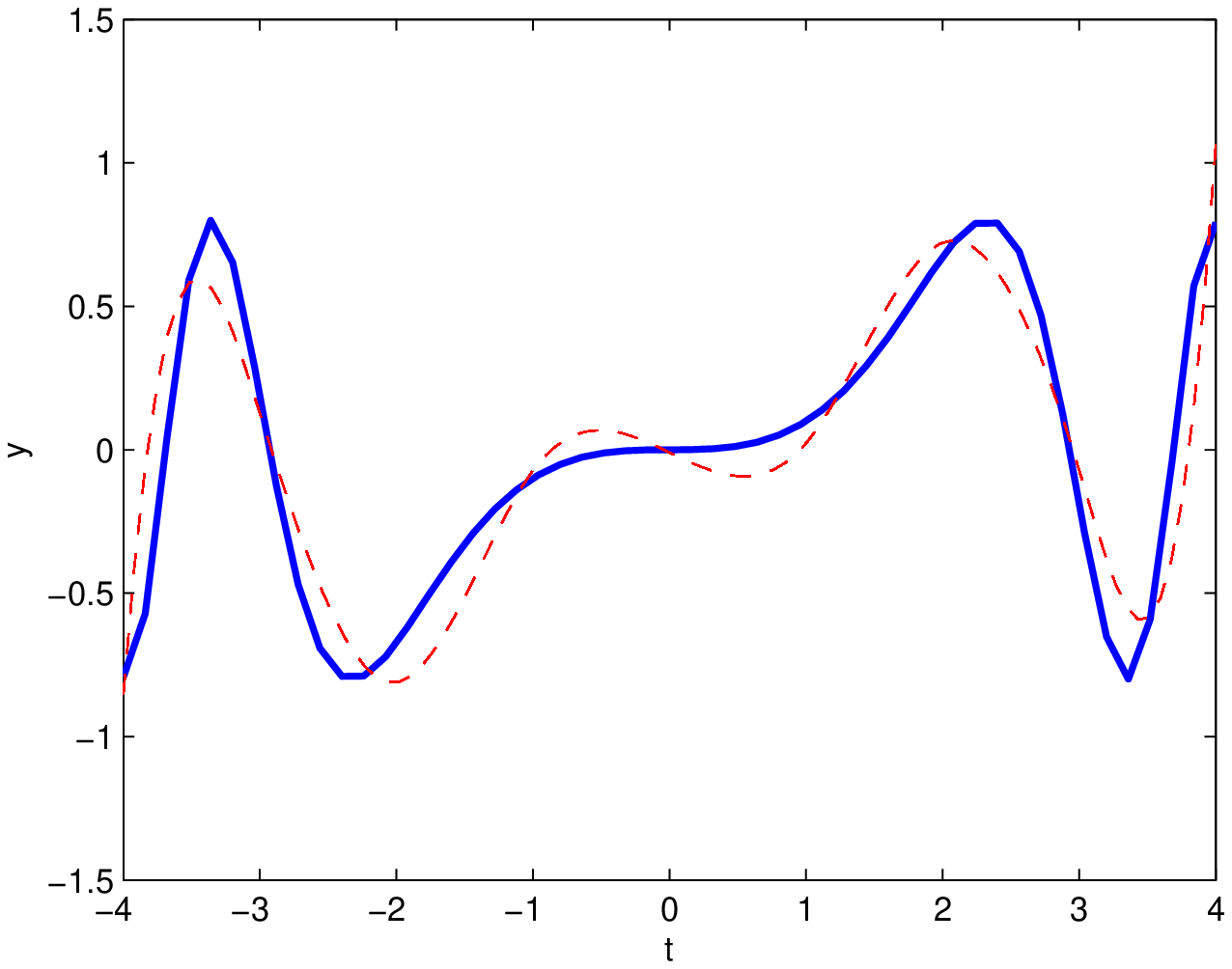}} 
\subfigure[\label{Bin_fig13} $N_m$=20]{\includegraphics[width=0.43\linewidth]{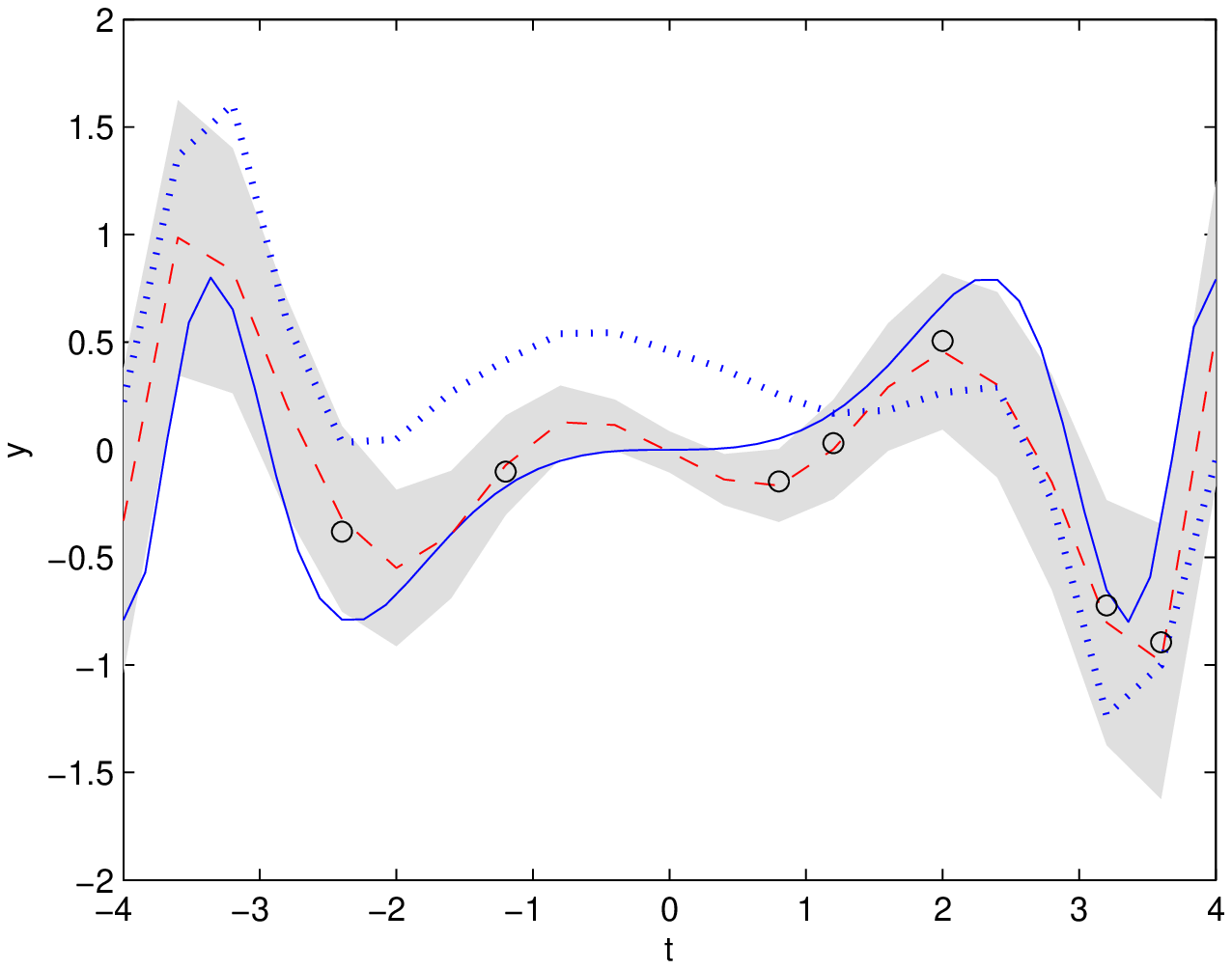}} \\
\subfigure[\label{Bin_fig21} $N_m$=40]{\includegraphics[width=0.43\linewidth]{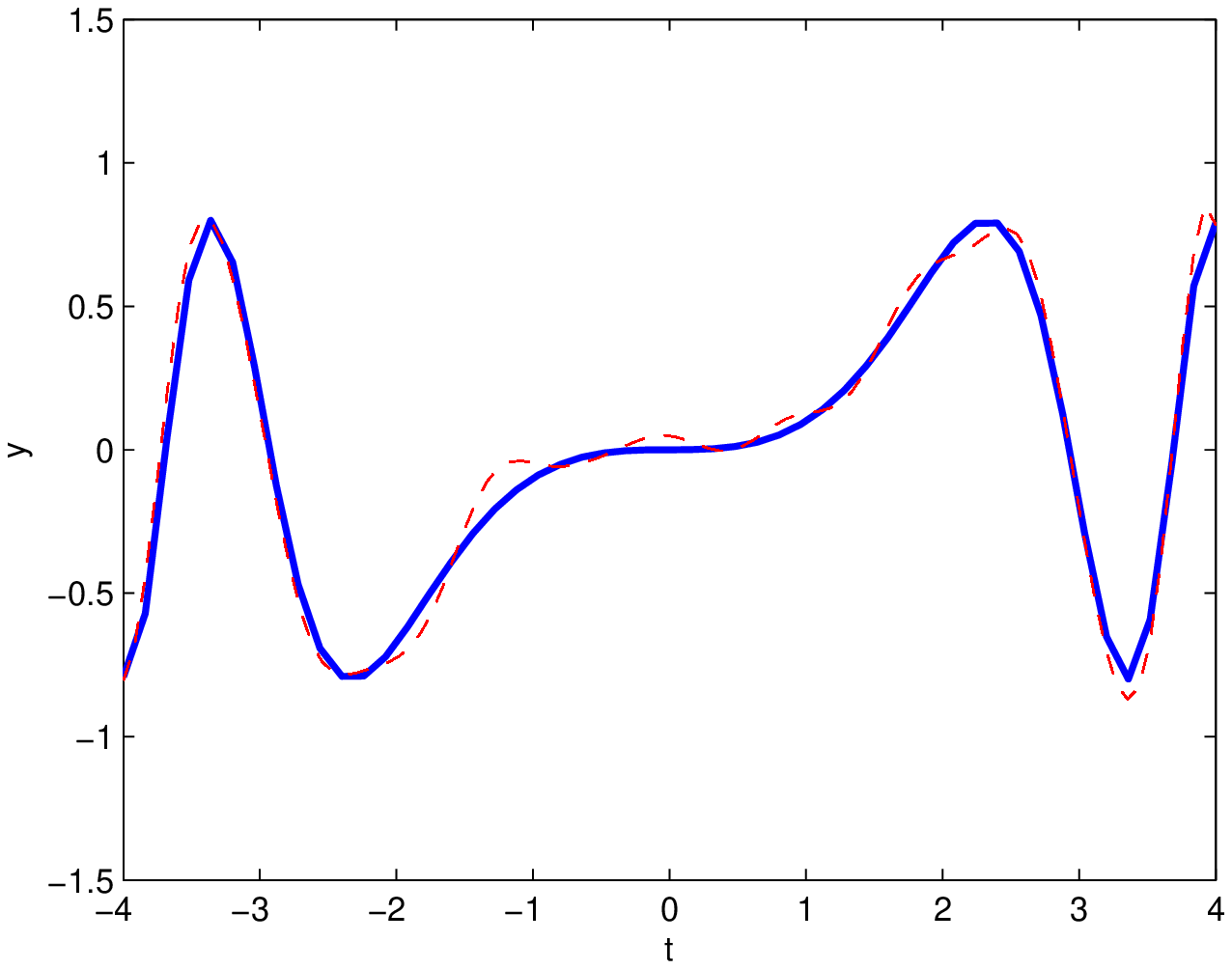}} 
\subfigure[\label{Bin_fig23} $N_m$=40]{\includegraphics[width=0.43\linewidth]{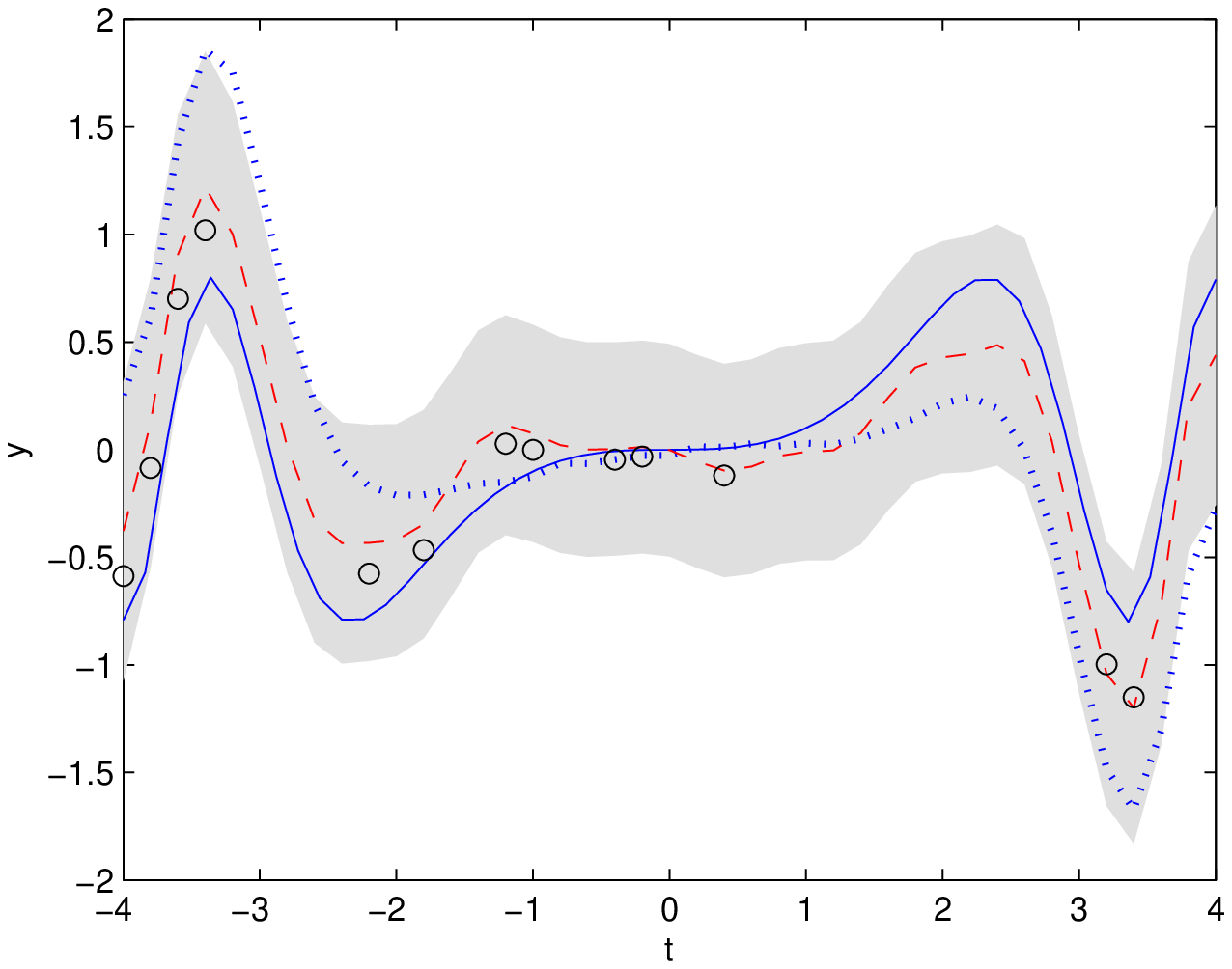}} \\
\subfigure[\label{Bin_fig31} $N_m$=60]{\includegraphics[width=0.43\linewidth]{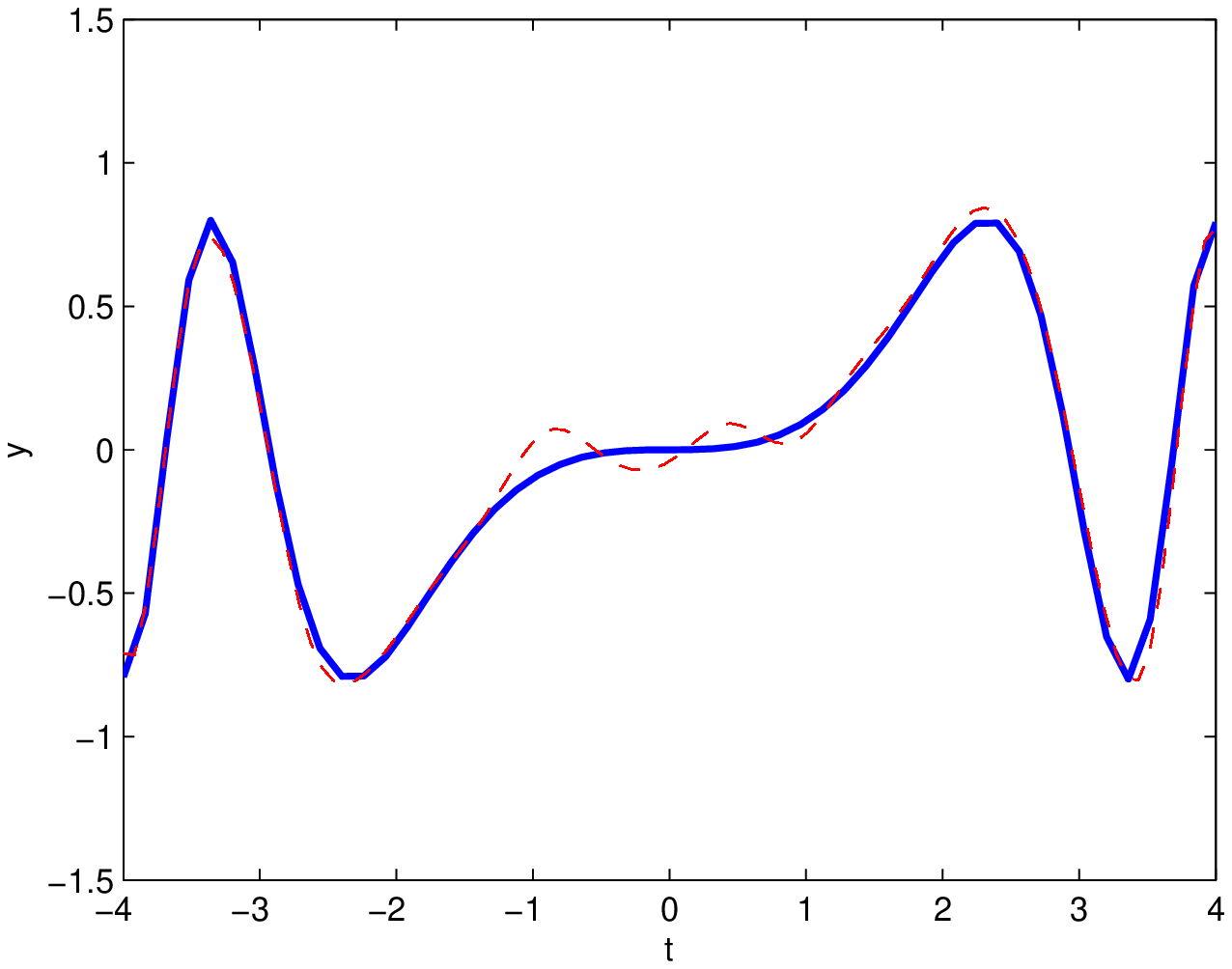}} 
\subfigure[\label{Bin_fig33} $N_m$=60]{\includegraphics[width=0.43\linewidth]{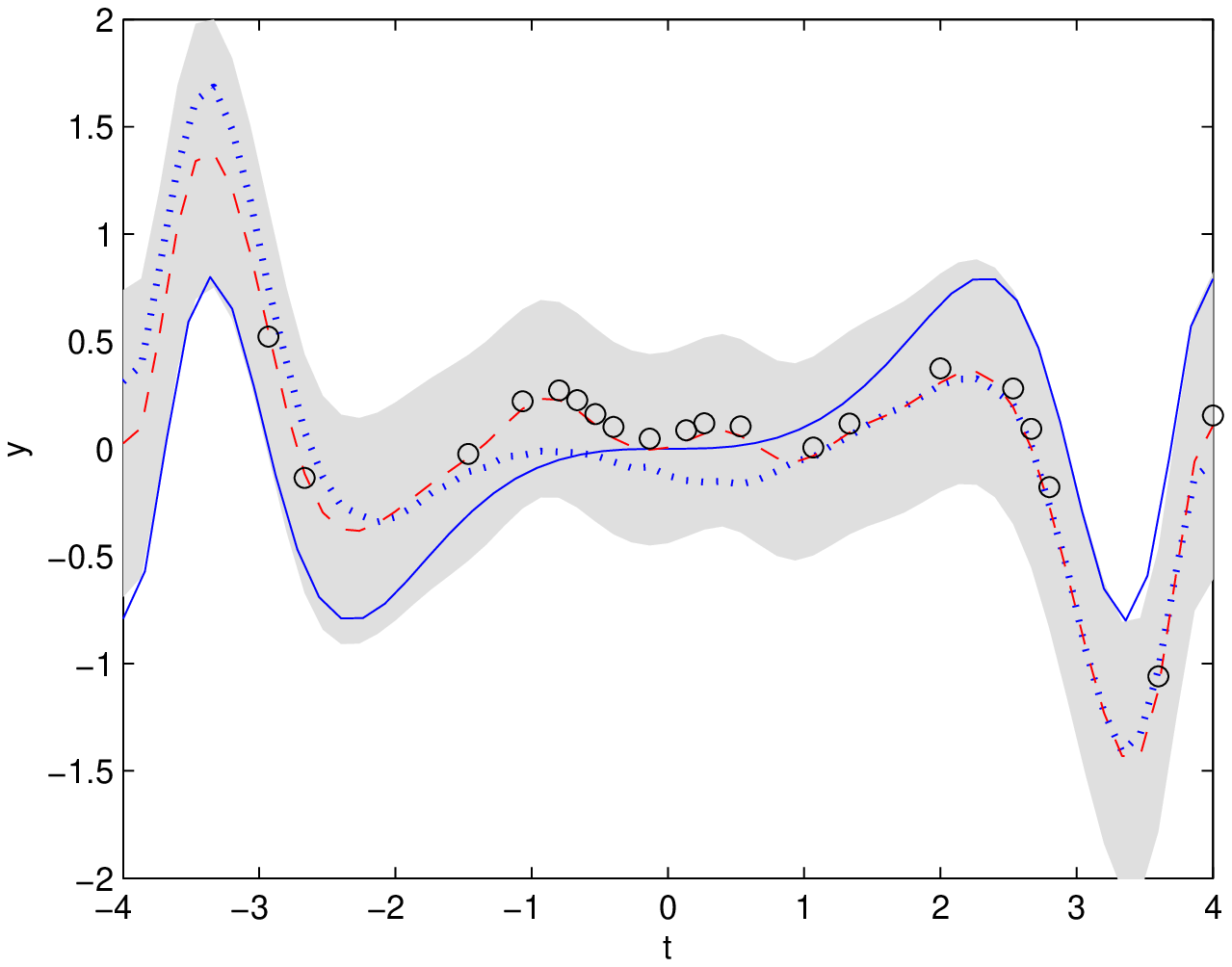}}
\end{center}
\caption{Binomial Data. For $N_m=20$, 40 and 60, left panel: the estimated mean curve $\hat{\beta}(t)$  (in dashed line) and the true mean curve (in solid line); right panel: the estimated underlying process $\hat{y}_m(t)$ (dashed line) with 95\% confidence band (shaded area), the true curve ${y}_m(t)$ (dotted line) and the true mean curve (solid line), and the circles are the estimated values of the underlying process at the test points.}
\label{simu_mean_curves}
\end{figure}

Sixty curves, each containing $N_m$ data points, are generated and used as training data. We use a GGPFR model with binomial distribution and logit link function: $\logit (\pi_m(t))=\beta(t)+\tau_m(t)$ where $\tau_m(t)$ follows a GPR model. Cubic B-spline approximation is used to estimate the mean curve $\beta(t)$, where the knots are placed at equally spaced points in the range 
and the number of basis functions is determined by BIC 
{which is given by BIC$=-2l(\hve B, \hve\theta)+G\log(M)$ with $G$ being the total number of parameters.} 
A Gaussian approximation method as specified around \eqref{nest} is used to calculate the empirical Bayesian estimates of $\ve B$ and $\ve \theta$.  Table \ref{table_parameter} lists the average estimates of the hyper-parameters $\ve \theta=\{w_1,v_1,a_1\}$ for $N_m=20$, 40 and 60 for ten replications. The empirical Bayesian estimates are closer to the true values with $N_m$ increasing. The estimates of mean curve $\hat{\beta}(t)$ for different $N_m$ along with the true mean curves are presented in the left panels of Figure \ref{simu_mean_curves}. As discussed in Section 3, the consistency of $\hat{\beta}(t)$ to $\beta(t)$ depends on the observations obtained from all training curves. The figures show that the estimated mean curves by the GGPFR method are very close to the true one even for the case of $N_m=20$.    

\begin{table}
\begin{center}
\caption{Estimates of the hyper-parameters}\label{table_parameter}
\begin{tabular}{ c  c c c  c c cc  c c}
\hline\hline
Parameter & True & $N_m=20$ & $N_m=40$ &  $N_m=60$   \\
\hline
$w_1$ & 1.0  &0.6927  & 1.1044       & 1.0660    \\
$v_1$ & 0.04 &0.0022  & 0.0691       & 0.0481    \\
$a_1$ & 0.1  &0.0992  & 0.0705       & 0.0816    \\
\hline 
\end{tabular}
\end{center}
\end{table}

One of the most important features of GGPFR is the ability to model each individual $z_m(t)$ or the underlying continuous process $y_m(t)$. The right panels in Figure \ref{simu_mean_curves} show the estimated underlying processes $\hat{y}_m(t)$, the true $y_m(t)$ as well as the true mean curves $\beta(t)$ for one replication. Although the underlying processes are similar to the mean curve, the samples of $y_m(t)$ are systemically different to $\beta(t)$, meaning that although $\hat{\beta}(t)$ is a consistent estimator of $\beta(t)$, it is not a good estimator of $y_m(t)$. The theoretical result in Section 3 shows that the use of GPR part $\tau_m(t)$ in the GGPFR model can overcome this drawback, resulting in the consistency of $\hat y_m(t)$ or $\hat z_m(t)$. This feature is demonstrated in the right panels of Figure \ref{simu_mean_curves}. A simulation study is conducted to illustrate this feature. We generate a new curve and its corresponding observations with $N_m$ data points, of which two thirds are randomly selected as observations to estimate the underlying process and the remaining points are used as test data to make prediction. The values of the root of mean squared errors ({\it rmse}) and the correlation coefficients ($r$) between $\hat y_m(t)$ and $y_m(t)$ at the test data points are calculated, and the average values based on 50 repetitions are given in Table \ref{table_rcc_gp}. The right panels of Figure \ref{simu_mean_curves} (in circles) presents the results for one replication. Both the table and the figure show that $\hat y_m(t)$ is a good estimate of $y_m(t)$, and the accuracy improves as $N_m$ increases. 

\begin{table}
\begin{center}
\caption{The values of rmse and correlation between $\hat y_m(t)$ and $y_m(t)$ by squared exponential covariance function} \label{table_rcc_gp}
\begin{tabular}{ c c  c  c  }
\hline\hline
 Value  & $N_m=20$ & $N_m=40$ &  $N_m=60$  \\
\hline
rmse   &0.3193 & 0.2639    & 0.2387    \\
r     & 0.8771 & 0.8886   &  0.9072    \\
\hline 
\end{tabular}
\end{center}
\end{table}

\begin{figure}[h]
\begin{center}
\subfigure[\label{Bin_fig_MC} MC]{\includegraphics[width=0.43\linewidth]{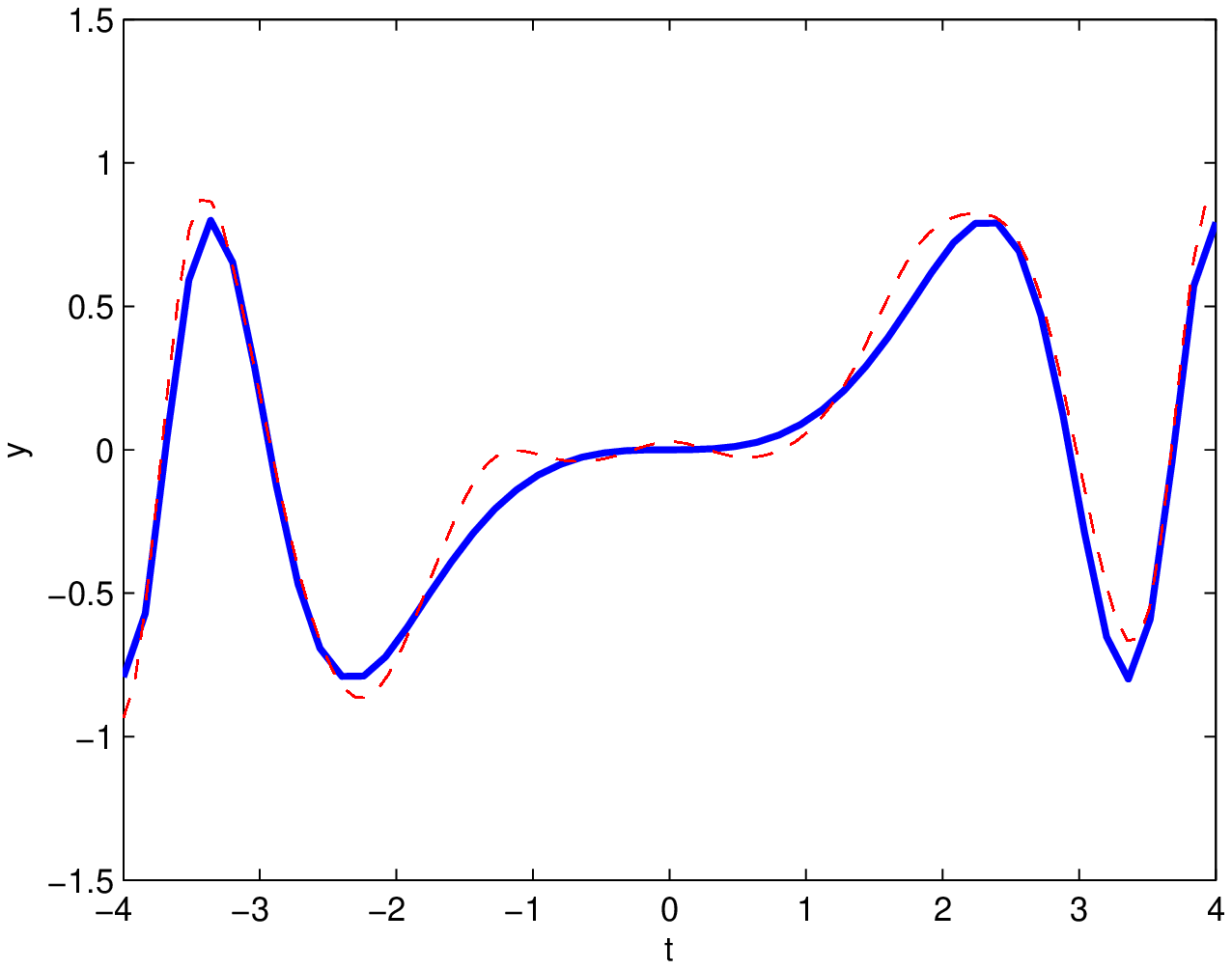}} 
\subfigure[\label{Bin_fig_RQ} RQ]{\includegraphics[width=0.43\linewidth]{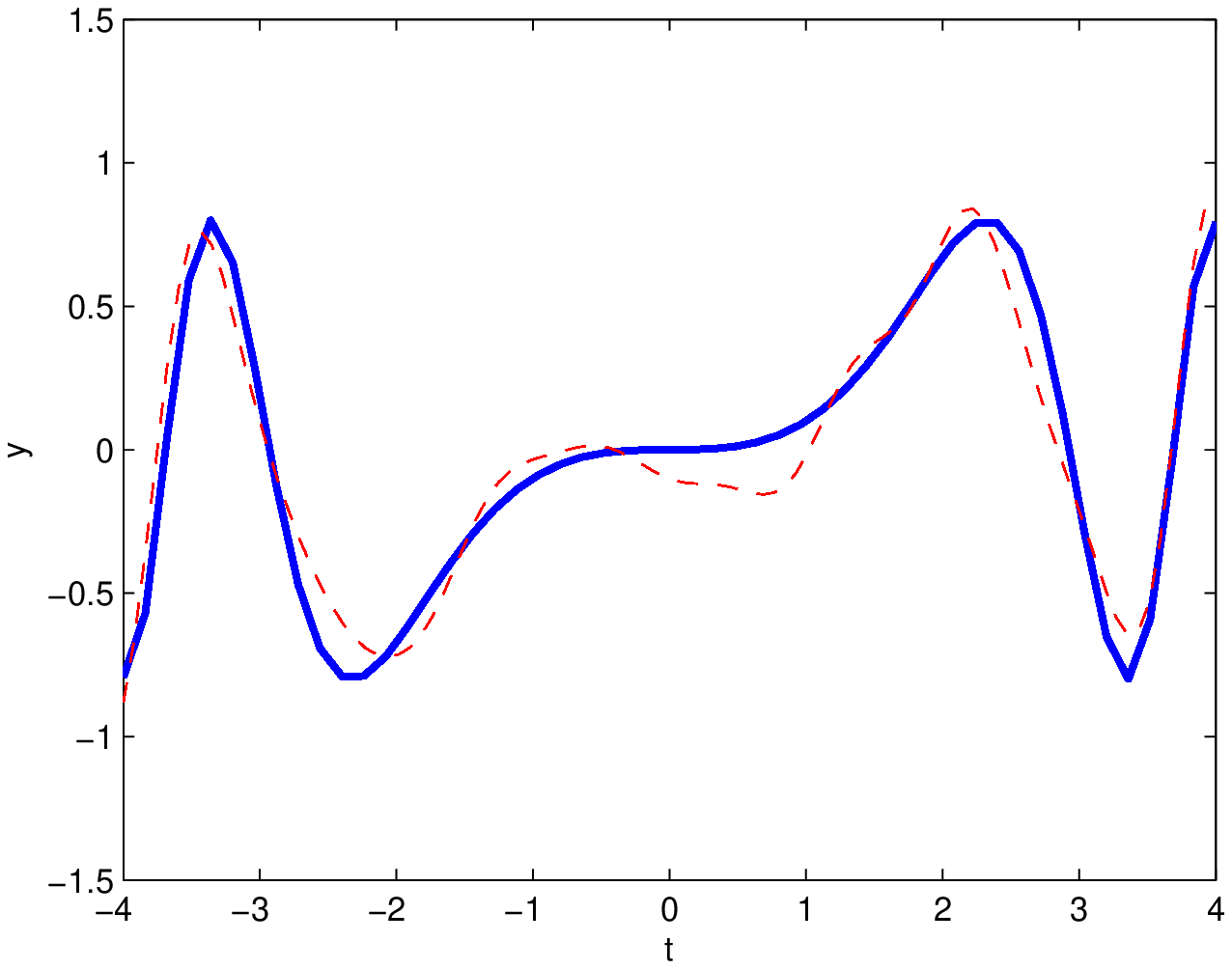}}  \\
\subfigure[\label{Bin_fig_PP} PP]{\includegraphics[width=0.43\linewidth]{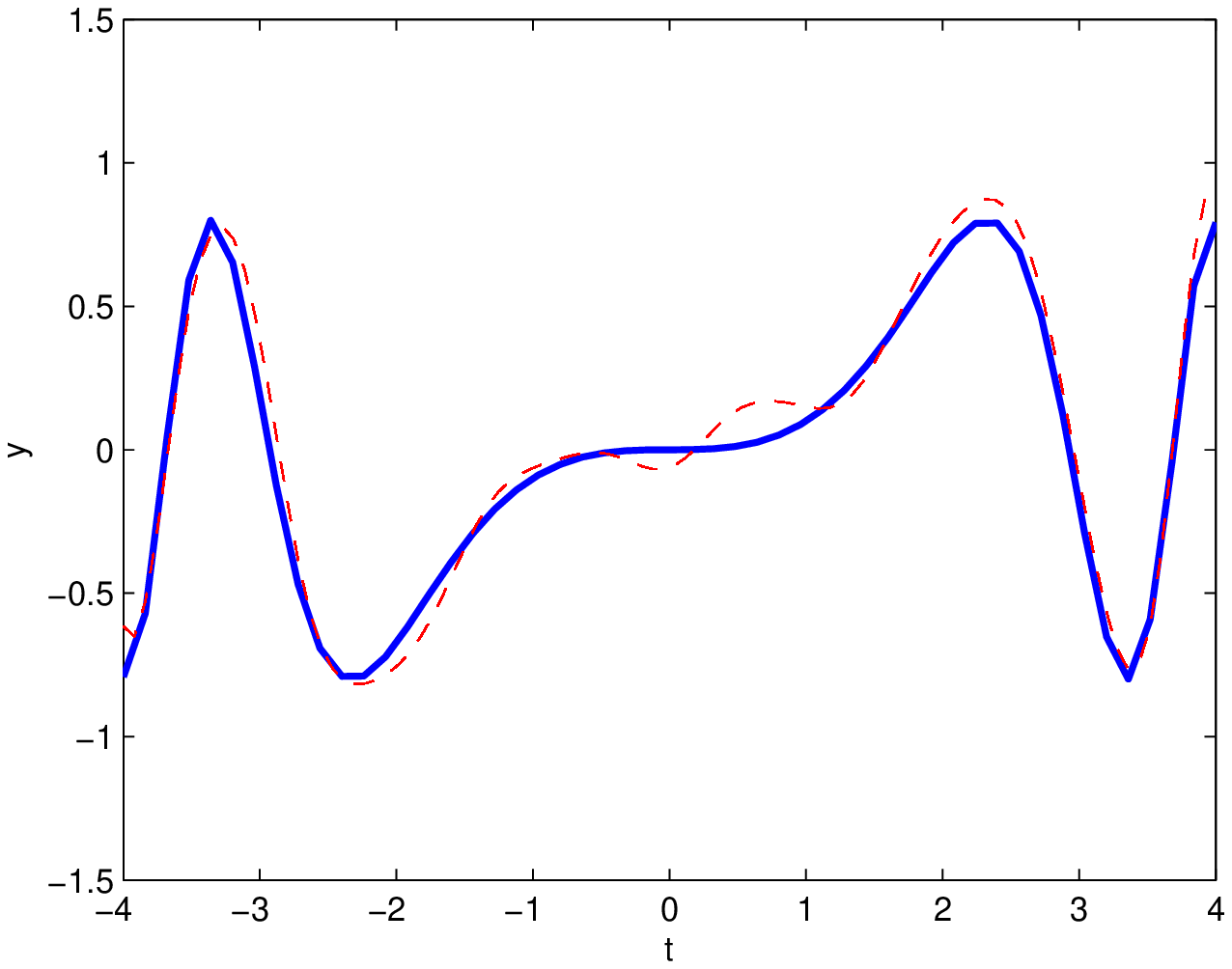}} 
\subfigure[\label{Bin_fig_NP} NP]{\includegraphics[width=0.43\linewidth]{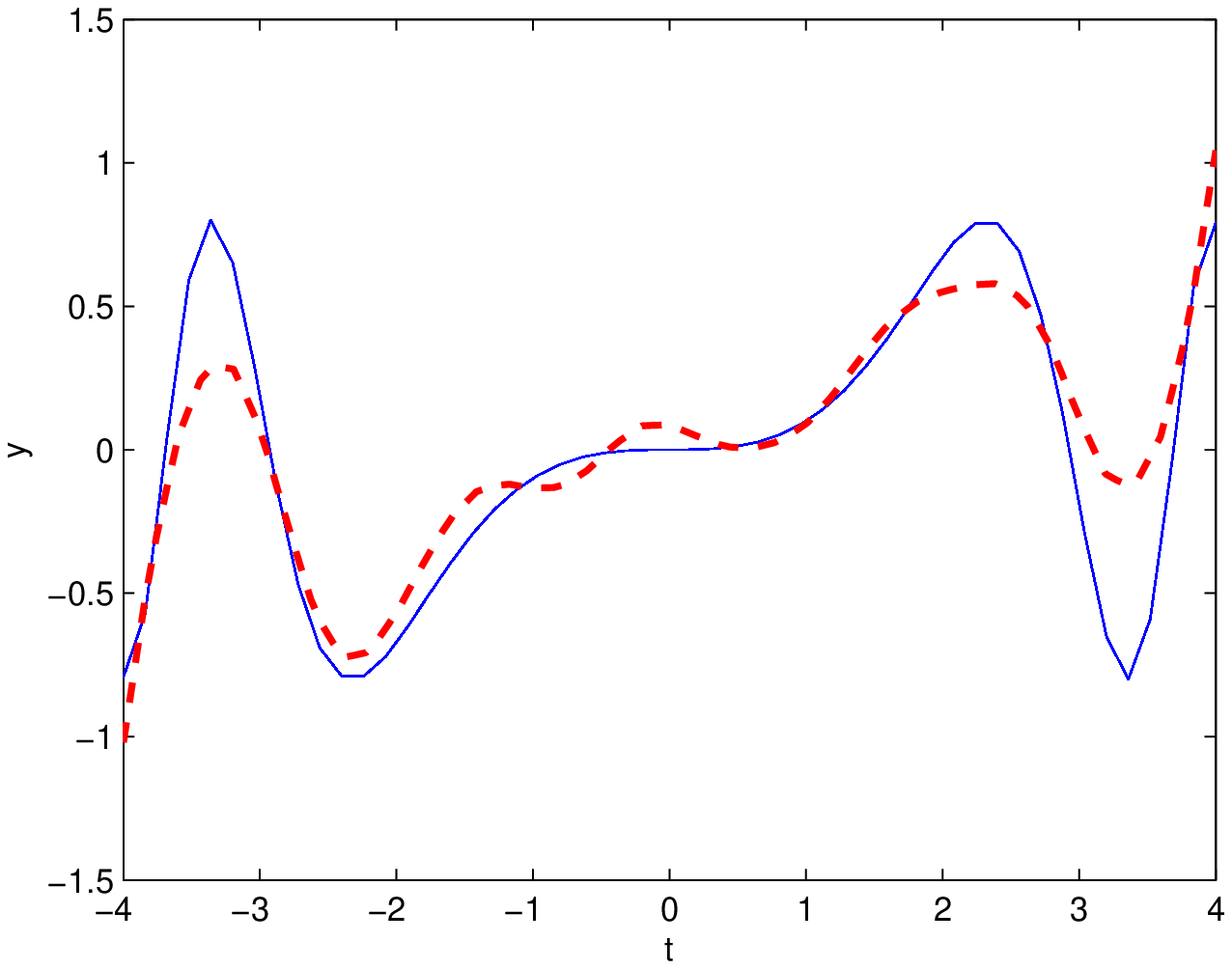}}
\end{center}
\caption{The estimated mean curves (dashed line) by different covariance functions and nonparametric covariance method. The solid lines are the true mean curve.}
\label{simu_mean_curves_others}
\end{figure}

{\bf (ii) Sensitivity on the choice of covariance kernels.} To test the sensitivity of the GGPFR model on different covariance functions, besides the squared exponential (SE) covariance function the above example for $N_m=40$ is further analyzed using three other covariance functions: Mat{\'e}rn class with $\nu=3/2$ (MC), rational quadratic (RQ) and piecewise polynomial with $q=2$ (PP); see \cite{Rasmussen06} for detailed description of these covariance functions. The results are also compared with the nonparametric covariance structure method (NP) as proposed by \cite{Hall08} {which is implemented using PACE package (http://www.stat.ucdavis.edu/PACE/)}. The estimated mean curves are presented in Figure \ref{simu_mean_curves_others}, and the values of {\it rmse} and the correlation coefficients ($r$) between the true underlying process $y_m(t)$ and the estimated curve $\hat y_m(t)$ are given in Table \ref{table_rcc_other}.

\begin{table}
\begin{center}
\caption{The values of rmse and correlation between $\hat y_m(t)$ and $y_m(t)$ by different covariance functions and nonparametric method} 
\label{table_rcc_other}
\begin{tabular}{ c c c c c c  }
\hline\hline
      & SE     & MC      & RQ       & PP      &  NP  \\
\hline
rmse & 0.2526  & 0.2692 & 0.2818  & 0.2940  & 0.3995    \\
r    & 0.9045  & 0.8830 & 0.8643  & 0.8540  & 0.7419    \\
\hline 
\end{tabular}
\end{center}
\end{table}

It can be seen from the figure and the table that the results by the GGPFR model with the misspecified covariance functions are comparable to those obtained by the true squared exponential covariance function, although the latter indeed provides the best results. Furthermore, the GGPFR models with different covariance kernels consistently outperform the nonparametric covariance method in terms of estimation of the mean function and the underlying process, despite the fact that the main advantage of Gaussian process covariance kernels is to deal with high-dimensional covariates.

{\bf (iii) A simulated example with a general covariance structure.} To test the performance of the GGPFR model for  data with more general covariance structure further simulation study is conducted as follows. The simulation is based on the latent process $y(t)$ with mean function $2\sqrt{0.4}\sin(0.4\pi t)$ and covariance function $\Cov (y(t),y(s))=\sum^{10}_{j=1}\alpha_j 
\phi_j(t)\phi_j(s)$, where $\alpha_j=j^{-3/2}$, $\phi_j(\cdot)$'s are discrete Chebyshev polynomials, and 
$0\leq t \leq 5$. Then 100 curves, each containing 50 equally spaced points in $[0,5]$, are simulated and
the binary observations $z_{mi}$ are consequently generated using the logit link function.

Same as above, various covariance functions, namely squared exponential (SE), Mat{\'e}rn class with $\nu=3/2$ (MC), rational quadratic (RQ) and piecewise polynomial with $q=2$ (PP), are used in the GGPFR model.  The estimated mean curves are presented in Figure \ref{simu_mean_curves_OP} of the supplementary materials, and the values of {\it rmse} and the correlation coefficients ($r$) between the true underlying process $y_m(t)$ and the estimated curve $\hat y_m(t)$ are given in Table \ref{table_rcc_OP}. The results are also compared with the nonparametric covariance structure method (NP).

The results show that the estimated mean functions by the GGPFR with different covariance functions are similar 
and all close to the true mean function, and the performance for estimation of the individual curves are comparable
to each other and the nonparametric method with RQ and PP giving slightly better estimation. 

\begin{table}
\begin{center}
\caption{The values of rmse and correlation between $\hat y_m(t)$ and $y_m(t)$ by different covariance functions and nonparametric method for the data with Chebyshev polynomials} 
\label{table_rcc_OP}
\begin{tabular}{ c c c c c c  }
\hline\hline
 & SE & MC   & RQ      & PP      &  NP  \\
\hline
rmse & 0.3141  &0.3151 & 0.2604  & 0.2682  & 0.3196    \\
r    & 0.9708 & 0.9595 & 0.9826  & 0.9840  & 0.9531    \\
\hline 
\end{tabular}
\end{center}
\end{table}

\subsection{Paraplegia Data}\label{para_data}

We now consider the example discussed in Section 1. This application involves the analysis of the standing-up manoeuvre for paraplegic patients, considering the body supportive forces as a potential feedback source in functional electrical stimulation (FES)-assisted standing-up. FES is a method of eliciting the action potential in the nerves innervating the paralysed muscles;  see \cite{Kamnik05} for more details. The analysis investigates the significance of arm, feet and seat reaction signals for the recognition of the human body's standing-up phases during rising from sitting position to standing position, i.e. sitting=0, seat unloading and ascending=1, stablising=2. The body position is usually difficult to measure unless some special equipments are employed in a designed laboratory. Therefore a number of easily measurable quantities such as the motion kinematics, reaction forces and other quantities are recorded in order to estimate the human body position. Here we select 8 input variables including the forces and torques under the patients' feet, under the arm support handle and under the seat while the body is in contact with it. In one standing-up, the output and the inputs were recorded for a few hundred time-steps, of which a quarter equally spaced time points are used in the example. The patients' heights are used as the scalar covariate $\vesub{u}{m}$.

Our data include 35 standings-ups, 5 repetitions for each of 7 patients. We randomly select 20 standing-ups as training data and the others as test data for prediction. Since the standing-up phases are ordered, we use a GGPFR model for ordinal data as considered in the Appendix E of the supplementary materials. The estimated functional coefficient $\hat\beta (t)$ from the selected 20 standing-ups is given in Figure \ref{fig_para_output}(e) of the supplementary materials. 


After the empirical Bayesian estimates are obtained, we consider the problem of prediction. We randomly select two thirds of the data from one standing-up  as observations and predict the remaining one third, and compare the predicted responses with the actual observations. This is the interpolation problem. The average error rate for the fifteen test standing-ups is 11.81\%. Taking into account the complexity of the problem, this is a very good result. Two randomly selected standing-ups and their predictions are shown in the top panels of Figure \ref{fig_para_output} in the supplementary materials.

We also consider the extrapolation problem by selecting the first two thirds of the data from one standing-up and predict the remaining data points.  On comparison of the predicted values with the actual observations, the average error rate for extrapolation is 19.23\%. This is a pretty good result for the difficult extrapolation problem. Two randomly selected standing-ups and their predictions are shown in the middle panel of Figure \ref{fig_para_output} in the supplementary materials.

For comparison, the data are also analyzed using the generalized varying coefficient model with the probit link function where the response variable is assumed to have a binomial distribution and the latent process is modeled by 
$$y_m(t) = \beta_0(t) + \sum^p_{i=1}\beta_i(t) x_{mi}(t),$$
with $x_{mi}(t)$'s representing the input functional covariates. The same prediction problems as discussed above are conducted and it is obtained that the average error rate for interpolation is 29.96\% and that for extrapolation is 21.01\%. It is obvious that the GGPFR performs significantly better than the generalized varying coefficient model for interpolation whilst the former is slightly better than the latter for extrapolation. 

{In the above analysis of the paraplegia data the observed standing-ups \jian{ are all treated as independent. However, the repeated curves collected from different patients may have a hierarchical structure.   To address this problem, the GGPFR model is extended to the case of clustered data; see Appendix G in the supplementary materials for details.} }

\section{Discussion}

We proposed a GGPFR model in this paper for \jian{concurrent} regression analysis of non-Gaussian functional data. The use of a GPR model enables us to deal with the relationship between multi-dimensional functional covariates and functional dependent variable nonparametrically. The GPR model for the latent process $\tau_m$ can be understood as nonlinear random effects. It can easily be integrated with parametric terms such as linear  mixed effects models; see \cite{Shi12} for detailed discussion on this type of models for Gaussian functional data\jian{, and an example of such models is also discussed in Appendix G.}  

We provided a general framework on how to use a Gaussian process to define a model for generalized nonparametric regression analysis for response variables from exponential families. The procedure of inference and implementation is provided and the asymptotic theory based on information consistency is established. Although the detailed formulae were given only for binomial and ordinal data with logit and porbit link functions, it is not difficult to extend them to other distributions in the exponential families. The GGPFR model assumes that the response variable follows a distribution from exponential family. This assumption can be avoided by using quasi-likelihood method. 


The GPR and the related methods have been used in numerous applications for many years, for example, in spatial statistics under the name of `kriging' \citep[see e.g.][]{Diggle03} and in machine learning as one type of `kernel machines' \citep[see e.g.][]{Rasmussen06}. Some recent developments in statistics can be found in \cite{Shi11}. This paper provides a useful extension to the existing GPR methods.

One of the main advantages of Gaussian process regression method is that it can be used
to address the problem with large dimensional covariates with a known covariance kernel. When
the dimension of covariates is small, nonparametric approaches can be applied to estimate the covariance structure; see for example \cite{Bosq00}, \cite{Yao05a, Yao05b} and \cite{Hall08}. The GGPFR model is also related to varying coefficient models which usually have some special structures such as linear forms in the covariates. The proposed model can be used to describe flexible structures between the response and the covariates and can be regarded as an extension of the varying coefficient nonparametric mixed effects model discussed in \cite{Wu2006} because in some sense the latter corresponds to the GPFR model with linear covariance kernel.

Related to the topics discussed in this paper, some interesting problems are worth further development. For example, how to address heterogeneity among different subjects \citep[see e.g.][]{Shi08}, how to build a more general asymptotic theory such as posterior consistency and covergence rate \citep[see e.g.][]{Choi05, Ghosal06}, {and how to deal with functional data in which predictors are contaminated by measurement errors \citep[see e.g.][]{Senturk08}}. 

The second interesting problem is related to computation. Gaussian approximation has been used in the paper. Although it has provided reasonably accurate results in most cases, it is of interest to develop more efficient and accurate computational methods; see for example  \cite{Shi2005} and \cite{Banerjee2013} or \cite{Andrieu09} and \cite{Andrieu10}.
As shown in Section 4.1, although the GGPFR model with a misspecified covariance kernel may still provide a reasonable result, how to choose a good covariance kernel remains an important and interesting topic. This article focuses on a special form of mean model  $\mu_m(t)=\vess{u}{m}{T}\ve{\beta(t)}$, but there should be no significant difficulty to extend it to other mean models such as varying coefficient models and standard functional regression models in the sense of \cite{Ramsay05}. However new computational methods and statistical theories may need to be developed 
if the GPR model is incorporated with these mean models.     

\jian{Finally, the model discussed in the paper is based on a concurrent regression framework. The idea can be extended to so-called ``function-on-function" regression framework, i.e. the functional response variable at each time point depends on the entire curve or the recent past values of functional predictors 
(see e.g. \cite{Ramsay05} and \cite{Senturk10}, among others). Some discussion on the connection between these models
can be found in \cite{Senturk10}.}

\section{Supplementary Materials}
Some technical details used in Sections 2 and 3 and more numerical examples as well as the GGPFR model for clustered
functional data are provided in the supplementary materials. (PDF file)

\newpage

\begin{center}{\LARGE Supplementary Materials}\end{center}
\bigskip

\appendix
\makeatletter   
 \renewcommand{\@seccntformat}[1]{APPENDIX~{\csname the#1\endcsname}.\hspace*{1em}}
 \makeatother
 \renewcommand{\thefigure}{A.\arabic{figure}}
 \renewcommand{\thetable}{A.\arabic{table}}

\section{Maximising the function (13) w.r.t. $\ve{\tau}_m$}

The function (\ref{def_Psi}) can be maximised by using the Newton-Raphson iteration $\ve{\tau}_m^{new} = \ve{\tau}_m - (\nabla^2\Psi)^{-1}\nabla\Psi$. In fact, we have
$$\nabla\Psi = V - \vess{C}{m}{-1}\ve{\tau}_m, \quad \nabla^2\Psi = W - \vess{C}{m}{-1},$$
where,
$$V = \left( \frac{d}{d\tau_{m1}}\log\big\{p(z_{m1}|\tau_{m1},\ve{B})\big\}, \ldots, \frac{d}{d\tau_{mN_m}}\log\big\{p(z_{mN_m}|\tau_{mN_m},\ve{B})\big\} \right)^T,$$
$$W = \mbox{diag}\left( \frac{d^2}{d(\tau_{m1})^2}\log\big\{p(z_{m1}|\tau_{m1},\ve{B})\big\}, \ldots, \frac{d^2}{d(\tau_{mN_m})^2}\log\big\{p(z_{mN_m}|\tau_{mN_m},\ve{B})\big\} \right).$$
Therefore we have the following iterative equation
\begin{align*}
\ve{\tau}_m^{new} &= \ve{\tau}_m - (W - \vess{C}{m}{-1})^{-1}(V - \vess{C}{m}{-1}\ve{\tau}_m)    \\
&= \ve{\tau}_m - (W\vesub{C}{m} - I)^{-1}(\vesub{C}{m}V - \ve{\tau}_m).
\end{align*}

\section{Derivation of equation (19)} 

 Since $p(\tau^*|\vesub{\tau}{k}) = N(\vesup{a}{T} \vesub{\tau}{k}, \sigma^{*2}),$ we have $\tau^*=\vesup{a}{T} \vesub{\tau}{k} + \varepsilon_1$ with 
$\varepsilon_1\sim N(0, \sigma^{*2})$. Since $\tilde{p}_G(\vesub{\tau}{k}|\D) = N(\vesub{\tilde{\tau}}{k}, \ve{\Omega})$, we have $\vesub{\tau}{k} = \vesub{\tilde{\tau}}{k} + \varepsilon_2$ with $\varepsilon_2\sim N(0,\ve{\Omega})$. Thus, 
$\tau^*=\vesup{a}{T} \vesub{\tilde{\tau}}{k} + \vesup{a}{T}\varepsilon_2 + \varepsilon_1$, 
so $p(\tau^*|\D)=N(\vesup{a}{T} \vesub{\tilde{\tau}}{k},\vesup{a}{T}\ve{\Omega}\ve{a}+\sigma^{*2})$.

\section{Prediction using Laplace approximation}

For convenience we denote $(\ve{\tau}^T_k,\tau^*)^T$ and its covariance
matrix $\ve{C}_{N+1,N+1}$ by $\ve{\tau}_+$ and $C_+$, respectively. The posterior mean in \eqref{pred_mean0} can be calculated by
\begin{align}
\E(z^*|\D) &= \E \big[\E(z^*|\tau^*,\D)\big] = \int
h(\vess{u}{k}{T}\hve{B}^T\ve{\Phi}(t^*) + \tau^*) p(\tau^*|\D)
d\tau^*  \nonumber \\
&= \int h(\vess{u}{k}{T}\hve{B}^T\ve{\Phi}(t^*) + \tau^*) p(\tau^*,\ve{\tau}_k|\D) d\tau^*d\ve{\tau}_k   \nonumber \\
&= \frac{1}{p(\ve{Z}_k)} \int h(\vess{u}{k}{T}\hve{B}^T\ve{\Phi}(t^*) + \tau^*)
p(\ve{Z}_k|\ve{\tau}_k)p(\tau^*,\ve{\tau}_k|\vesup{x}{*},\vesub{X}{k})d\tau^*d\ve{\tau}_k   \nonumber \\
&= \frac{1}{p(\ve{Z}_k)} \int h(\vess{u}{k}{T}\hve{B}^T\ve{\Phi}(t^*) + \tau^*)\Big\{ \prod^{N}_{i=1}p(z_{ki}|\tau_{ki},\hve{B}) \Big\}  \nonumber \\
&\qquad \cdot
(2\pi)^{-(N+1)/2}|\vesub{C}{+}|^{-1/2}\exp\Big\{-\half\ve{\tau}_+^T
\vess{C}{+}{-1} \ve{\tau}_+\Big\} d \ve{\tau}_+.
 \label{pred_mean}
\end{align}
The calculation of the integral is not tractable, since the dimension of $\ve{\tau}_+$ is usually very large. We now use
Laplace approximation. Denoting
\begin{align*}
\tilde\Psi(\ve{\tau}_+) =& \log h(\vess{u}{k}{T}\hve{B}^T\ve{\Phi}(t^*) + \tau^*) + \sum^{N}_{i=1}\log\big\{ p(z_{ki}|\tau_{ki},\hve{B}) \big\}  \nonumber \\
& - \frac{N+1}{2}\log (2\pi) - \half\log|\vesub{C}{+}| -\half \ve{\tau}_+^T\vess{C}{+}{-1}\ve{\tau}_+,
\label{def_tilde_Psi}
\end{align*}
the integral \eqref{pred_mean} can be expressed as
\begin{equation*}
\E(z^*|\D) = \frac{1}{p(\ve{Z}_k)} \int
\exp\{\tilde\Psi(\ve{\tau}_+)\}d\ve{\tau}_+.
\end{equation*}

Let $\ve{\hat\tau}_+$ be the maximiser of $\tilde\Psi(\ve{\tau}_+)$,
then by using Laplace approximation we have
\begin{equation}
\int \exp\{\tilde\Psi(\ve{\tau}_+)\}d\ve{\tau}_+ = \exp\Big\{
\tilde\Psi(\ve{\hat\tau}_+) + \frac{N+1}{2}\log (2\pi) -
\half\log|\vess{C}{+}{-1}+\vesub{K}{+}| \Big\},
\label{int_tilde_psi}
\end{equation}
where $\vesub{K}{+}$ is the second order derivative of
$$\log h(\vess{u}{k}{T}\hve{B}^T\ve{\Phi}(t^*) + \tau^*)+\sum^{N}_{i=1}\log\big\{p(z_{ki}|\tau_{ki},\hve{B})\big\}$$ with
respect to $\ve{\tau}_+$ and evaluated at $\ve{\hat\tau}_+$.

The calculation of $p(\vesub{Z}{k})$ is the same as
\eqref{laplace}:
\begin{equation}
p(\ve{Z}_k)= \int \exp\{\Psi(\ve{\tau}_k)\}d\ve{\tau}_k =
\exp\Big\{ \Psi(\ve{\hat\tau}_k) + \frac{N}{2}\log (2\pi) -
\half\log|\vess{C}{k}{-1}+\vesub{K}{k}| \Big\}, \label{p_z_k}
\end{equation}
where $\ve{\hat\tau}_k$ and $\vesub{K}{k}$ are defined similarly
as in \eqref{laplace}. If $\vesub{Z}{k}$ is part of the training data,
the calculation of $p(\vesub{Z}{k})$ is a by-product of
calculating the maximum likelihood estimates by Laplace approximation. The related value
obtained in the final iteration can be used here.
 Thus $\E(z^*|\D)$ follows from
\eqref{p_z_k} and \eqref{int_tilde_psi}.

We can also use \eqref{var1}-\eqref{var3} to calculate $\Var (z^*|\D)$. Here, 
 $\E[\Var(z^*|\tau^*,\D)]$ and $\Var[\E(z^*|\tau^*,\D)]$ can be obtained by Laplace approximation similarly to $\E(z^*|\D)$.

\section{Some technical details for consistency} 

{\bf Lemma 1:} Suppose $z_i$'s are independent samples from an exponential family given in \eqref{exfamily} and $\tau_0\in \CF$ has a Gaussian process prior with zero mean and bounded covariance function $k(\cdot,\cdot; \ve\theta)$ for any covariate values in $\mathcal{X}$. Suppose that $k(\cdot,\cdot; \ve\theta)$ is continuous in $ \ve\theta$ and the estimator $\hat{\ve\theta}_n \rightarrow  \ve\theta$ almost surely as $n\rightarrow \infty$.  If there exists a positive number $\kappa$ such that $|b''(\alpha)|\leq e^{\kappa\alpha}$, then 
\begin{align}
-\log p_{gp}(z_1,\ldots,z_n) + \log p_0(z_1,\ldots,z_n) \leq \half \lVert \tau_0\rVert^2_k + \half \log|\ve I + \delta \vesub{C}{nn}| + K,  \label{up:bound}
\end{align}
where $\lVert \tau_0\rVert_k$ is the reproducing kernel Hilbert space (RKHS) norm of $\tau_0$ associated with $k(\cdot,\cdot; \ve\theta)$, $\vesub{C}{nn}$ is the covariance matrix of $\tau_0$ over the covariates $\ve{X}_n$, $\ve I$ is the $n\times n$ identity matrix and $\delta$ and $K$ are some positive constants.

\medskip

{\it Proof.} Let $\mathcal{H}$ be the {\it Reproducing Kernel Hilbert Space} (RKHS) associated with the covariance function $k(\cdot,\cdot; \ve\theta)$, and $\mathcal{H}_n$ the span of $\{k(\cdot,\ve{x}_i; \ve\theta)\}$, i.e. $\mathcal{H}_n=\{f(\cdot): \; f(\ve{x})=\sum^n_{i=1}\alpha_i k(\ve{x},\ve{x}_i; \ve\theta),$ for any $\alpha_i\in \mathbb{R}\}$. We first assume the true underlying function $\tau_0\in \mathcal{H}_n$, then $\tau_0(\cdot)$ can be expressed as
$$\tau_0(\cdot)=\sum^n_{i=1}\alpha_i k(\cdot,\ve{x}_i; \ve\theta)\triangleq K(\cdot)\boldsymbol{\alpha} ,$$
where ${K(\cdot)}=(k(\cdot,\ve{x}_1; \ve\theta),\ldots,k(\cdot,\ve{x}_n; \ve\theta))$ and $\boldsymbol{\alpha}=(\alpha_1,\ldots,\alpha_n)^T$. By the properties of RKHS, $\lVert \tau_0 \rVert^2_k = \boldsymbol{\alpha}^T\vesub{C}{nn}\boldsymbol{\alpha}$, and $(\tau_0(\ve{x}_1),\ldots,\tau_0(\ve{x}_n))^T=\vesub{C}{nn}\boldsymbol{\alpha}$, where $\vesub{C}{nn}=(k(\ve{x}_i,\ve{x}_j; \ve\theta))$ is the covariance matrix over $\ve{x}_i,\; i=1,\ldots,n$.

Let $P$ and ${\bar P}$ be any two measures on $\mathcal{F}$, then it yields by Fenchel-Legendre duality relationship that, for any functional $g(\cdot)$ on $\mathcal{F}$,
\begin{equation} E_{\bar P}[g(\tau)] \leq \log E_P[ e^{g(\tau)}] + D[{\bar P},P] . \label{F-L} \end{equation}

Now in the above inequality let
\begin{description}
\item[(A1)] $g(\tau)$ be $\log p(z_1,\ldots,z_n|\tau)$ for any $z_1,\ldots,z_n$ in $\mathcal{Z}$ and $\tau\in\mathcal{F}$;
\item[(A2)] $P$ be the measure induced by $GP(0,k(\cdot,\cdot;\hat{\ve\theta}_n))$, hence its finite dimensional distribution at $z_1,\ldots,z_n$ is $\tilde{p}(z_1,\ldots,z_n) = \CN(0,{\boldsymbol{\hat C}}_{nn})$, and
\begin{equation} E_P[e^{g(\tau)}] = E_P[p(z_1,\ldots,z_n|\tau)] = \int_{\mathcal{F}} p(z_1,\cdots,z_n|\tau) d p_n(\tau) = p_{gp}(\ve{z}_n), \label{A2} \end{equation}
where $\boldsymbol{\hat C}_{nn}$ is defined in the same way as $\vesub{C}{nn}$ but with $ \ve\theta$ being replaced by its estimator $\hat{\ve\theta}_n$;
\item[(A3)] ${\bar P}$ be the posterior distribution of $\tau(\cdot)$ on $\mathcal{F}$ which has a prior distribution $GP(0,k(\cdot,\cdot; \ve\theta))$ and normal likelihood $\prod^n_{i=1}\CN(\hat{z}_i;\tau(\ve{x}_i),\sigma^2)$, where 
\begin{equation} 
\hat{\ve{z}}\triangleq \begin{pmatrix}\hat{z}_1 \\ \vdots \\ \hat{z}_n \end{pmatrix} = (\vesub{C}{nn}+\sigma^2 \ve{I})\boldsymbol{\alpha}, 
\label{pseudo-obs}
\end{equation}
and $\sigma^2$ is a constant to be specified. In other words, we assume a model $z=\tau(\ve{x})+e$ with $e\sim \CN(0,\sigma^2)$ and $\tau(\cdot)\sim GP(0,k(\cdot,\cdot; \ve\theta))$, and $\hat{\ve{z}}$ defined by \eqref{pseudo-obs} is a set of observations at $\ve{x}_1,\ldots,\ve{x}_n$. Thus, ${\bar P}(\tau)=p(\tau|\hat{\ve{z}},\ve{X}_n)$ is a probality measure on $\mathcal{F}$. Therefore, by Gaussian process regression, the posterior of $(\tau_1,\ldots,\tau_n)\triangleq (\tau(\ve{x}_1),\ldots,\tau(\ve{x}_n))$ is
\begin{align}
{\bar p}(\tau_1, \cdots, \tau_n) & \triangleq p(\tau_1, \cdots, \tau_n |\hat{\ve{z}},\ve{X}_n)  \nonumber \\
&=\CN(\vesub{C}{nn}(\vesub{C}{nn}+\sigma^2 \ve{I})^{-1}\hat{\ve{z}},\vesub{C}{nn}(\vesub{C}{nn}+\sigma^2 \ve{I})^{-1}\sigma^2) \nonumber \\
& =\CN(\vesub{C}{nn}\boldsymbol{\alpha}, \vesub{C}{nn}(\vesub{C}{nn}+\sigma^2 \ve{I})^{-1}\sigma^2) \\
& =\CN(\vesub{C}{nn}\boldsymbol{\alpha}, \vesub{C}{nn}B^{-1}) ,  \label{Q-finite}
\end{align}
where $B=\ve{I}+\sigma^{-2}\vesub{C}{nn}$.
\end{description}

\medskip

It follows that
\begin{align}
D[{\bar P},P] &= \int_{\CF} \log \big( \frac{d{\bar P}}{dP}\big) d{\bar P}  \nonumber \\
& = \int_{R^n} {\bar p}(\tau_1,\ldots,\tau_n) \log  \frac{{\bar p}(\tau_1,\ldots,\tau_n)}{\tilde{p}(\tau_1,\ldots,\tau_n)}  d\tau_1\ldots d\tau_n    \nonumber \\
&= \half \big[ \log |\hC| - \log |\vesub{C}{nn}| + \log |B| + \text{tr}(\hC^{-1}\vesub{C}{nn} B^{-1}) + (\vesub{C}{nn}\boldsymbol{\alpha})^T\hC^{-1}(\vesub{C}{nn}\boldsymbol{\alpha}) - n \big]   \nonumber \\
&= \half \big[ -\log |\hC^{-1}\vesub{C}{nn}| +\log |B| + \text{tr}(\hC^{-1}\vesub{C}{nn}B^{-1}) + \lVert \tau_0\rVert^2_k   \nonumber  \\
& \quad + \boldsymbol{\alpha}^T\vesub{C}{nn}(\hC^{-1}\vesub{C}{nn}-\ve{I})\boldsymbol{\alpha} - n  \big].  \label{K-L}
\end{align}

On the other hand,
$$E_{\bar P}[g(\tau)] = E_{\bar P}[\log p(z_1,\ldots,z_n|\tau)] = \sum^n_{i=1}E_{\bar P}[\log p(z_i|\tau(\ve{x}_i))].$$
By Taylor's expansion, expanding $\log p(z_i|\tau(\ve{x}_i))$ to the second order at $\tau_0(\ve{x}_i)$ yields
\begin{align*}
\log p(z_i|\tau(\ve{x}_i)) =&  \log p(z_i|\tau_0(\ve{x}_i)) + \frac{d \big[\log p(z_i|\tau(\ve{x}_i)) \big]} {d\tau(\ve{x}_i)}\Big |_{\tau(\ve{x}_i)=\tau_0(\ve{x}_i)} \big( \tau(\ve{x}_i)-\tau_0(\ve{x}_i) \big)  \\
& + \half \frac{d^2 \big[\log p(z_i|\tau(\ve{x}_i)) \big]} {[d\tau(\ve{x}_i)]^2}\Big |_{\tau(\ve{x}_i)=\tilde{\tau}(\ve{x}_i)} \big( \tau(\ve{x}_i)-\tau_0(\ve{x}_i) \big)^2,
\end{align*}
where $\tilde{\tau}(\ve{x}_i)=\tau_0(\ve{x}_i) + \lambda (\tau(\ve{x}_i)-\tau_0(\ve{x}_i))$ for some $0\leq \lambda\leq 1$.

For canonical link function, we have
$$p(z_i|\tau(\ve{x}_i)) = \exp\left\{ \frac{z_i\tau(\ve{x}_i)-b(\tau(\ve{x}_i))}{a(\phi_i)} + c(z_i,\phi_i)  \right\} ,$$
thus
$$ \frac{d^2 \big[\log p(z_i|\tau(\ve{x}_i)) \big]} {[d\tau(\ve{x}_i)]^2}\Big |_{\tau(\ve{x}_i)=\tilde{\tau}(\ve{x}_i)} 
= -\frac {b''(\tilde{\tau}(\ve{x}_i))}{a(\phi_i)}. $$
It follows that
\begin{align*}
E_{\bar P}\big[\log p(z_i|\tau(\ve{x}_i))\big] = &  \log p(z_i|\tau_0(\ve{x}_i)) + \frac{d \big[\log p(z_i|\tau(\ve{x}_i)) \big]} {d\tau(\ve{x}_i)}\Big |_{\tau(\ve{x}_i)=\tau_0(\ve{x}_i)} E_{\bar P}\big[ \tau(\ve{x}_i)-\tau_0(\ve{x}_i) \big]  \\
& - \frac{1}{2a(\phi_i)} E_{\bar P} \big[ b''(\tilde{\tau}(\ve{x}_i)) \big( \tau(\ve{x}_i)-\tau_0(\ve{x}_i) \big)^2 \big].
\end{align*}
Since ${\bar P}(\cdot)$ is the posterior of $\tau(\cdot)$ which has prior $GP(0,k(\cdot,\cdot; \ve\theta))$ and normal likelihood $\prod^n_{i=1}\CN(\hat{z}_i;\tau(\ve{x}_i),\sigma^2)$, $\tau(\ve{x}_i)$ is normally distributed under ${\bar P}$ and it follows from \eqref{Q-finite} that
\begin{align*}
\tau(\ve{x}_i) &\sim \CN(C^{(i)}_{nn}\boldsymbol{\alpha}, (\vesub{C}{nn}B^{-1})_{ii})  \\
& = \CN(\tau_0(\ve{x}_i), (\vesub{C}{nn}B^{-1})_{ii}) \triangleq \CN(\tau_{0i},c_{ii}) ,
\end{align*} 
where $C^{(i)}_{nn}$ denotes the $i$th row of $\vesub{C}{nn}$ and $(\vesub{C}{nn}B^{-1})_{ii}$ is the $i$th diagonal element of $\vesub{C}{nn}B^{-1}$. Therefore,
$E_{\bar P}\big[ \tau(\ve{x}_i)-\tau_0(\ve{x}_i) \big]=0$ and 
\begin{align*}
& E_{\bar P} \big[ b''(\tilde{\tau}(\ve{x}_i)) \big( \tau(\ve{x}_i)-\tau_0(\ve{x}_i) \big)^2 \big] 
\leq E_{\bar P} \big[ e^{\kappa \tilde{\tau}(\ve{x}_i)} \big( \tau(\ve{x}_i)-\tau_0(\ve{x}_i) \big)^2 \big]  \\
= & \int^{+\infty}_{-\infty} (\tau_i-\tau_{0i})^2 e^{\kappa\tau_{0i} + \kappa\lambda (\tau_i-\tau_{0i})}  \CN(\tau_{0i}, c_{ii}) d\tau_i  \\
= & \;e^{\kappa\tau_{0i}+\half \kappa^2\lambda^2 c_{ii}} (\kappa^2\lambda^2c_{ii}+1)c_{ii} \leq \tilde{\delta} c_{ii},
\end{align*}
since the covariance function is bounded. Here $\tilde\delta$ is a generic positive constant.

Thus, we have
\begin{align*}
E_{\bar P}\big[\log p(z_i|\tau(\ve{x}_i))\big] \geq  \log p(z_i|\tau_0(\ve{x}_i)) - \half \tilde\delta (\vesub{C}{nn}B^{-1})_{ii} ,
\end{align*}
and
\begin{align*}
\sum^n_{i=1} E_{\bar P}\big[\log p(z_i|\tau(\ve{x}_i))\big] \geq  \sum^n_{i=1}\log p(z_i|\tau_0(\ve{x}_i)) - \half \tilde\delta \; \text{tr}(\vesub{C}{nn}B^{-1}) ,
\end{align*}
i.e.
\begin{equation} \log p_0(z_1,\ldots,z_n) \leq E_{\bar P}[g(\tau)] + \half  \tilde\delta \; \tr(\vesub{C}{nn}B^{-1}) . \label{E-Q} \end{equation}

Combining the bounds \eqref{A2}, \eqref{K-L}, \eqref{E-Q} and applying \eqref{F-L} gives
\begin{align}
& -\log p_{gp}(z_1,\ldots,z_n) + \log p_0(z_1,\ldots,z_n)   \nonumber \\
\leq & -\log E_P[e^{g(\tau)}] + E_{\bar P}[g(\tau)] + \half  \tilde\delta \;\tr(\vesub{C}{nn}B^{-1})  \nonumber  \\
\leq & \; D[{\bar P},P] + \half  \tilde\delta \;\tr(\vesub{C}{nn}B^{-1})  \nonumber  \\
= & \half \lVert \tau_0\rVert^2_k + \half \big[ -\log |\hC^{-1}\vesub{C}{nn}| +\log |B| + \text{tr}(\hC^{-1}\vesub{C}{nn}B^{-1}+ \tilde\delta \vesub{C}{nn}B^{-1})   \nonumber  \\
& + \boldsymbol{\alpha}^T\vesub{C}{nn}(\hC^{-1}\vesub{C}{nn}-\ve{I})\boldsymbol{\alpha} - n  \big].  \label{RHS}
\end{align}

Since the covariance function is continuous in $ \ve\theta$ and $\hve\theta_n \rightarrow  \ve\theta$ we have $\hC^{-1}\vesub{C}{nn}-\ve{I} \rightarrow 0$ as $n\rightarrow\infty$. Therefore there exist some positive constants $K$ and $\epsilon$ such that
$$ -\log |\hC^{-1}\vesub{C}{nn}| < K, \quad  \boldsymbol{\alpha}^T\vesub{C}{nn}(\hC^{-1}\vesub{C}{nn}-\ve{I})\boldsymbol{\alpha} < K, $$
$$\tr( \hC^{-1}\vesub{C}{nn}B^{-1}) < \tr((\ve{I}+\epsilon \vesub{C}{nn})B^{-1}),$$
since the covariance function is bounded.

Thus
$$ \text{ RHS of \eqref{RHS}} < \half \lVert \tau_0\rVert^2_k + \half \big[ 2K +\log |B| 
+ \tr\big( (\ve{I} + (\epsilon+\tilde\delta)\vesub{C}{nn})B^{-1} \big) -n  \big ]. $$

Note that the above inequality holds for all $\sigma^2>0$, thus letting $\sigma^2=(\epsilon+\tilde\delta)^{-1}$ 
and $\delta=\epsilon+\tilde\delta$ yields that the RHS becomes
$$\half \lVert \tau_0\rVert^2_k + \half \log|\ve{I} + \delta \vesub{C}{nn}| + K.$$
Thus, we have
\begin{equation}
-\log p_{gp}(z_1,\ldots,z_n) \leq -\log p_0(z_1,\ldots,z_n) + \half \lVert \tau_0\rVert^2_k + \half \log|\ve{I} + \delta \vesub{C}{nn}| + K, 
\label{H_n_case}
\end{equation}
for any $\tau_0(\cdot) \in \mathcal{H}_n $.

Taking infimum on RHS of \eqref{H_n_case} over $\tau_0$ and applying {\it Representer Theorem} (see Lemma 2 in Seeger et al. (2008)) we obtain
$$ -\log p_{gp}(z_1,\ldots,z_n) + \log p_0(z_1,\ldots,z_n) \leq \half \lVert \tau_0\rVert^2_k + \half \log|\ve{I} + \delta \vesub{C}{nn}| + K$$
for all $\tau_0(\cdot) \in \mathcal{H}$. The proof is complete.  \hfill $\Box$

\bigskip
{\it Proof of Theorem 1.} It follows from the definition of information consistency that
\begin{align*}
D[p_0(\ve{z}_n),p_{gp}(\ve{z}_n)] & = \int_{\mathcal{Z}^n} p_0(z_1,\cdots,z_n)\log \frac{p_0(z_1,\cdots,z_n)}{p_{gp}(z_1,\cdots,z_n)} dz_1\cdots dz_n \\
&= \int_{\mathcal{Z}^n} p_0(z_1,\cdots,z_n)[-\log p_{gp}(z_1,\ldots,z_n) + \log p_0(z_1,\ldots,z_n)]dz_1\cdots dz_n.
\end{align*}
Applying Lemma 1 we obtain that
\begin{equation}\frac 1 n  E_{\ve{X}_n}\Big(  D[p_0(\ve{z}_n),p_{gp}(\ve{z}_n)] \Big) \leq \frac 1 {2n} \lVert \tau_0\rVert^2_k + \frac 1 {2n} E_{\ve{X}_n}\Big(\log|\ve I + \delta \vesub{C}{nn}|\Big) + \frac K n, \label{errorb}
\end{equation}
where $\delta$ and $K$ are two postive constants. Theorem 1 follows from \eqref{errorb}.   \hfill $\Box$

\bigskip
{\bf Remark A.1.} Lemma 1 requires that the estimator of the hyperparameter $ \ve\theta$ is consistent. We now prove that the estimator
by maximizing the marginal likelihood based on Laplace approximation \eqref{laplace} satisfies this condition when the number of curves and the number of observations on each curve are sufficiently large. The method of proof is similar to \cite{Vonesh1996}. We still assume that the mean function is known and consider the estimation of the hyperparameter $ \ve\theta$ only. Suppose that we have $M$ independent curves and, for simplicity, there are equal number $n$ of observations on each curve. Then the marginal log-likelihood is given by
\[
l({ \ve\theta}) = \sum^M_{m=1}\log \int
\exp\{\Psi(\ve{\tau}_m)\}d\ve{\tau}_m,
\]
where $\Psi(\ve{\tau}_m)$ is defined as
\begin{equation*}
\Psi(\ve{\tau}_m) = \sum^{n}_{i=1}\log\big\{p(z_{mi}|\tau_{mi})\big\} - \frac{n}{2}\log (2\pi) - \half\log|\vesub{C}{m}| -\half \ve{\tau}_m^T\vess{C}{m}{-1}\ve{\tau}_m.
\end{equation*}
Its Laplace approximation is 
\begin{equation*}
l^*( \ve\theta) = \sum^M_{m=1}\Psi(\ve{\hat\tau}_m) + \frac{nM}{2}\log (2\pi) -
\half\sum^M_{m=1}\log|\vess{C}{m}{-1}+\vesub{K}{m}|,
\end{equation*}
where $\ve{\hat\tau}_m$ is the maximiser of $\Psi(\ve{\tau}_m)$. \cite{Evangelou2011} proved that
$$l( \ve\theta) = l^*( \ve\theta) + O(nM^{-1}). $$

Now let $U( \ve\theta) = \partial l( \ve\theta)/\partial \ve\theta$ and $U^*( \ve\theta) = \partial l^*( \ve\theta)/\partial \ve\theta$ and let $\hve\theta$ be the maximum likelihood estimator based on the Laplace approximation, i.e. $\hve\theta$ satisfying $U^*(\hve\theta)=0$. Under usual regularity conditions on $l( \ve\theta)$ and assuming $\hve\theta$ is an interior point in a neighbourhood of the true hyperparameter $ \ve\theta$, by Taylor expansion about $ \ve\theta$ we have
$$ M^{-1}U(\hve\theta) = M^{-1}U( \ve\theta) + M^{-1}H( \ve\theta)(\hve{\theta}-  \ve\theta) + O_p(1)( \lVert \hve{\theta}-  \ve\theta \rVert^2) ,  $$
where $H( \ve\theta)$ is the Hessian matrix of $l( \ve\theta)$.

Given sufficient regularity conditions on $l( \ve\theta)$, we have 
$$M^{-1}H( \ve\theta) = O_p(1), \quad M^{-1}U( \ve\theta)=O_p(M^{-\half}) \quad\text{and}\quad   
M^{-1}U(\hve\theta) = M^{-1} U^*(\hve\theta) + O(nM^{-2}) . $$ 

It follows that
$$ M^{-1}U(\hve\theta) = M^{-1}U( \ve\theta) + O_p(1)( \hve{\theta}-  \ve\theta), $$
and hence
\begin{align*}
\hve\theta -  \ve\theta &= \frac{M^{-1}U(\hve\theta) - M^{-1}U( \ve\theta)}{O_p(1)} = M^{-1} U^*(\hve\theta) + O(nM^{-2}) +  O_p(M^{-\half})  \\
& = O_p(\max\{nM^{-2},M^{-\half} \}).
\end{align*}
Therefore, the estimator $\hve\theta \rightarrow  \ve\theta$ almost surely if $M$ tends to infinity and $n = o(M^{2})$.

For Gaussian process regression model (where $M=1$) the consistency of the empirical Bayesian estimator of the hyper-parameters as $n\rightarrow\infty$ is proved in \cite{Yi11} under certain regularity conditions.  

\medskip
{\bf Remark A.2.} The consistency considered in Theorem 1 assumes the mean function is known. If the mean function is unknown and is estimated from the observations, its uncertainty needs to be taken into account. In fact, denote by $\hat\mu(t)$ the estimator of the mean function $\mu(t)$ and let
$$\hat p_{gp}(\ve{z}_n) = \int_{\mathcal{F}} \hat p(z_1,\cdots,z_n|\tau(\ve{X}_n))dp_n(\tau)$$
where $\hat p(z_1,\cdots,z_n|\tau(\ve{X}_n))$ is the conditional distribution of $z_1,\cdots,z_n$ with the estimated mean function $\hat\mu(t)$. It follows from Lemma 1 that
\begin{align*}
& -\log \hat p_{gp}(z_1,\ldots,z_n) + \log p_0(z_1,\ldots,z_n)  \\
=& \log p_{gp}(z_1,\ldots,z_n) - \log \hat p_{gp}(z_1,\ldots,z_n) -\log p_{gp}(z_1,\ldots,z_n) + \log p_0(z_1,\ldots,z_n) \\
\leq & \log p_{gp}(z_1,\ldots,z_n) - \log \hat p_{gp}(z_1,\ldots,z_n) + \half \lVert \tau_0\rVert^2_k + \half \log|\ve I + \delta \vesub{C}{nn}| + K.  
\end{align*}
For canonical link function, we have
$$\hat p(z_1,\cdots,z_n|\tau(\ve{X}_n)) = \exp\left\{ \sum^n_{i=1}\frac{z_i(\hat\mu+\tau(\ve{x}_i))-b(\hat\mu+\tau(\ve{x}_i))}{a(\phi_i)} + \sum^n_{i=1}c(z_i,\phi_i)  \right\} 
\triangleq e^{g(\hat\mu+\tau)},$$
$$ p(z_1,\cdots,z_n|\tau(\ve{X}_n)) = \exp\left\{ \sum^n_{i=1}\frac{z_i(\mu+\tau(\ve{x}_i))-b(\mu+\tau(\ve{x}_i))}{a(\phi_i)} + \sum^n_{i=1}c(z_i,\phi_i)  \right\} \triangleq e^{g(\mu+\tau)} .$$
If $z_i$ has finite first two moments and its variance is bounded away from zero, there exist positive constants $K_1$, $K_2$ and $K_3$ such that $|b'(\cdot)|<K_1$ and $K_2<a(\cdot)<K_3$. It follows that
$$ b(\hat\mu+\tau) - b(\mu+\tau) \leq K_1 \lVert \hat\mu - \mu \rVert, \;\text{ or, }\; 
 - b(\mu+\tau) \leq K_1 \lVert \hat\mu - \mu \rVert - b(\hat\mu+\tau). $$
Hence,
\begin{align*}
g(\mu+\tau) &\leq \sum^n_{i=1}\frac{z_i(\mu-\hat\mu) + K_1 \lVert \hat\mu - \mu \rVert }{a(\phi_i)} + g(\hat\mu+\tau)  \\ 
&\leq \frac{\sum^n_{i=1}(|z_i|+K_1)}{K_2} \lVert \hat\mu - \mu \rVert + g(\hat\mu+\tau) . 
\end{align*}
It yields that
\begin{align*}
\log p_{gp}(z_1,\ldots,z_n) - \log \hat p_{gp}(z_1,\ldots,z_n)  
= & \log \frac{\int_{\mathcal{F}} e^{g(\mu+\tau)} dp_n(\tau) }{ \int_{\mathcal{F}} e^{g(\hat\mu+\tau)} dp_n(\tau) }  \\
\leq &  \frac{\sum^n_{i=1}(|z_i|+K_1)}{K_2} \lVert \hat\mu - \mu \rVert . 
\end{align*}
Therefore, following the same argument as in \eqref{errorb} we obtain
\begin{equation*}\frac 1 n  E_{\ve{X}_n}\Big(  D[p_0(\ve{z}_n),\hat p_{gp}(\ve{z}_n)] \Big) 
\leq \tilde{K}\lVert \hat\mu - \mu \rVert + \frac 1 {2n} \lVert \tau_0\rVert^2_k + \frac 1 {2n} E_{\ve{X}_n}\Big(\log|\ve I + \delta \vesub{C}{nn}|\Big) + \frac K n, 
\end{equation*}
where $\tilde{K}$, $\delta$ and $K$ are postive constants.

It is obvious that $\hat p_{gp}(\ve{z}_n)$ is information consistent if the mean function $\mu(t)$ is consistent. Therefore the information consistency of $\hat z(\cdot)$ also depends on the convergence of the mean function in this case. The problem of consistency of mean function in functional data analysis has been studied under various circumstances by a number of authors,
see for example \cite{Li2010} and \cite{Cai2011}, among others. Particularly, \cite{Li2010} proved that the local linear estimator of the mean function is consistent and the convergence rate depends on both the number of curves and the
number of observations on each curve, and \cite{Cai2011} studied the minimax convergence rate of the mean function and revealed the phase transition phenomena. However, the consistency of the mean function for generalized Gaussian process functional regression is still an open problem and worth further investigation.

\section{Ordinal Data}
We further demonstrate the proposed method using simulated ordinal data.
The true model used to generate the latent process is 
$y_{mi}(x_{mi}) = 1/ {(1+\exp(-1.5x_{mi}))} + \tau_{mi},$
where, for each $m$,  $x_{mi}(=t_{mi})$ are equally spaced points in $(-4,4)$ and $\{ \tau_{mi} \}$  is a Gaussian process with zero mean and the squared exponential covariance function defined in \eqref{covfun0} 
with $v_1=0.0049$, $w_1=0.33$ and $a_1=0.01$. The observations $z_{mi}$ are generated as follows:
\begin{equation}
z_{mi} = \left\{
\begin{array}{ll}
0 & \text{ if } y_{mi}\leq 0.2,  \\
1 & \text{ if } 0.2< y_{mi}\leq 0.7,  \\
2 & \text{ if } y_{mi}> 0.7 . 
\end{array}
\right.
\end{equation}

\begin{figure}
\begin{center}
\subfigure[\label{probita}]{\includegraphics[width=0.43\linewidth]{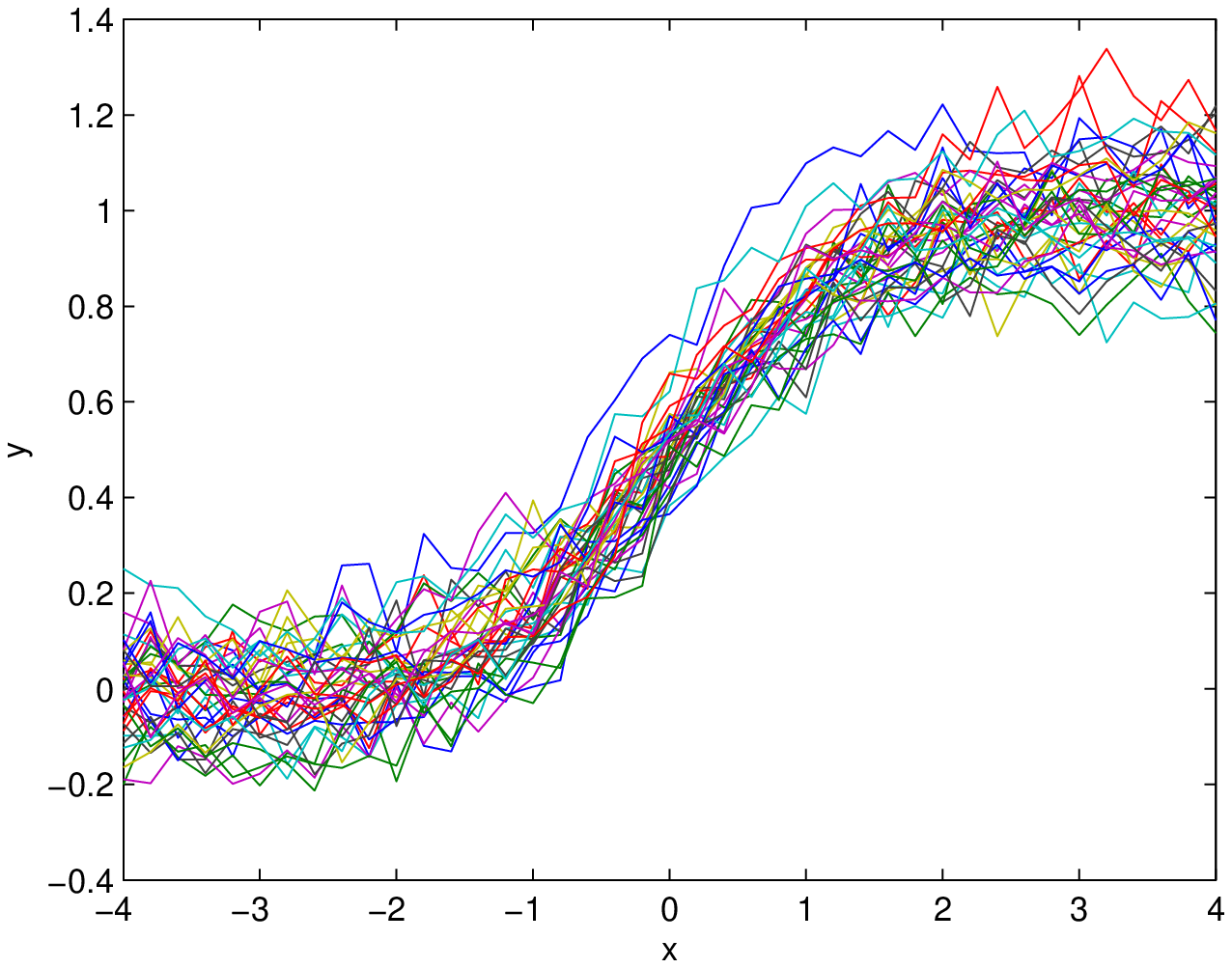}}
\subfigure[\label{probitb}]{\includegraphics[width=0.43\linewidth]{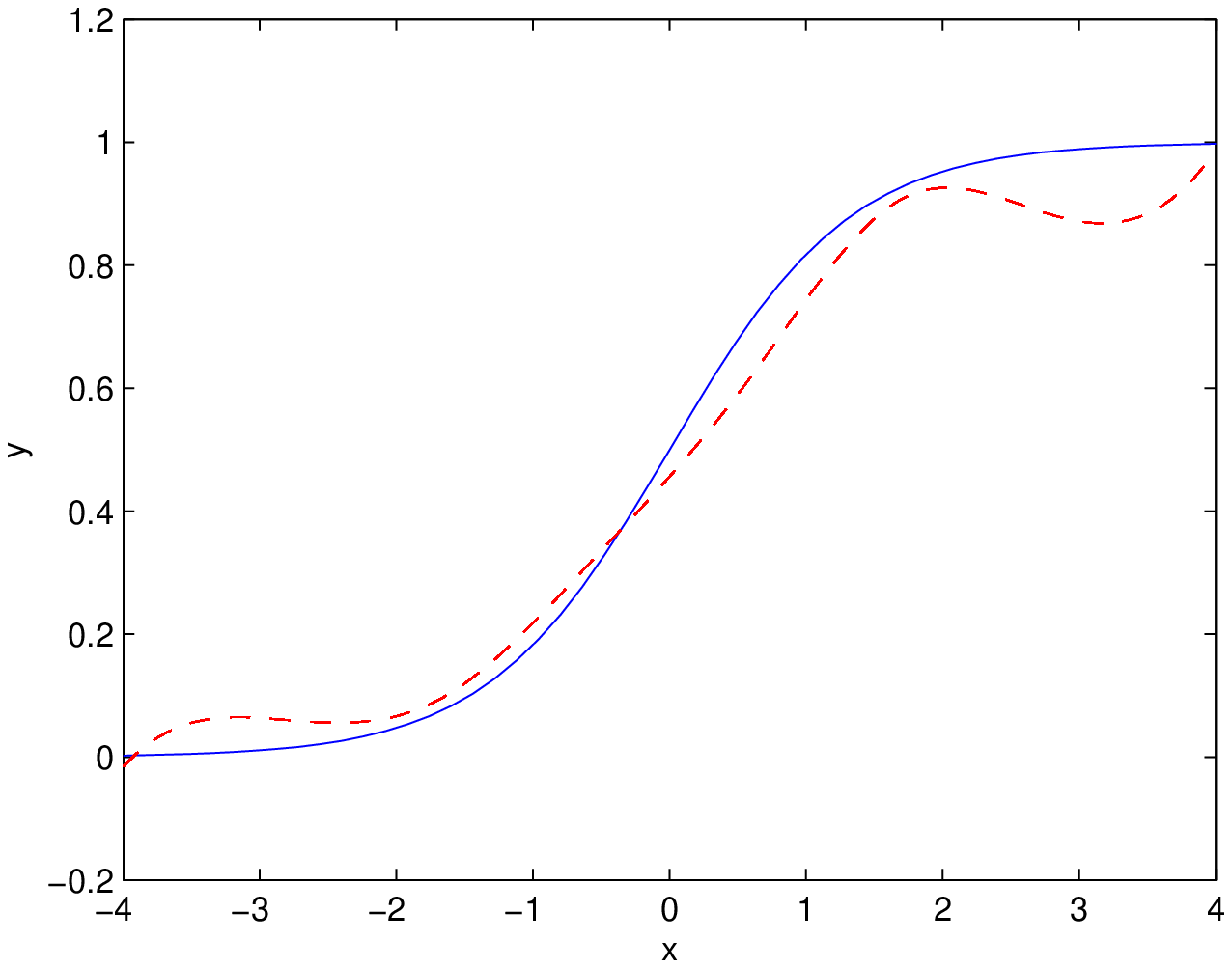}}
\end{center}
\caption{Ordinal Data. (a) Forty sample underlying curves; and (b) the estimated (dashed) and true (solid) mean curves of $\beta(t)$. }
\label{latent_curves}
\end{figure}

A sample of forty underlying curves, each containing 40 data points, is shown in Figure \ref{latent_curves}(a). Note that as commonly used in Gaussian process regression methods a small amount of ``jitter" (noise) is added in order to avoid the singularity of the covariance matrix and to make the matrix computations better conditioned. We use a generalized GPFR model with probit link function to model these ordinal data. That is, for a data set with $r$ ordered categories, we define $y_m(t)=\beta(t)+\tau_m(t)$ where $\tau_m(t)$ follows a GPR model and
\[
z_m(t)=j \ \mbox{ if } \ b_{j} < y_m(t) \leq b_{j+1} \ \mbox{ for } \ j=0,1,2, \ldots,  r-1,
\]
where $b_0=-\infty$, $b_{r}=\infty$, and $b_j$ for $ j=1,\ldots, r-1$ are the thresholds to be estimated. The density function for $\vesub{z}{m}=\{{z}_{m1}, \ldots,{z}_{mN_m}\}$ is given by
\[
p(\vesub{z}{m}|\vesub{y}{m})=\prod_{i=1}^{N_m}p(z_{mi}|{y}_{mi})=\prod_{i=1}^{N_m}p(b_{z_{mi}} < {y}_{mi} \leq b_{z_{mi}+1} ).
\]
The marginal log-likelihood is calculated by \eqref{eq:loglike}, and the empirical Bayesian estimates of the B-spline coefficients, the hyper-parameters and the thresholds can then be obtained. 

In this example $r=3$ and the thresholds $b_1$ and $b_2$ are unknown parameters. 
The estimated mean curve is shown in Figure \ref{latent_curves}(b) along with the true mean curve. The estimates of the hyper-parameters $(\hat v_1,\hat w_1,\hat a_1)=(0.0053,0.3310,0.0100)$, and the thresholds $(\hat b_1,\hat b_2) = (0.2875,0.6351)$.

\begin{figure}[h]
\begin{center}
\subfigure[\label{Ord_fig11} Interpolation]{\includegraphics[width=0.43\linewidth]{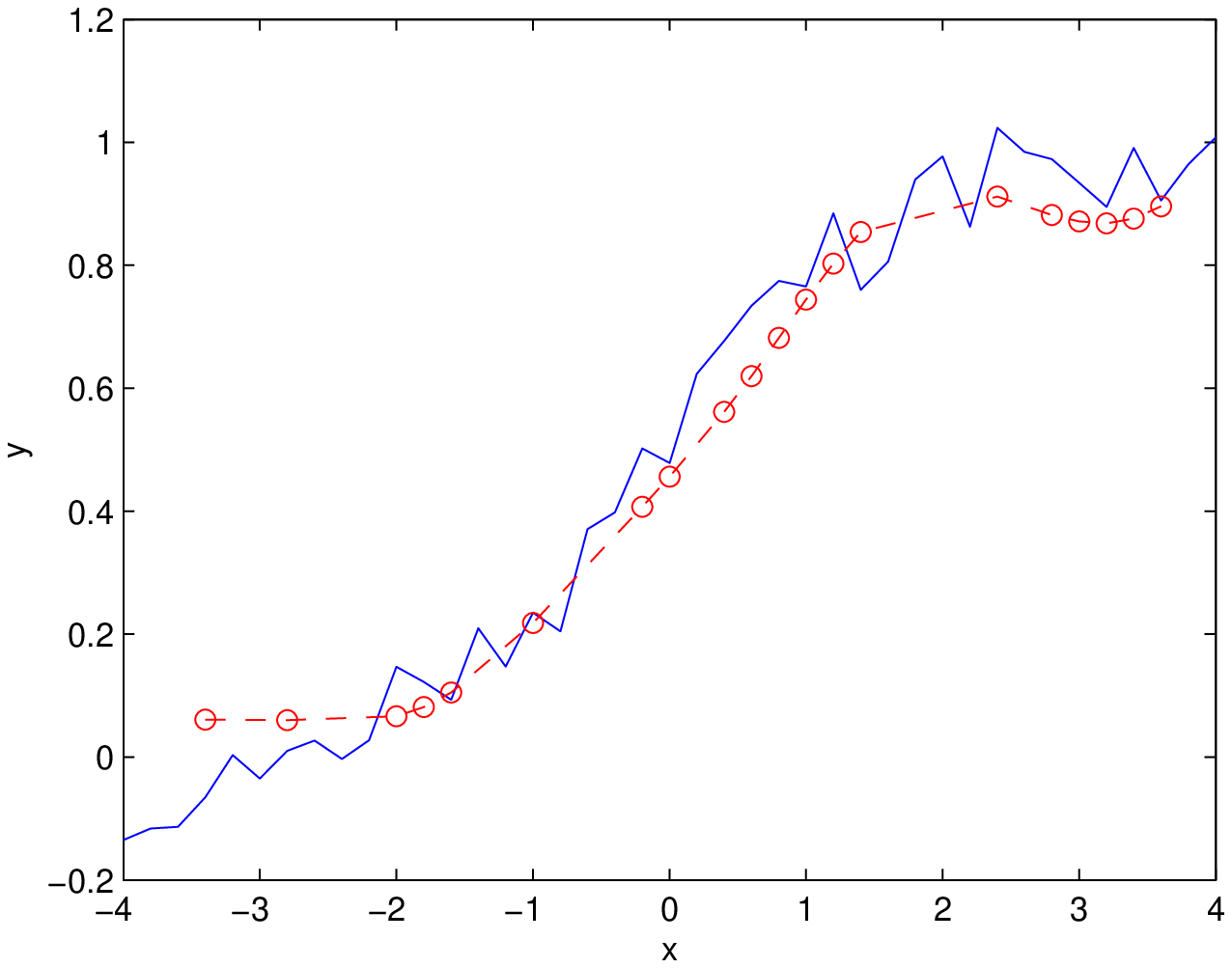}} 
\subfigure[\label{Ord_fig12} Interpolation]{\includegraphics[width=0.43\linewidth]{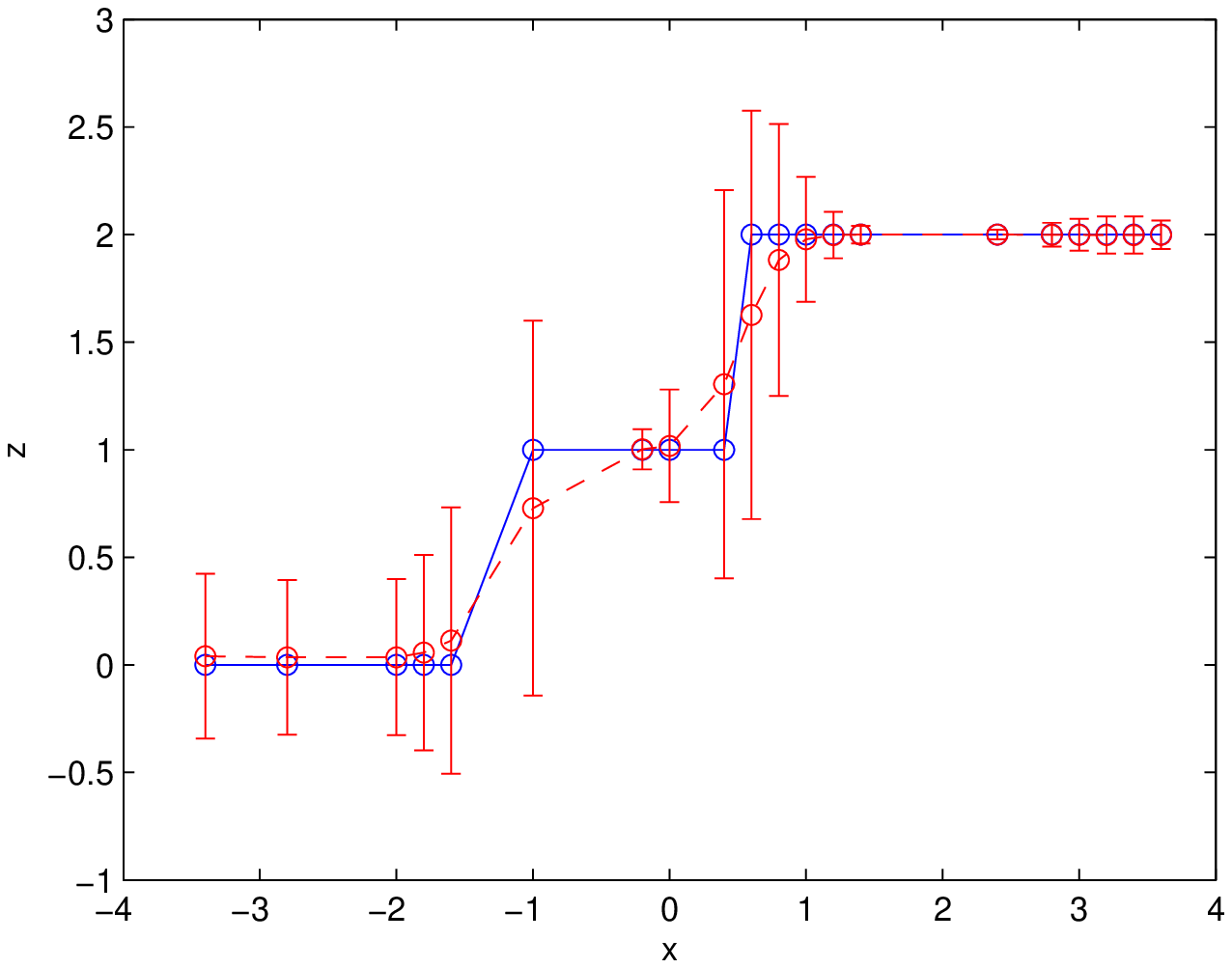}} \\
\subfigure[\label{Ord_fig13} Extrapolation]{\includegraphics[width=0.43\linewidth]{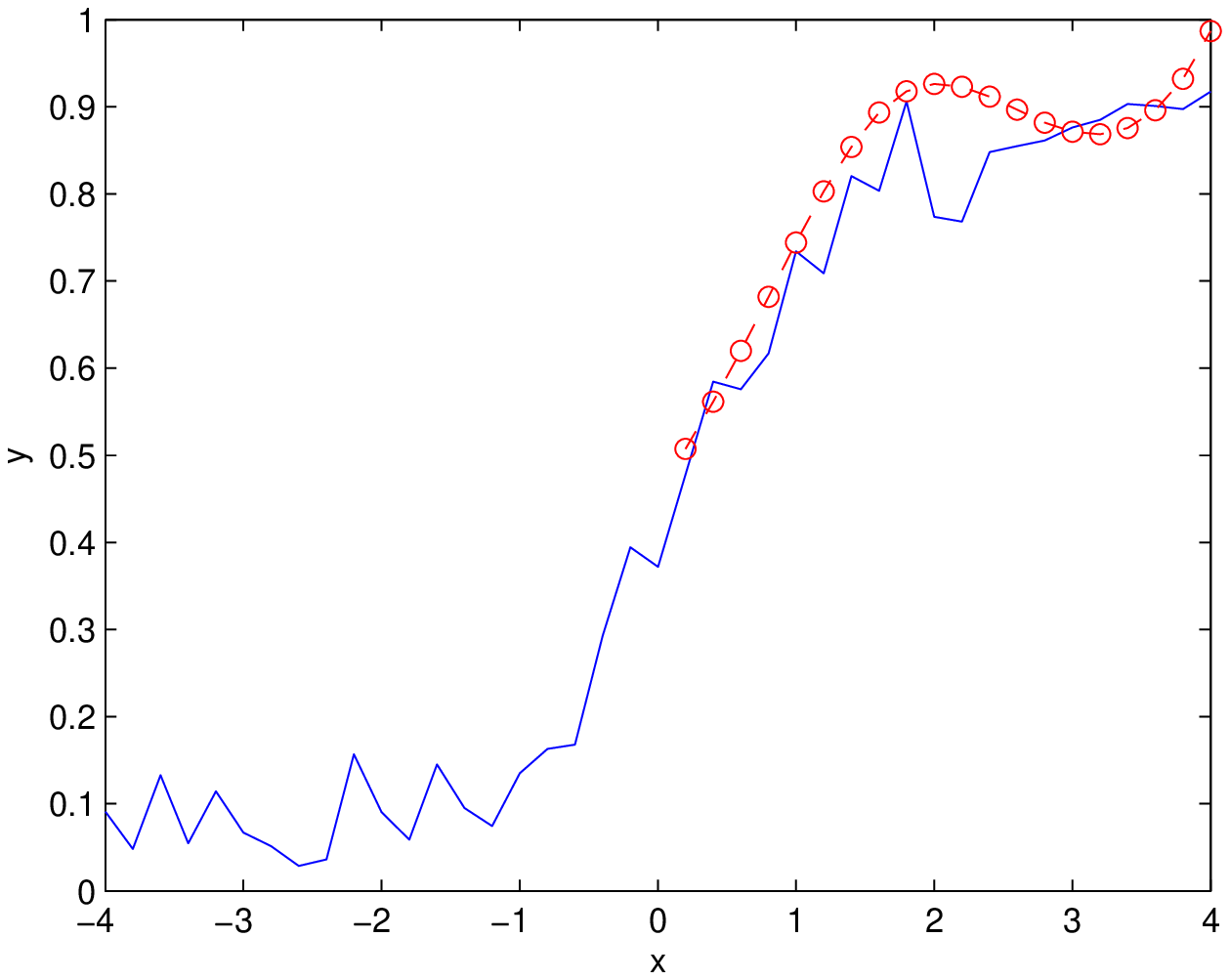}}
\subfigure[\label{Ord_fig14} Extrapolation]{\includegraphics[width=0.43\linewidth]{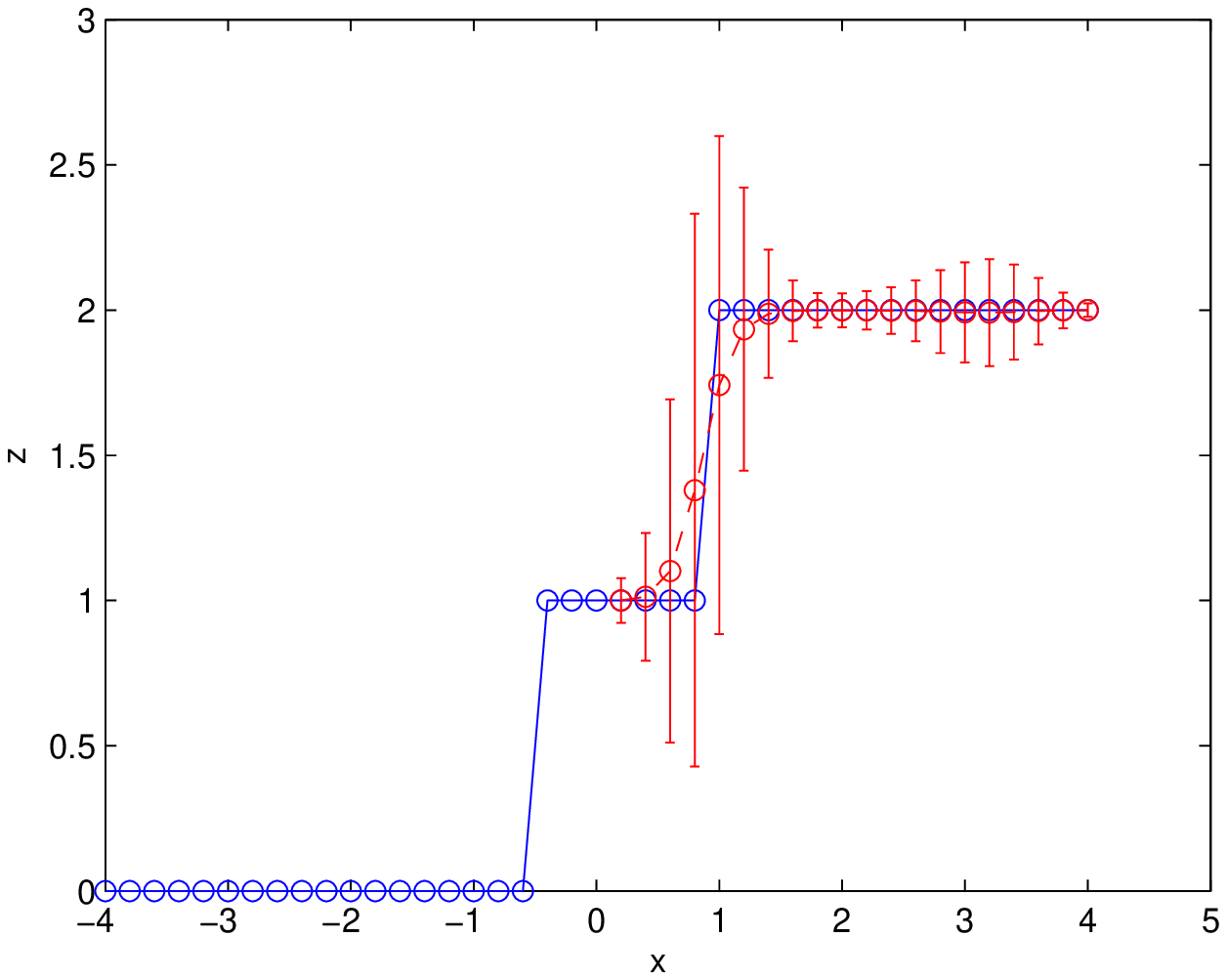}}
\end{center}
\caption{Ordinal Data. Left panels: the true (solid) and estimated (dashed) underlying processes $y_m(t)$. Right panels: the true observations (solid),  the predicted values (dashed) and 95\% error bars for $z_{m}(t)$. The circles represent the data points. }
\label{fig_ord_output}
\end{figure}

We now consider the problem of predictions. We generate a new curve with a total number of 40 data points, of which half are randomly selected as observations to estimate the underlying process and the remaining points are used as test data to make prediction. This is an interpolation problem. The predictive means and variances of the response at the test points are calculated by the formulae \eqref{pred_mean0} and \eqref{var1}, and the results are then compared with the true response values. The average error rate based on 30 repetitions is 5\%, a pretty good result. A randomly selected sample of observations and the predictions are shown in the top panels of Figure \ref{fig_ord_output}. 

Next we consider the problem of extrapolation, i.e., select the first half of the data as observations and predict the remaining half, and compare the predicted responses with the actual observations. The average error rate based on 30 repetitions is 5.75\%, which is also a very good result. A randomly selected sample of observations and the predictions are shown in the bottom panels of Figure \ref{fig_ord_output}.

\section{Primary Biliary Cirrhosis Data}

Primary biliary cirrhosis (PBC) is a rare but fatal chronic liver disease for which there is no totally effective treatment other than liver transplantation \citep{Murtaugh94}. The data used in this paper were from a study of the progression of PBC in 312 patients who were seen at the Mayo Clinic between January 1974 and May 1984 and a follow-up to April 30 1988. The patients were scheduled to have measurements of blood characteristics at 6 months, 1 year and annually thereafter post diagnosis and generated 1945 patient visits.

\begin{figure}
\begin{center}
\subfigure[\label{fig_mean}]{\includegraphics[width=0.33\linewidth]{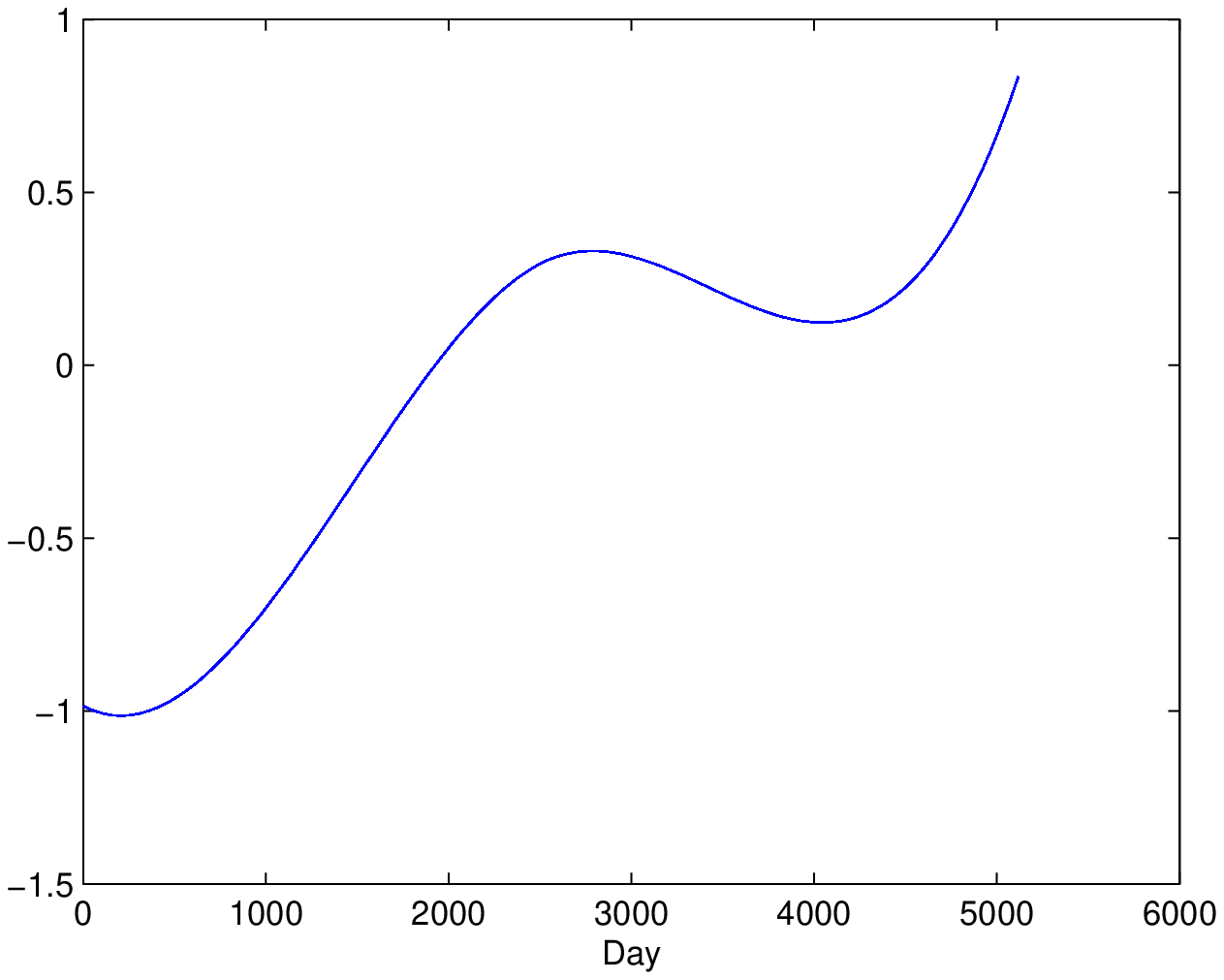}} \\
\subfigure[\label{fig22}]{\includegraphics[width=0.33\linewidth]{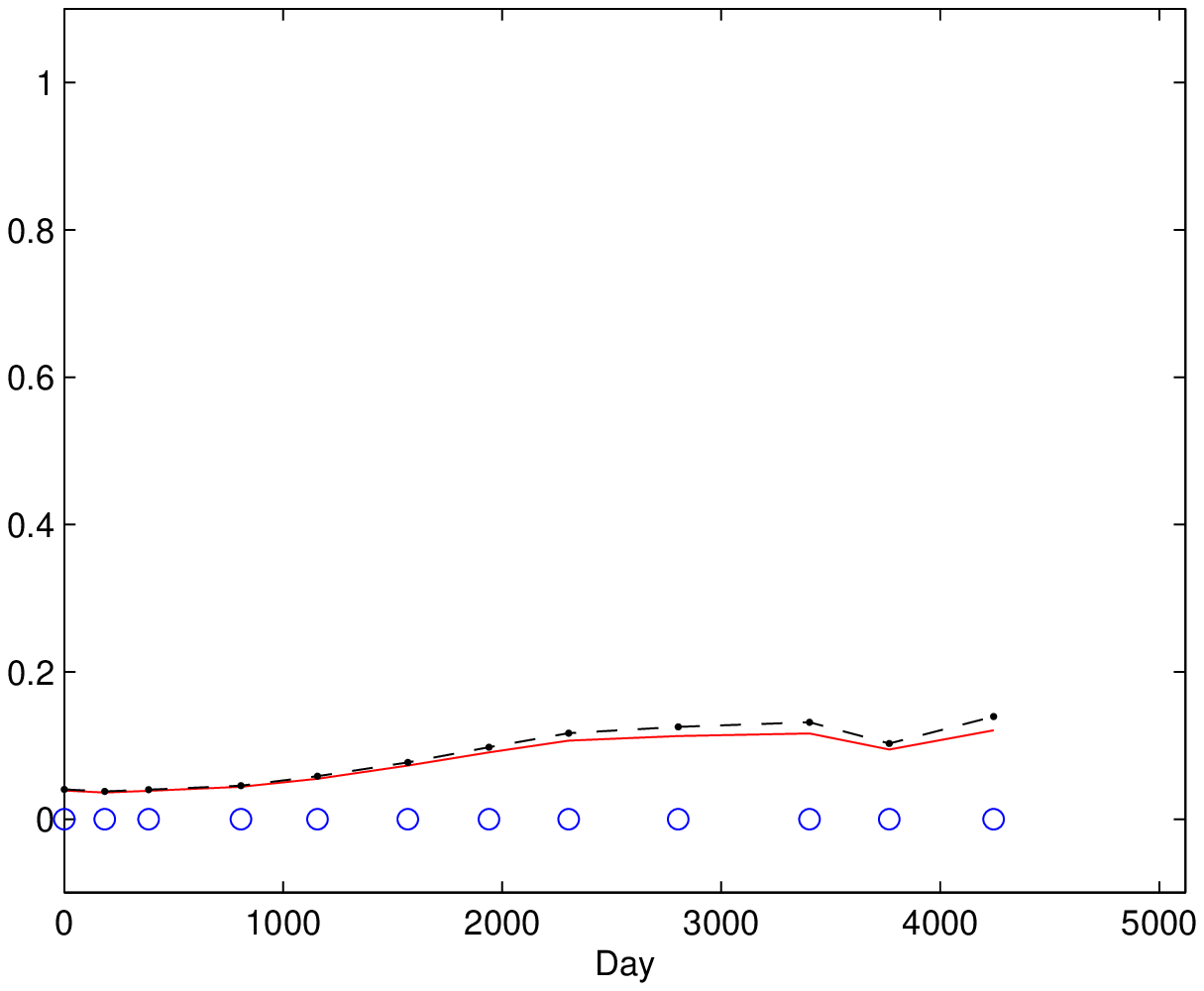}} 
\subfigure[\label{fig32}]{\includegraphics[width=0.33\linewidth]{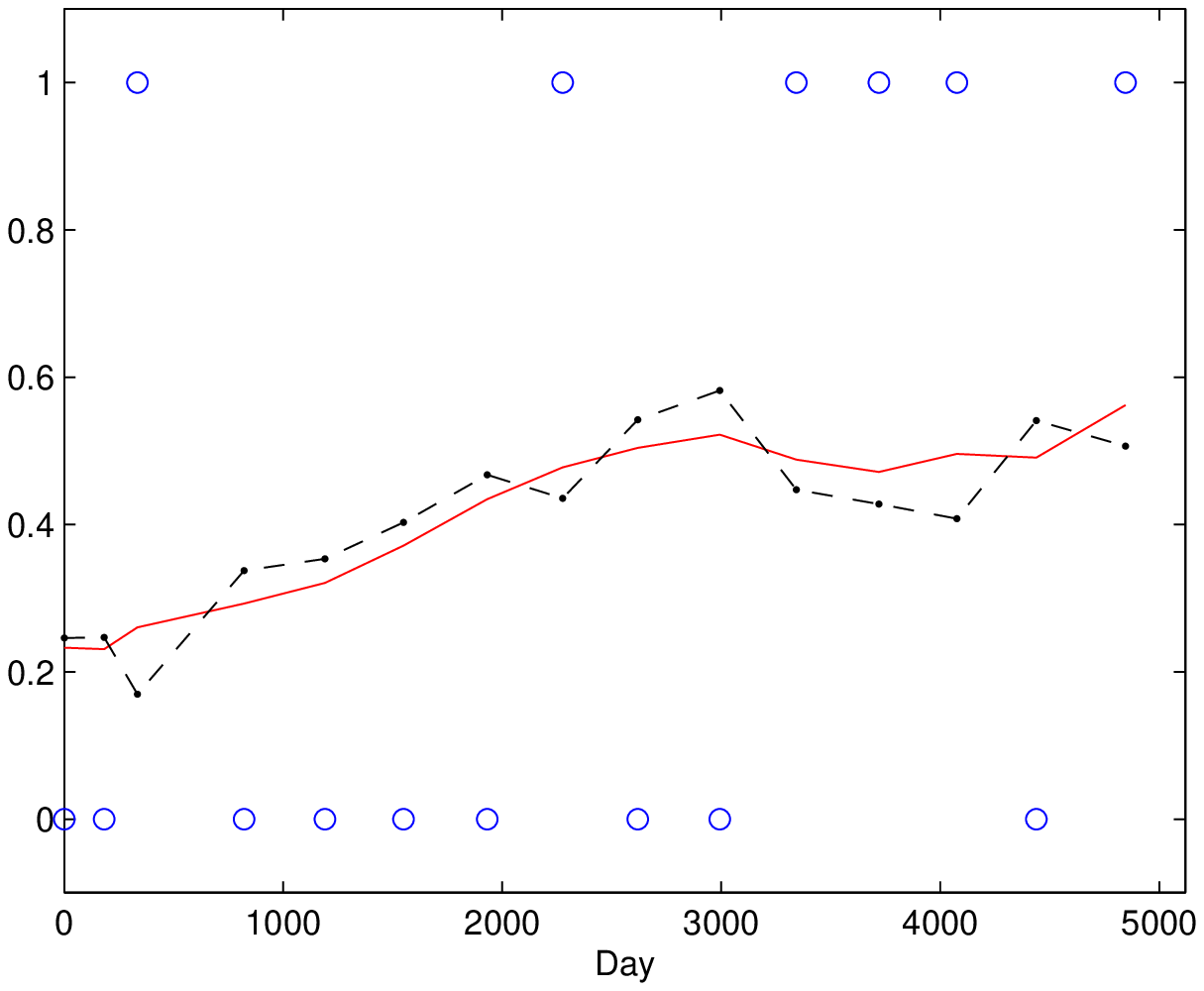}}  
\subfigure[\label{fig42}]{\includegraphics[width=0.33\linewidth]{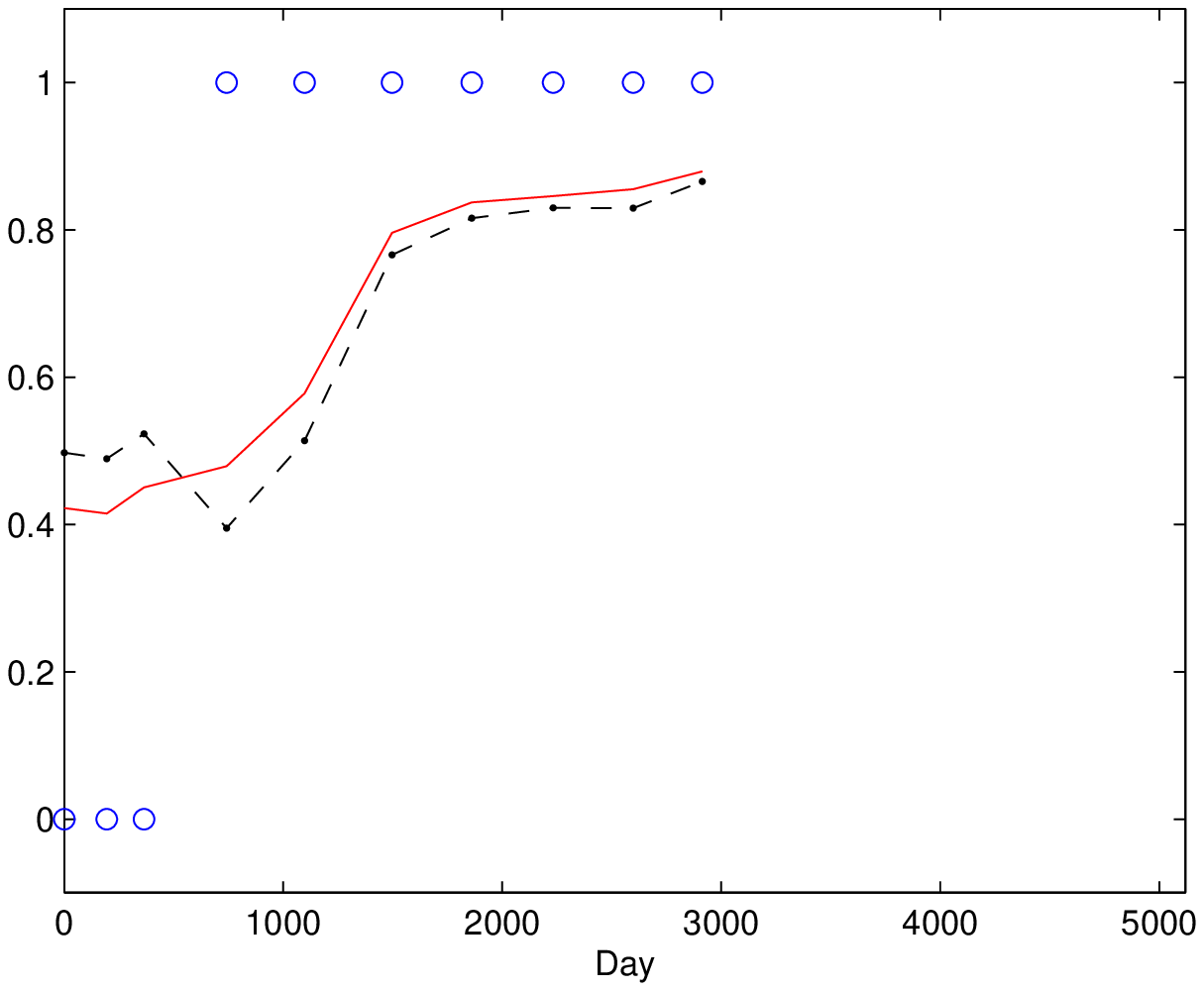}}  \\
\subfigure[\label{fig21}]{\includegraphics[width=0.33\linewidth]{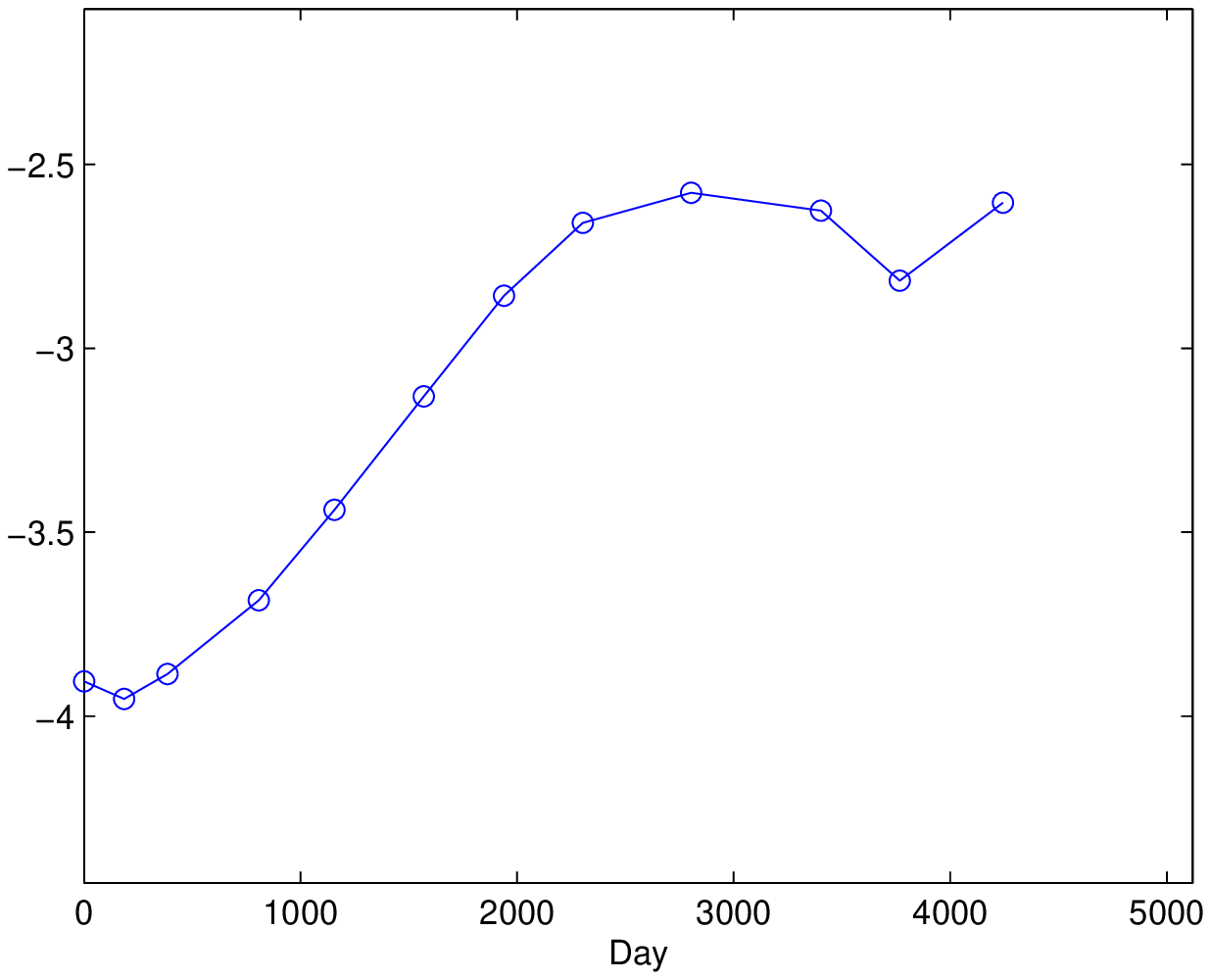}} 
\subfigure[\label{fig31}]{\includegraphics[width=0.33\linewidth]{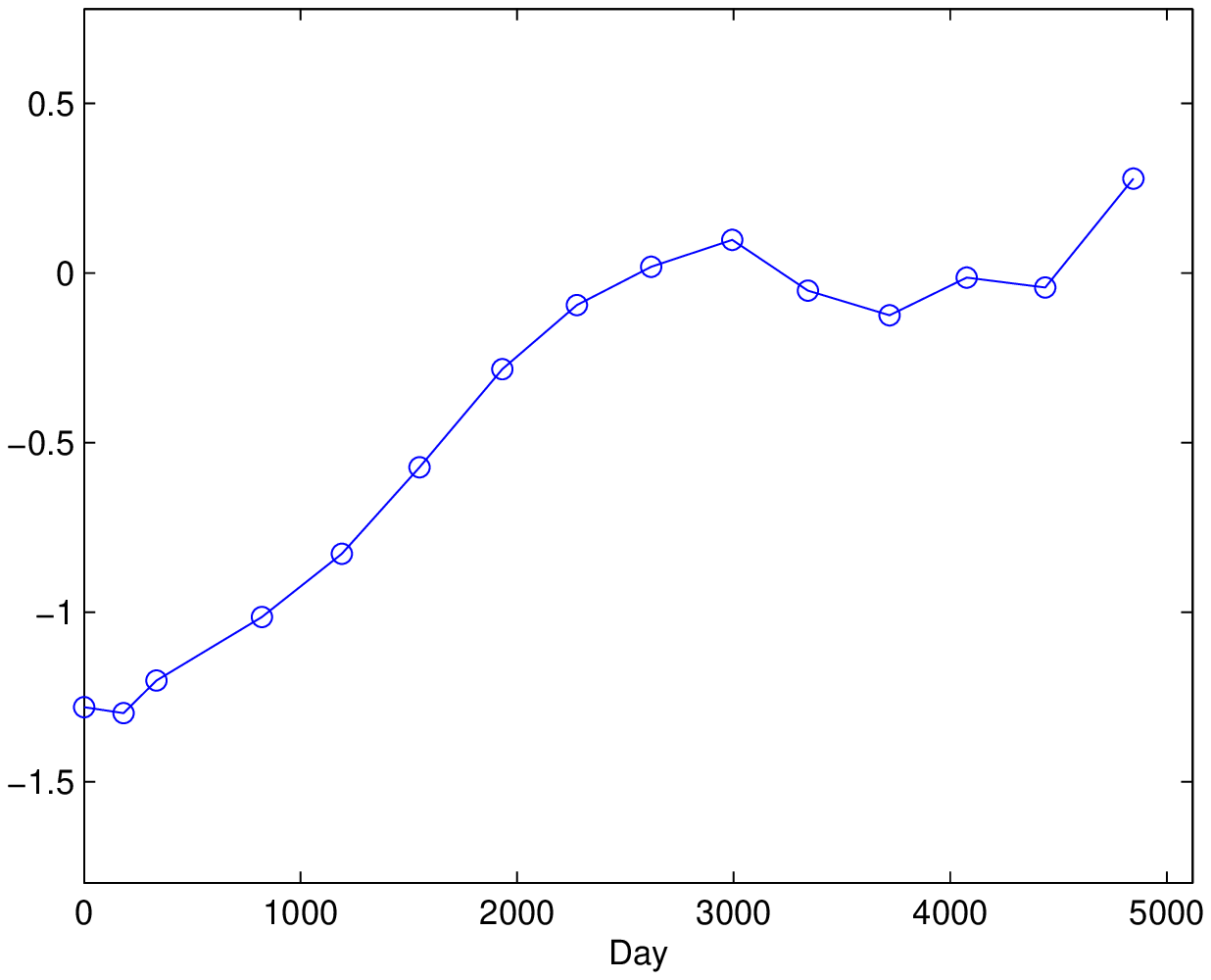}} 
\subfigure[\label{fig41}]{\includegraphics[width=0.33\linewidth]{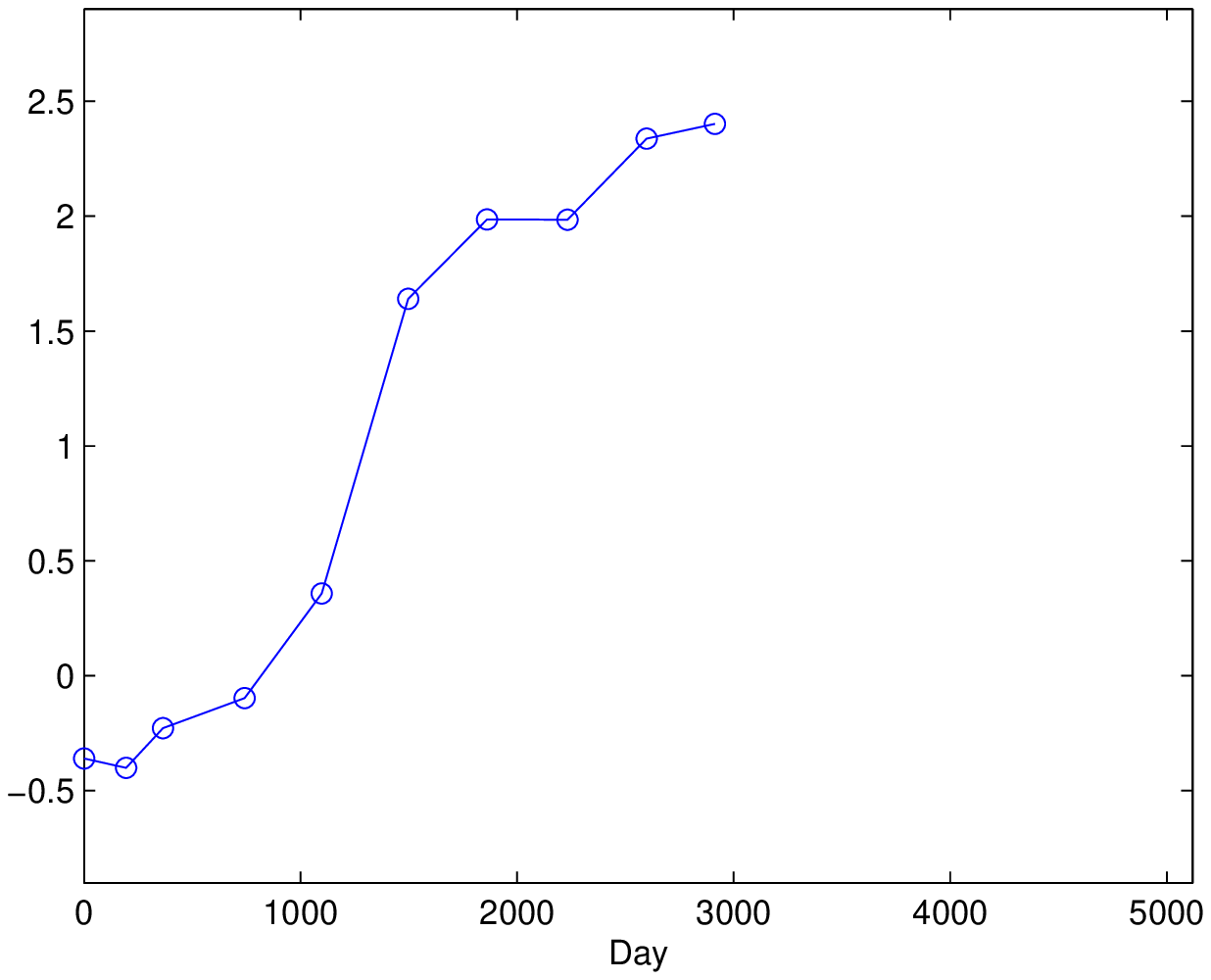}}
\end{center}
\caption{PBC data. (a): The estimated common mean curve $\hat \beta(t)$; (b), (c), (d): the observed responses $z_{mi}$ (circles), predicted trajectories $\hat \pi_m(t)$ obtained from complete data (solid lines) and leave-one-point-out predicted values (dashed lines) for three randomly selected patients; (e), (f), (g): the corresponding underlying processes $\hat y_{m}(t)$.}
\label{pbc_patients}
\end{figure}

To demonstrate the usefulness of our methods, in this example we restrict the analysis to the patients who survived at least 3 years (1095 days) since they entered the study and were alive and had not had a transplant at the end of the 3rd year, and for whom no data were missing. As a result, 185 patients with a total of 1334 observations were obtained. We investigate the dynamic behaviour of the presence of hepatomegaly (0=no, 1=yes), which is a longitudinally measured Bernoulli variable with sparse and irregular measurements. As considered in \citet{Murtaugh94}, four longitudinal measurements (the number of days since enrollment, serum bilirubin in mg/dl, albumin in gm/dl, and prothrombin time in seconds) are used as input variables. We use a GGPFR model for binomial distribution with logit link to deal with these data. Although the covariate $\ve{x}(t)$ in this example is four-dimensional, the procedure is the same as the one considered in Section 4.1. The estimated mean curve for latent Gaussian process is given in Figure \ref{pbc_patients}(a). Figures \ref{pbc_patients} (b)-(g) present the predicted trajectories $\hat \pi_m(t)$ obtained from the complete data, the leave-one-point-out predicted values as well as the patient-specific underlying processes $\hat y_m(t)$ for three randomly selected patients. These predicted trajectories describe the dynamic relationship between the probability of the presence of hepatomegaly and the covariates over time, and reasonably coincide with the observed longitudinal binary responses. We note that the estimate of $\hat y_m(t)$ for each individual patient is quite different to the common mean curve, which is the evidence that the GGPFR model can cope with individual characteristics.

\section{GGPFR for clustered functional data}

Let $\{z_{ij}(t), t\in \CT\}$ be a functional or longitudinal response variable for the $j$-th subject in the $i$-th cluster for $i=1,\ldots,N$ and $j=1,\ldots,N_i$. We assume that $z_{ij}(t)$'s are independent for different clusters,
but dependent within clusters. $z_{ij}(t)$ has a distribution as given by \eqref{exfamily}.

We define the following mixed effect GGPFR (ME-GGPFR)
\begin{eqnarray*}
\E(z_{ij}(t)|y_{ij}(t))& =& h(y_{ij}(t)),  \\
 y_{ij}(t) &=& \vess{u}{ij}{T} \ve{\beta}(t) + \vess{w}{ij}{T}(t)\vesub{v}{i} + \tau_{ij}(t),  \\
\tau_{ij}(t)= \tau_{ij}(\vesub{x}{ij}(t)) & \sim & GPR(0, k(\cdot,\cdot;\ve \theta)|\vesub{x}{ij}(t)),
\end{eqnarray*}
where, $\vesub{w}{ij}(t)$ is a $r$-dimensional vector of functional covariates,
$\vesub{v}{1},\ldots,\vesub{v}{N}$ are i.i.d $N(0,\Gamma)$, and the others are defined
similarly as in Section 2. Hence the unobserved latent variable
$y_{ij}(t)$ consists of three parts: the first term represents the overall mean, the second the random cluster effect, and the third the variation of the individual curves. For simplicity and without loss of generality, 
we assume $\Gamma = \diag (\gamma_1,\ldots,\gamma_r)$.

Now suppose that we have $N_{ij}$ observations in the $j$-th subject of the $i$-th cluster, collected at
$\vesub{T}{ij}=\{t_{ij1},\ldots,t_{ijN_{ij}}\}$. We denote $z_{ij}(t_{ijk})$, $y_{ij}(t_{ijk})$,
$\tau_{ij}(t_{ijk})$, $\ve{x}_{ij}(t_{ijk})$ and $\ve{w}_{ij}(t_{ijk})$ by $z_{ijk}$, $y_{ijk}$, $\tau_{ijk}$,
$\ve{x}_{ijk}$ and $\ve{w}_{ijk}$, respectively, for $k=1,\ldots,N_{ij}$, $j=1,\ldots,N_{i}$ and
$i=1,\ldots,N$. Then the discrete form of the model is
\begin{eqnarray}
z_{ijk}|\alpha_{ijk}, \phi, y_{ijk} & \sim &  \ EF(\alpha_{ijk}, \phi),  \label{cluster_EF1}\\
\E(z_{ijk}|y_{ijk}) &=& b'(\alpha_{ijk}) = h(y_{ijk}),  \label{cluster_EF2} \\
y_{ijk} &=& \vess{u}{ij}{T} \ve{\beta}(t_{ijk}) + \vess{w}{ijk}{T}\vesub{v}{i} + \tau_{ijk} .  \nonumber
\end{eqnarray}

Let
$\vesub{z}{ij}=(z_{ij1},\ldots,z_{ijN_{ij}})^T$, $\vesub{y}{ij}=(y_{ij1},\ldots,y_{ijN_{ij}})^T$,
$\ve{\beta}(t_{ij}) =(\ve{\beta}(t_{ij1}),\ldots,\ve{\beta}(t_{ijN_{ij}}))^T$,
$\vesub{w}{ij}=(\ve{w}_{ij1},\ldots,\ve{w}_{ijN_{ij}})^T$, 
$\vesub{x}{ij}=(\ve{x}_{ij1},\ldots,\ve{x}_{ijN_{ij}})^T$, and
$\ve{\tau}_{ij}=(\tau_{ij1},\ldots,\tau_{ijN_{ij}})^T$, then the latent variable can be written as
\begin{equation*}
\vesub{y}{ij} = \ve{\beta}(t_{ij})\vesub{u}{ij} + \vesub{w}{ij}\vesub{v}{i} + \ve{\tau}_{ij},
\end{equation*}
where the $N_{ij}$-dimensional random vector $\ve{\tau}_{ij} \ \sim \ N(0,\vesub{C}{ij})$ and the elements of $\vesub{C}{ij}$ are given by \eqref{cmij} if we use covariance kernel (\ref{covfun0}). 

As discussed in Section 2, the functional coefficient $\ve{\beta}(t)$ is approximated by B-spline so that
$\ve{\beta}(t)=\vesup{B}{T} \ve{\Phi}(t)$. Thus, at the observation  
point $\vesub{T}{ij}$, we have
$\ve{\beta}(t_{ij}) = \vesub{\Phi}{ij}\ve{B}$, 
where $\vesub{\Phi}{ij}$ is an $N_{ij} \times D$ matrix with the $(k,d)$-th
element $\Phi_d(t_{ijk})$. 

Denote $\vesub{Z}{i}=(\vess{z}{i1}{T},\ldots,\vess{z}{iN_i}{T})^T$
 and $\ve{Z}=\{\vesub{Z}{1},\ldots,\vesub{Z}{N}\}$,
and define $\vesub{Y}{i}$, $\ve{Y}$, $\ve{\tau}_{i}$, $\ve{\tau}$, $\vesub{X}{i}$, $\ve{X}$, 
$\vesub{W}{i}$, $\ve{W}$ in the similar way, then the model for $\vesub{y}{ij}$ can be collectively 
written as
\begin{equation*}
\vesub{Y}{i} = \vesub{\Phi}{i}\ve{\tilde{B}}\vesub{U}{i} + \vesub{W}{i}\vesub{v}{i} + \ve{\tau}_{i},
\end{equation*}
where $\vesub{\Phi}{i}=\diag(\vesub{\Phi}{i1}, \ldots, \vesub{\Phi}{iN_i})$,
$\ve{\tilde{B}}= \diag(\ve{B},\ldots,\ve{B})$, $\vesub{U}{i}=(\vess{u}{i1}{T},\ldots,\vess{u}{iN_i}{T})^T$.
As $\tau_{ij}(t)$'s are independent random samples from $GP(0, k(\cdot,\cdot;\ve \theta))$, 
$\ve{\tau}_{ij}$'s are independently normal and it follows that $\ve{\tau}_{i} \ \sim \ N(0,\vesub{\tilde{C}}{i})$ 
with $\vesub{\tilde{C}}{i} = \diag(\vesub{C}{i1},\ldots,\vesub{C}{iN_i})$. 
Define $\ve{\tilde\tau}_{i} = \vesub{W}{i}\vesub{v}{i} + \ve{\tau}_{i}$, then
$$\vesub{\tilde\tau}{i} \sim N(0,\Sigma_{i}), 
\quad \Sigma_i = \vesub{W}{i}\Gamma\vess{W}{i}{T} + \vesub{\tilde{C}}{i}. $$

The marginal density of $\ve{Z}$ is therefore given by
\begin{eqnarray*}
p(\ve Z|\ve{B}, \Gamma, \ve{\theta}, \ve X, \ve W)&=&\prod^N_{i=1}p(\vesub{Z}{i}|\ve{B},\Gamma,\ve{\theta},\vesub{X}{i},\vesub{W}{i}) \nonumber \\
&=& \prod^N_{i=1}\int p(\vesub{Z}{i}|\ve{Y}_i)p(\ve{Y}_i|\ve{B},\ve{\theta},\Gamma,\vesub{X}{i},\vesub{W}{i})d\ve{Y}_i   \nonumber \\
&=& \prod^N_{i=1}\int p(\vesub{Z}{i}|\ve{\tilde\tau}_i,\ve{B})p(\ve{\tilde\tau}_i|\ve{\theta},\Gamma,\vesub{X}{i},\vesub{W}{i})d\ve{\tilde\tau}_i   \nonumber 
\end{eqnarray*}
and the log-likelihood is 
\begin{align*}
& l(\ve{B},\Gamma,\ve{\theta}) = \sum^N_{i=1}\log\{p(\vesub{Z}{i}|\ve{B},\Gamma,\ve{\theta},\vesub{X}{i},\vesub{W}{i})\} \nonumber \\
&= \sum^N_{i=1}\log\int p(\vesub{Z}{i}|\ve{\tilde\tau}_i,\ve{B})p(\ve{\tilde\tau}_i|\ve{\theta},\Gamma,\vesub{X}{i},\vesub{W}{i})d\ve{\tilde\tau}_i   \nonumber
\\
&= \sum^N_{i=1}\log \int \Big\{ \prod^{\tilde N_i}_{l=1}p(\vess{Z}{i}{(l)}|\ve{\tilde\tau}^{(l)}_i,\ve{B})\Big\} 
(2\pi)^{-\frac{\tilde N_i}{2}}|\Sigma_i|^{-\frac 1 2}\exp\Big\{-\half\ve{\tilde\tau}^T_i\Sigma_i^{-1}\ve{\tilde\tau}_i\Big\}
d\ve{\tilde\tau}_i,
\end{align*}
where $\tilde N_i=\sum^{N_i}_{j=1}N_{ij}$, and $\vess{Z}{i}{(l)}$ and $\ve{\tilde\tau}^{(l)}_i$
are the (scalar) elements of $\vesub{Z}{i}$ and $\ve{\tilde\tau}_i$ respectively. 
The conditional distribution $p(\vess{Z}{i}{(l)}|\ve{\tilde\tau}^{(l)}_i,\ve{B})$ is derived from the exponential family as defined in \eqref{cluster_EF1} and \eqref{cluster_EF2}. The above log-likelihood function is similar to \eqref{eq:loglike} except that
the latent process now becomes a long curve by joining all the curves in the same cluster together, therefore the estimation
of the parameters and the prediction can be carried out in the same way as described in Section 2.

The above model is applied to the paraplegia data discussed in Section 4, where each patient is treated as a cluster. 
The same response and input variables are used and the random effect covariates $\vesub{w}{ij}(t)$ are the same as 
$\vesub{x}{ij}(t)$. We randomly select 4 standing-ups from each of 7 patients as training data and the remaining ones are used
for prediction. Both interpolation and extrapolation problems are conducted after the empirical Bayesian estimates are obtained, 
and the same dataset is also analyzed using the method described in Section 4 for comparison. The above experiment is repeated 
five times and the average error rates are reported in Table \ref{para_ME_GGPFR}. It can be seen that the mixed effect GGPFR
which takes the cluster effect into account 
outperforms the GGPFR method where all curves are regarded as independent samples, especially for interpolation problem.

\begin{table}
\begin{center}
\caption{The average error rates (\%) by mixed-effect GGPFR and GGPFR for paraplegia data}\label{para_ME_GGPFR}
\begin{tabular}{c c | c c}
\hline\hline 
 \multicolumn{2}{c|}{ME-GGPFR} & \multicolumn{2}{c}{GGPFR} \\ 
\hline 
 Interpolation & Extrapolation & Interpolation & Extrapolation \\ 
\hline 
 5.31 & 20.12  & 14.99 & 23.57 \\ 
\hline 
\end{tabular}
\end{center}
\end{table}


\begin{figure}[h]
\begin{center}
\subfigure[\label{OP_fig_SE} SE]{\includegraphics[width=0.43\linewidth]{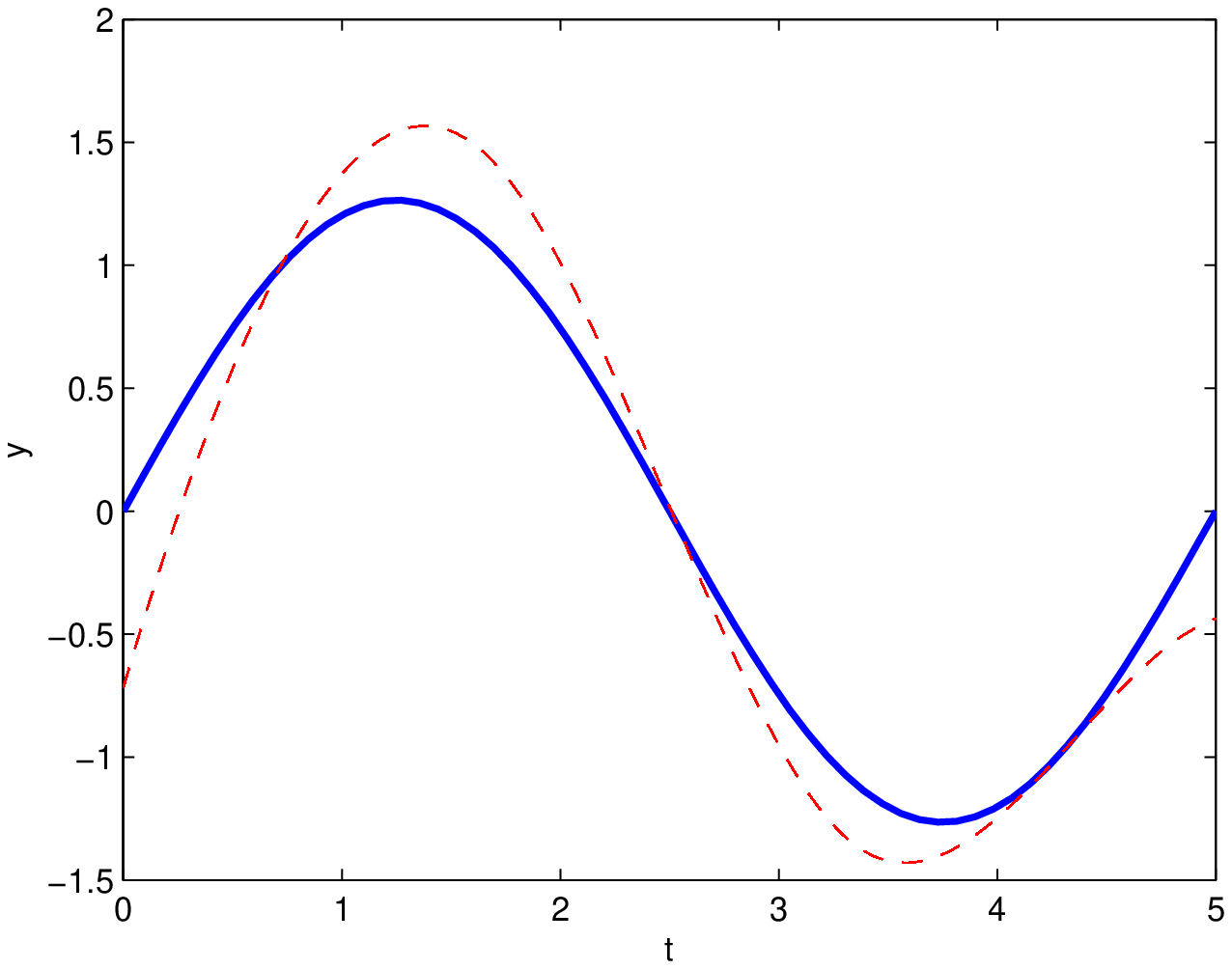}}
\subfigure[\label{OP_fig_MC} MC]{\includegraphics[width=0.43\linewidth]{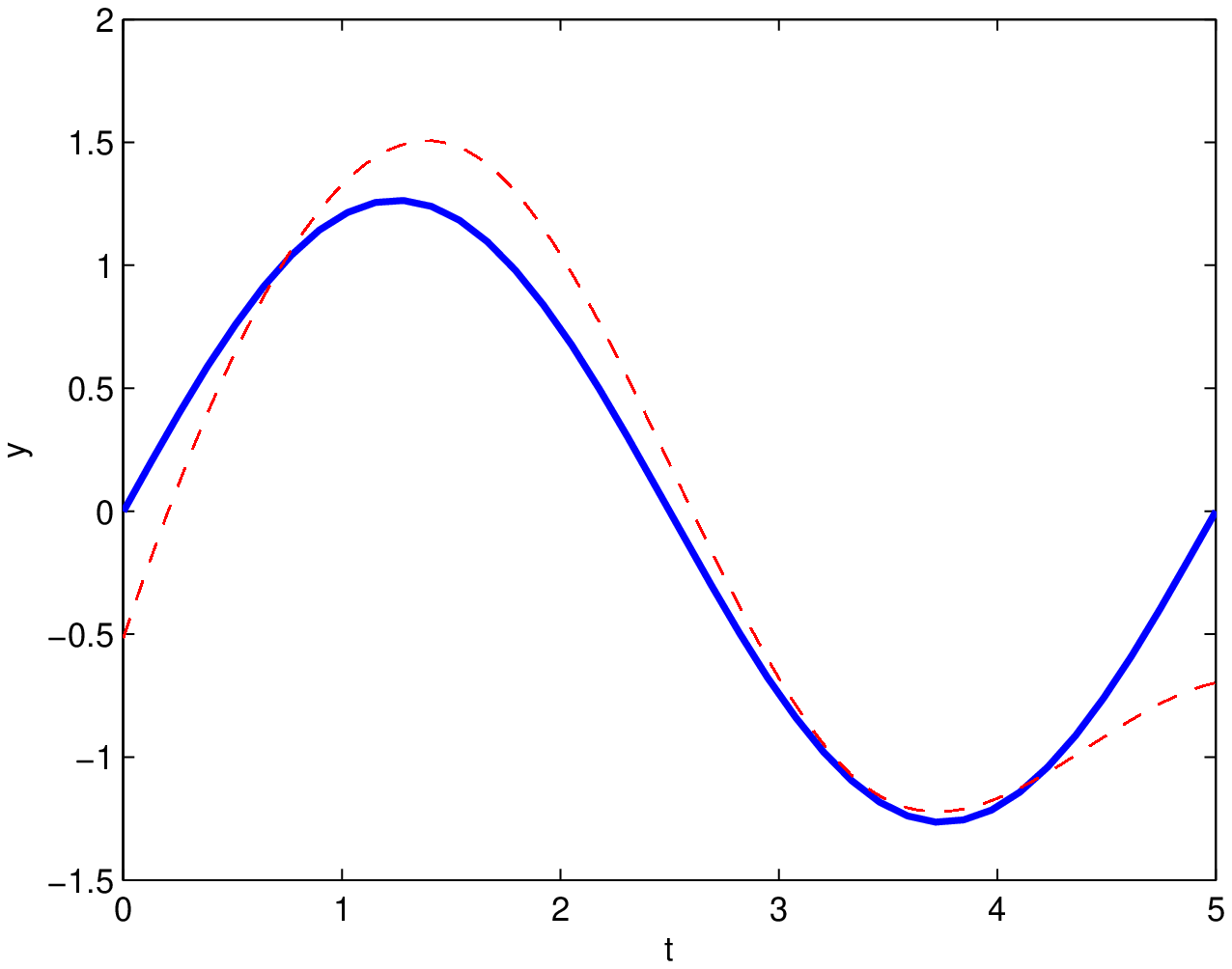}} \\
\subfigure[\label{OP_fig_RQ} RQ]{\includegraphics[width=0.43\linewidth]{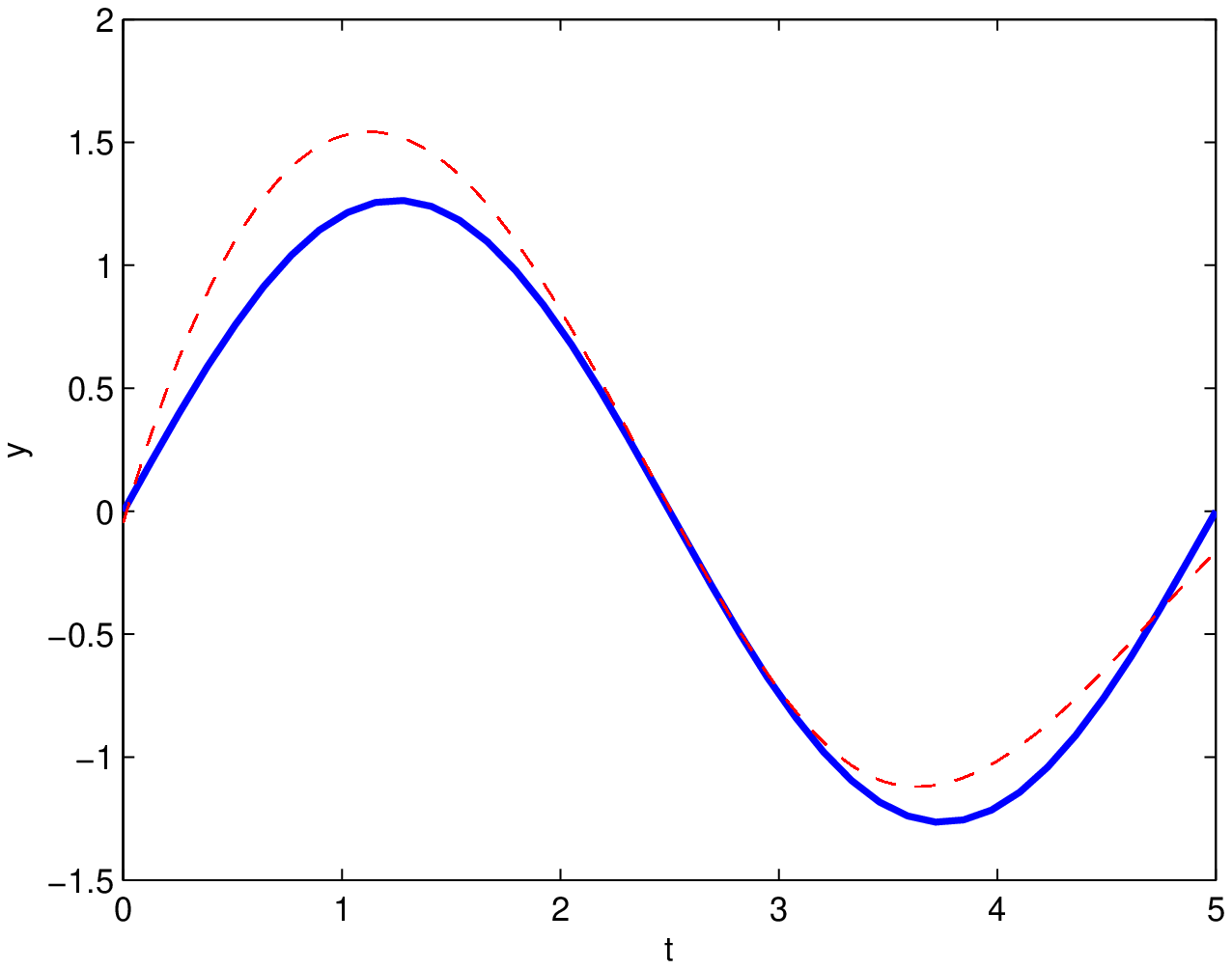}} 
\subfigure[\label{OP_fig_PP} PP]{\includegraphics[width=0.43\linewidth]{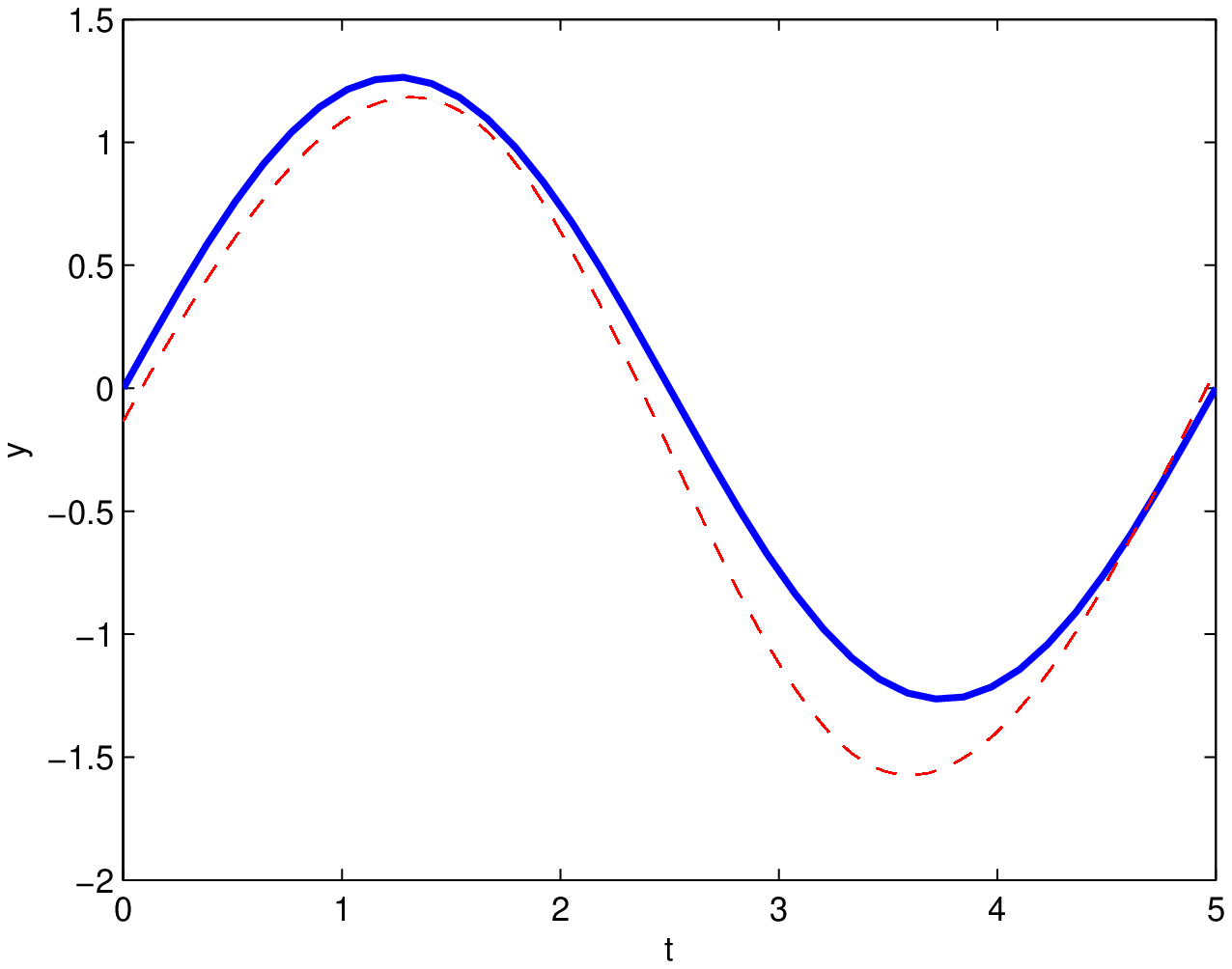}} \\
\subfigure[\label{OP_fig_NP} NP]{\includegraphics[width=0.43\linewidth]{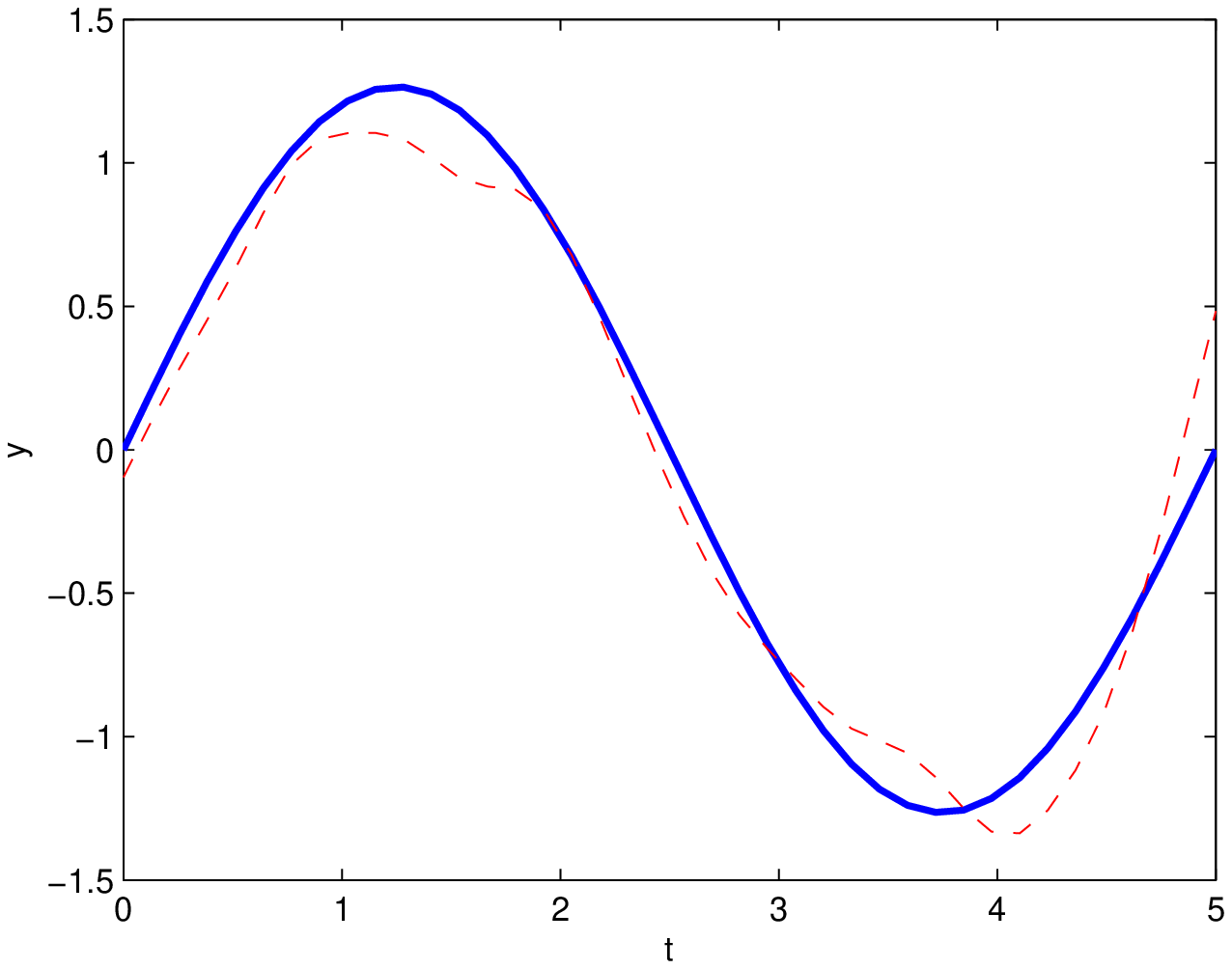}}
\end{center}
\caption{The estimated mean curves (dashed line) by different covariance functions and nonparametric covariance method for the data with Chebyshev polynomials. The solid lines are the true mean curve.}
\label{simu_mean_curves_OP}
\end{figure}

\begin{figure} 
\begin{center}
\subfigure[\label{Para_fig11}Interpolation]{\includegraphics[width=0.46\linewidth]{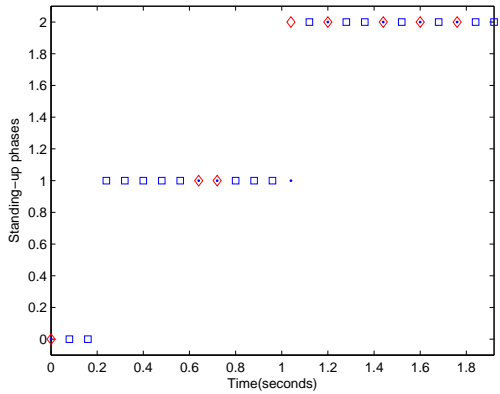}}
\subfigure[\label{Para_fig12}Interpolation]{\includegraphics[width=0.46\linewidth]{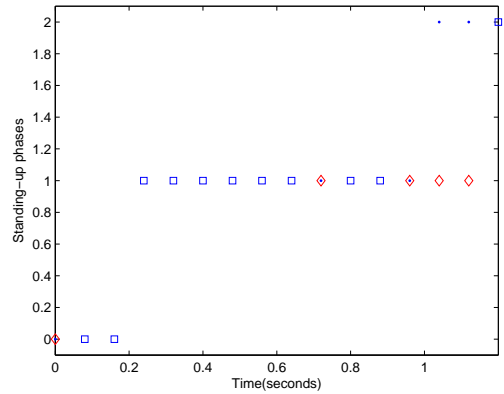}} \\
\subfigure[\label{Para_fig13}Extrapolation]{\includegraphics[width=0.46\linewidth]{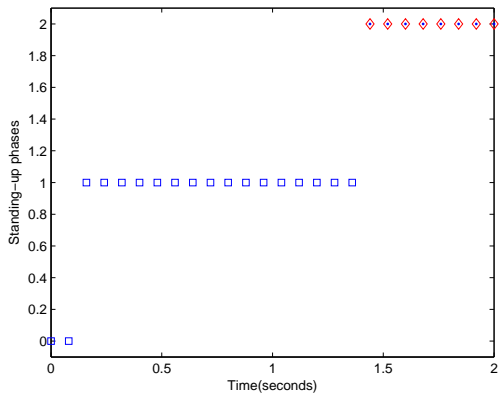}}
\subfigure[\label{Para_fig14}Extrapolation]{\includegraphics[width=0.46\linewidth]{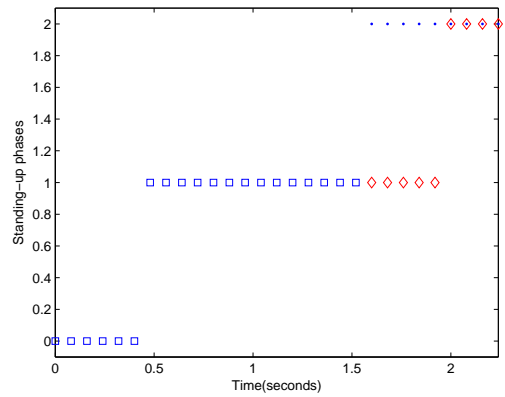}} \\
\subfigure[\label{figbetat}]{\includegraphics[width=0.46\linewidth]{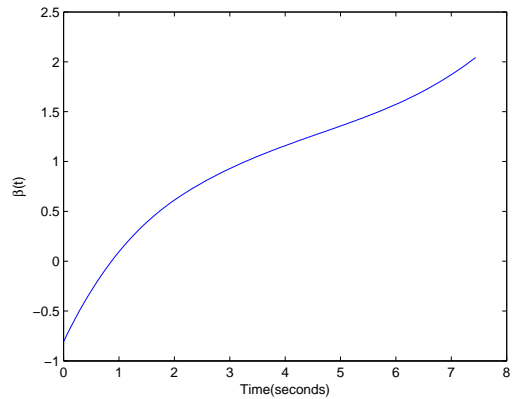}}
\end{center}
\caption{Paraplegia Data. (a)-(d): Randomly selected standing-ups and their predictions by interpolation and extrapolation, two standing-ups for each. The squares are observations, the diamonds are predicted responses and the points are the actual response values. (e): The estimated functional coefficient $\hat \beta (t)$.}
\label{fig_para_output}
\end{figure}

\end{document}